# A plug-in energy-balance freezing module for transformer frost nowcasting without leaf-sensor inputs

**Authors:** Juyeon Kim[b,c,†]; Chaeyeon Yi[a,b,†]; Yun Am Seo[b,c,*]

† These authors contributed equally to this work (co-first authors). * Corresponding author.

**Affiliations:**
[a] Research Center for Atmospheric Environment, Hankuk University of Foreign Studies, 81 Oedae-ro, Mohyeon-eup, Cheoin-gu, Yongin-si, Gyeonggi-do 17035, Republic of Korea
[b] NAVI Hyper-Tropicalization Research Institute, Jeju National University, 102 Jejudaehak-ro, Jeju-si, Jeju-do 63243, Republic of Korea
[c] Department of Data Science, Jeju National University, 102 Jejudaehak-ro, Jeju-si, Jeju-do 63243, Republic of Korea

**ORCID:** Juyeon Kim 0009-0007-8147-4994; Chaeyeon Yi 0000-0003-2802-2431; Yun Am Seo 0000-0001-9283-4376

**E-mail:** Juyeon Kim, kj7951511@stu.jejunu.ac.kr; Chaeyeon Yi, prpr2222@hufs.ac.kr; Yun Am Seo, seoya@jejunu.ac.kr

**Corresponding author:** Yun Am Seo, Department of Data Science / NAVI Hyper-Tropicalization Research Institute, Jeju National University, 102 Jejudaehak-ro, Jeju-si, Jeju-do 63243, Republic of Korea. **Telephone:** +82-64-754-3592. **Fax:** +82-64-725-2579. **E-mail:** seoya@jejunu.ac.kr. **ORCID:** 0000-0001-9283-4376

## Abstract

Frost protection requires local warnings where leaf-sensor inputs are unavailable. In the frost literature we surveyed, physics entered machine-learning models through inputs, numerical-model coupling, or losses, rather than model structure. We develop a transformer-compatible module that embeds leaf energy-balance physics in freezing-probability forecasts. An energy-balance temperature reference and a learned residual predict leaf temperature, which drives a multiplicative gate imposing a temperature-dependent probability cap. We mathematically prove conditional monotonicity and an alarm boundary, and test whether placing the same energy-balance information in the output pathway or in attention yields different benefits. Using observations from 19 Korean stations over five frost seasons, we evaluated forecasts for the next 12 hourly intervals with and without leaf-sensor inputs. In the 2025–2026 hold-out season without these inputs, the proposed time-token transformer (PET-E) achieved a lead-mean critical success index (CSI) of 0.516, compared with 0.478 for fixed-configuration, joint-lead LightGBM, with significant gains at all 12 leads. Relative to the unmodified backbone, the fraction of all alarms issued for intervals with an observed minimum leaf temperature ≥ 2 °C decreased from 0.087 to 0.051. Across the hold-out season and a retrospective 2024–2025 evaluation, under both sensor conditions, the module reduced Brier scores and warm-surface alarm fractions in both the time-token transformer and iTransformer, with no statistically significant CSI decrease at any lead. At held-out stations without static or leaf-sensor inputs, both module-equipped backbones exceeded LightGBM at all 12 leads (CSI 0.468 and 0.467 versus 0.437). In the time-token backbone, adding the temperature pathway beyond the gate alone reduced this alarm fraction in all four settings, whereas adding the attention prior increased CSI in three. These findings link an explicit temperature-dependent probability constraint to improved forecast performance and distinguish the effects of output-side and attention-based embedding.

---

[1] **Abbreviations:** AAFOS, Automated Agricultural Frost Observation System; AP, average precision; AUC, area under the receiver operating characteristic curve; BASE, same-capacity transformer without physics mechanisms; BSS, Brier skill score; CSI, critical success index; GBM, gradient boosting model (LightGBM); GIS, geographic information system; iT-EB, iTransformer with the energy-balance freezing module; KMA, Korea Meteorological Administration; LOSO, leave-one-station-out; LSTM, long short-term memory; MLP, multilayer perceptron; NWP, numerical weather prediction; PET-A, physics-embedded transformer with an additive gate; PET-E, physics-embedded transformer with the energy-balance freezing module; PET-M, physics-embedded transformer with a multiplicative gate and a learned attention favorability; POD, probability of detection; REV, relative economic value.

# 1. Introduction

## 1.1 The need for station-level frost warning

Frost-related freezing can cause irreversible damage to flowering-stage crops after only brief exposure (Snyder and de Melo-Abreu, 2005). In Korea, after the spring frost of 2023, the fruit set of apple and pear fell from the previous year by 16.5% and 31.8%, respectively, and in the first quarter of 2024 apple and pear prices were 109% and 148% higher than a year earlier (Kim and Chae, 2024). Warming has not removed the risk: the number of frost days has not decreased consistently across regions (Kim and Chae, 2024; Chun et al., 2021), and earlier flowering increases the chance of overlap with low-temperature exposure (Kim and Chae, 2024).

Frost-protection measures such as sprinkling, wind machines, and covering (Drepper et al., 2022) must be prepared and activated in time according to orchard temperature and humidity (Snyder and de Melo-Abreu, 2005), and false alarms also incur cost. The value of a warning therefore depends on the lead time, on station-level accuracy, and on timely detection of the transition into freezing (onset) and out of it (cessation). In two orchard frost episodes, Weather Research and Forecasting (WRF) model forecasts placed minimum temperatures on the same side of a threshold as observations in most comparisons (Prabha and Hoogenboom, 2008). However, temperatures on a frost night differ within a few hundred meters and cold air collects in low spots (Snyder and de Melo-Abreu, 2005), so gridded products are difficult to use directly for station-level warning. Data-driven nowcasting from station observations can capture these local differences. Leaf sensors, however, cannot be installed at every growing site, and equipped stations lose data during sensor failure and maintenance; a practical warning system must therefore also perform without leaf-sensor inputs.

## 1.2 The physics of frost — the structure of the prediction problem

At night the surface loses heat by longwave radiation. On clear nights with weak wind, turbulent heat exchange is suppressed and the leaf surface temperature can fall several °C below the air temperature (Jordan and Smith, 1994; Snyder and de Melo-Abreu, 2005). The depression is approximated by the balance between the net longwave radiation lost by the leaf and the sensible heat it receives from the air, and its magnitude is set by the sky emissivity (water vapor and clouds), the wind speed, and the leaf size. Surface water can then freeze, but not as a step at exactly 0 °C: water can remain supercooled (Rogers and Yau, 1989), and the supercooling of plant parts varies with the concentration of ice-nucleation-active bacteria (Snyder and de Melo-Abreu, 2005). We therefore treat freezing as a soft threshold whose probability increases as the temperature falls, and model radiative frost as (i) the accumulation of conditions favorable to radiative cooling (clear sky, weak wind, and night) and (ii), when surface water is present, a freezing propensity with a soft temperature threshold.

A frost prediction model should accordingly (a) identify times when weather conditions favor radiative cooling and (b) represent the threshold behavior in which the probability changes sharply near the surface freezing temperature. Both are functions of observed variables, so frost physics can be built into the architecture instead of being imposed as a penalty in the loss function. We implement (a) as an energy-balance depression that enters the reference for the predicted leaf temperature and the attention prior, and (b) as the output gate (§3.5).

## 1.3 Three gaps in existing approaches

Machine-learning methods applied to frost prediction include station-based classification for Internet of Things (IoT) warning systems (Diedrichs et al., 2018), causality-oriented modeling (Ding et al., 2019), multi-lead-time comparison of deep networks and random forests (Talsma et al., 2023), post-processing of numerical-model output (Rozante et al., 2023), a voting ensemble (de Almeida et al., 2025), a graph neural network with spatiotemporal attention (Lira et al., 2022), and satellite-image-based spatialization (Kang et al., 2025); Korean studies have used the national frost observation record as labels (Chun et al., 2021; Kim and Kim, 2022; Kim et al., 2017, 2024a). Three gaps remain in the literature we surveyed (including the review of 253 papers by Hua et al., 2025).

**First, frost physics has not been embedded in the model structure.** Physics entered through input-variable selection (Diedrichs et al., 2018; Talsma et al., 2023), serial coupling with numerical models (Rozante et al., 2023), or task-derived loss terms such as a daily-minimum air-temperature error (Barooni et al., 2023), and attention weights, where used, were learned from data alone (Lira et al., 2022). In the taxonomy of physics-informed machine learning (Karniadakis et al., 2021), these are observational and learning biases; **inductive-bias** studies that impose physics on the output activation form or the attention weighting were not found, although structural embedding has shown gains in adjacent fields, for example an advection operator in precipitation nowcasting (Zhang et al., 2023), a monotonicity-preserving lake temperature model (Daw et al., 2020), and an attention bias in subseasonal forecasting (Lyu et al., 2026).

**Second, verification against strong baselines is lacking.** Because a frozen surface is likely to remain frozen in the next hour, even persistence, a no-learning baseline that keeps the current state, scores highly at short leads, yet skill over persistence was rarely reported; Pérez Tárraga et al. (2025), for example, reported an area under the receiver operating characteristic curve (AUC) of 0.99 from 10-fold cross-validation without a persistence comparison. Deep models were rarely compared, under identical inputs and lead times, with gradient boosting, which exploits state persistence from flattened features; validation was mainly temporal extrapolation at training stations, and few studies tested models at stations without leaf sensors or at stations absent from the network, although station dependence has been noted as a limitation (Talsma et al., 2023).

**Third, existing labels did not reflect the temperature and moisture state of the leaf surface.** Labels relied on screen-level air-temperature thresholds (e.g., minimum air temperature ≤ 0 °C) or visual frost records; a recent Korean deep model used the visual records of 24 synoptic stations of the Korea Meteorological Administration (KMA) and excluded ground and grass-minimum temperatures because they are not provided as forecasts (Kim et al., 2024a). Frost, however, is a leaf-surface phenomenon, and the leaf surface falls up to several °C below the air temperature under radiative cooling (Jordan and Smith, 1994); no study was found that derived labels from a leaf-wetness sensor and an infrared leaf-temperature sensor.

## 1.4 Proposal and contributions

We develop a transformer-compatible **physics module** that connects leaf energy-balance physics directly to the structure of freezing-probability forecasts. It combines an energy-balance temperature reference, a learned residual, and a multiplicative gate with an explicit temperature-dependent probability bound. Controlled comparisons on two transformer backbones distinguish the effects of this output-side construction from those of placing the same physical prior in attention. The weather inputs come only from an observation-based objective analysis grid, the station leaf-sensor histories are used only in the sensor-equipped condition, and no products of numerical weather prediction (NWP) models are used. The physics-based freezing label is defined from observations at 19 stations over five frost seasons (2021–2026) of the Automated Agricultural Frost Observation System (AAFOS; Kim et al., 2024b) of the National Institute of Meteorological Sciences of the KMA. The contributions are as follows.

**(1) Physics module.** (i) The **energy-balance leaf-temperature path** computes, from air temperature, dew point, wind speed, and relative humidity, the clear-sky longwave emissivity (Brutsaert, 1975), its increase due to clouds (Crawford and Duchon, 1999), and the sensible heat exchange of the leaf (Schuepp, 1993), giving the leaf–air temperature depression $D$ in °C; the predicted leaf temperature is the air temperature minus $D$ plus a learned residual (§3.5), and no leaf sensor is required. (ii) The **bounded freezing gate** passes the predicted leaf temperature through a learned soft threshold, $P = \sigma(u)\,\sigma(-\hat{T}/\delta)$, so that the probability cannot exceed $\sigma(-\hat{T}/\delta)$ whatever the data path outputs (Proposition 4). Combined, with the residual and the other arguments fixed, the air-temperature bound for an alarm rises by the depression, which is non-decreasing toward weaker wind, drier air, and faster cooling (Proposition 5); the structure thus allows an alarm on clear nights with weak wind even above 0 °C air temperature. The same depression is also tested as a key-wise attention prior in the time-token backbone (ATT and FULL).

**(2) Attachability and the placement of physics.** The module is attached to a time-token transformer (PET-E) and to the variate-token iTransformer (Liu et al., 2024), forming iT-EB, and each is compared with its plain form under the same conditions. The gate, the energy-balance path, and the attention prior are added one at a time and compared with variants that put a proxy favorability (learned or direction-constrained) in attention, to determine at which position the same physics takes effect.

**(3) A physics-based label and dataset.** The freezing label combines moisture detection by the leaf-wetness sensors with the 0 °C threshold of the infrared leaf temperature, freezing = (wetness > 0) ∧ ($T_{\text{leaf}} \leq 0$ °C); it is consistent across firmware versions that report frozen surface water as different sensor states, and a sensitivity analysis over thresholds of 0/−0.5/−1/−2 °C quantifies the definitional uncertainty due to supercooling (§3.2).

Instrumented multi-station, multi-year leaf-surface networks are rare, and the definition can be applied to other leaf-sensor networks.

**(4) Evaluation that separates state persistence from deployment conditions.** Under identical conditions we compare persistence and network climatology, a two-stage “leaf-temperature regression → threshold” rule on the same backbone, a classical rule ($\hat{T} < 0 \wedge$ wet), logistic regression and gradient boosting (LightGBM) on flattened features, loss-embedding variants, and the external backbone iTransformer. Deployment without leaf sensors is evaluated by excluding the leaf-sensor inputs (sensor-free), another season by a rolling-origin repetition that moves the test window, and new stations by a 19-fold leave-one-station-out (LOSO) evaluation that separates stations and time simultaneously. Onset and cessation are measured separately, and the fraction of alarms issued at above-freezing observed leaf temperature is computed for all models with the same criterion.

**(5) Static terrain inputs and station generalization.** Validation favored retaining static inputs for temporal extrapolation and excluding them for sensor-free new-station evaluation; this distinction is used to evaluate input configurations by deployment setting and the transfer of the physics module (§5.7, §6).

**Research questions.** Q1: Without leaf-sensor inputs, can a model achieve a higher CSI than network climatology, GBM, and iTransformer? Q2: Does the model discriminate the onset and cessation of freezing? Q3: Does the skill hold in another season? Q4: Does the skill hold at stations not used for training, and how do static terrain inputs affect it? Q5: Does the physics module lower the fraction of alarms issued at above-freezing observed leaf temperatures? Q6: Does the physics module act in the same direction in another backbone? Q7: At which position, attention or the leaf-temperature path, does the same physics take effect?

The prediction target is the hourly **surface freezing state**, the physical basis of an agricultural frost “event” but not identical to one; it includes persistent winter freezing and shoulder-season radiative frost. Results are therefore reported separately for onset and cessation samples and for station strata (§5), comparisons with leaf-sensor inputs are also reported by lead because the t+1 interval is dominated by state inertia (§5.1, §6), and the module is judged on four criteria: forecast skill, the agreement of alarms with the observed surface temperature, transfer to another backbone, and transfer to held-out stations. Accordingly, the reported scores characterize surface-freezing forecasts over October–April and do not by themselves establish warning performance during crop-specific flowering periods.

## 1.5 Terminology and notation

Frost refers to the phenomenon and surface freezing to the label of this study. The **physics module** refers to the combination of the energy-balance leaf-temperature path and the bounded freezing gate. The proposed model **PET-E** (physics-embedded transformer, energy balance) is the validation-selected time-token transformer with the physics module attached (TB configuration); **FULL** additionally includes the energy-balance attention prior, and **iT-EB** is iTransformer with the physics module attached. **PET-M** is a comparison model with a multiplicative gate and learned favorability attention, **PET-A** is its additive-gate variant, and **BASE** is the same-capacity transformer without physics mechanisms. Nowcasting here predicts the next 12 hourly intervals, indexed by leads 1–12 (§3.4), and LOSO denotes leave-one-station-out evaluation. CSI and POD denote the critical success index and probability of detection (Schaefer, 1990). AP is the average precision, which summarizes the precision–recall curve. GBM denotes the gradient boosting model (LightGBM) trained on flattened features as a single model for all leads. The condition without leaf-sensor inputs is called sensor-free and the condition with them sensor-equipped. Stations with data in the training period are called trained stations. Stations first observed in the test season, with no training-period data, are called unseen stations. Stations used for neither training nor validation in a LOSO fold are called held-out stations, and results at held-out stations are referred to as performance at new stations. All stations belong to the AAFOS network. The nominal issue timestamp t is the start-time label of the last input hour. The input window closes at t + 1 h, and lead $\ell$ denotes the $\ell$-th hourly target interval after that closure (§3.4). The freezing state at issue time refers to the state in the last completed input hour. Samples whose target differs from the freezing state at issue time are called transition samples; those changing from non-freezing to freezing are onsets, and the reverse are cessations. Inductive bias is an assumption imposed on the model structure before training (Karniadakis et al., 2021). The bounded freezing gate is the multiplicative gate with the physics cap of Proposition 4. The main symbols are defined in §3.5–§3.9: data-path logit $u$, predicted leaf temperature $\hat{T}$, energy-balance depression $D$, learned residual $r$, transition scale $\delta$, attention prior $R_j$, decision threshold $\theta_\ell^*$, isotonic recalibration function $\mathcal{I}$, and cost–loss ratio $\alpha$.

---

# 2. Related Work

## 2.1 Machine learning for frost prediction

Machine-learning studies of frost based on station observations have used methods ranging from classical classifiers to deep neural networks. Diedrichs et al. (2018) predicted next-day minimum temperature and frost from five meteorological stations in Mendoza, Argentina, as part of an IoT frost-warning system for peach orchards, and Ding et al. (2019) used support vector machines to capture possible causal relations between environmental factors and frost, beyond mere association. Talsma et al. (2023) compared deep neural networks, convolutional neural networks, and random forests (RF) over 6–48 h lead times, reported a 6-h minimum-air-temperature root-mean-square error of 1.53–1.72 °C, and explicitly noted that the models are station-dependent and generalize poorly to new stations.

Among approaches coupled to numerical models, Rozante et al. (2023) used neural networks to post-process Eta model forecasts, which then outperformed the operational frost index of the Brazilian National Institute for Space Research (INPE), and de Almeida et al. (2025) added a voting committee that selects between a neural-network forecast and that index. Lira et al. (2022) proposed GRAST-Frost, which applies spatiotemporal attention to a multi-station graph. Satellite-image-based spatialization (Kang et al., 2025) is complementary to station forecasting.

A recent study in this journal (Zhou et al., 2026) classified late-spring frost damage records during apple flowering in China's four major apple-growing regions in 2018 and 2020 from daily meteorological and geographic variables; RF performed best (AUC 0.96), and Shapley additive explanations (SHAP) identified minimum air temperature and latitude as the main factors. That study identified the difference between frost thresholds based on air temperature at a height of 2 m and tissue-level temperature as a source of uncertainty; the label of this study (§3.2) uses leaf-surface measurements rather than a screen-level air-temperature proxy.

Korean studies have compared logistic regression, RF, and LSTM (Chun et al., 2021) and classical machine learning such as RF, support vector machines, and XGBoost (Kim et al., 2017; Kim and Kim, 2022) using the national frost observation record as labels, and a multi-input deep model has also been proposed (Kim et al., 2024a). The literature is extensive (Hua et al., 2025), but the studies described above used physics knowledge for input selection, post-processing, serial coupling, or task-derived loss terms (§1.3).

## 2.2 Physics-informed machine learning — from loss to structure

Physics-informed machine learning is distinguished by where the physics is imposed (Karniadakis et al., 2021). The family of physics-informed neural networks (PINNs; Raissi et al., 2019), which penalizes the residual of a governing equation in the loss, is a representative example of learning bias. Methods that impose constraints directly on the model structure have also been developed. In climate emulation, conservation laws were enforced as an architecture layer (Beucler et al., 2021), and PGA-LSTM preserved the monotonic increase of water density with depth in a lake-temperature LSTM (Daw et al., 2020). NowcastNet placed an advection evolution operator inside the nowcasting network (Zhang et al., 2023; independent evaluation by Das et al., 2024), and ClimODE modeled the advection dynamics as a neural ordinary differential equation (Verma et al., 2024). Other approaches incorporate prior knowledge into attention mechanisms. TelePiT added a teleconnection prior to the attention logits (Lyu et al., 2026). In satellite air-temperature estimation, attention was guided by a land-cover- and distance-based spatial prior while physics was imposed as loss regularization (Bouaziz et al., 2026). In agricultural AI, mechanistic knowledge has also been used to guide feature selection (selection of key factors by grasshopper developmental stage; Su et al., 2026). Additive attention bias has also been used in relative position representations (Shaw et al., 2018).

Within this research area, additive attention biases and soft thresholds are established components; this study contributes the physics they encode, the way that physics is computed, the coupling that caps the output probability, and its application to station frost nowcasting. We compute the leaf–air temperature depression in °C from an energy balance of observed variables, use it to form the energy-balance reference for the predicted leaf temperature, and couple this to a multiplicative freezing gate. The multiplicative gate imposes a physics cap that the data path cannot exceed. The same depression is also tested as an attention prior. Because the physics module connects only to the output of the backbone, it can be attached to other backbones to assess its generality; the attention prior connects to the time-token attention scores. We compare this structural embedding under identical conditions with loss embedding, the two-stage rule, and GBM. In the frost literature we surveyed, neither such an output gate nor structural embedding of frost physics was found. Heat-exchanger frosting, soil freezing, road freezing, and aircraft icing, which share terminology, are excluded from our comparison scope because their target physics and prediction task differ.

## 2.3 Adjacent surface tasks

Leaf-wetness-duration prediction treats the liquid-phase process on the same sensor surface and is an active area of machine-learning application (Solís and Rojas-Herrera, 2021; Alsafadi et al., 2024) but does not address the freezing phase transition. The label and gate of this study address the freezing phase transition of surface moisture.

---

# 3. Materials and Methods

## 3.1 Study area and observations

The whole process from data construction to evaluation is summarized in Figure 1. We used 1-min raw data from 19 AAFOS stations in South Korea, located in Gangwon, the Seoul metropolitan area, Gyeongbuk, and other provinces (August 2021–May 2026, with station-specific start times). Each of these stations has an infrared leaf-temperature sensor (TargetTC) and two leaf-wetness sensors (LWM) and observes the leaf-surface state directly (Kim et al., 2024b). These observations were used to construct the freezing label (§3.2) and the leaf-sensor history features (§3.4). Of the 21 AAFOS stations, we used the 19 equipped with both types of frost sensor (leaf temperature + wetness). To allow for equipment stabilization, we excluded the first 7 days after each station's observation start. Physical outliers (leaf temperature < −50 °C or > 60 °C) were removed. The provenance of the raw files is given in Supplementary Note S6. We did not interpolate across observation gaps (§3.4). Station names and identifiers, coordinates, elevations, sample periods, valid sample counts, and freezing fractions are summarized in Supplementary Table S10.

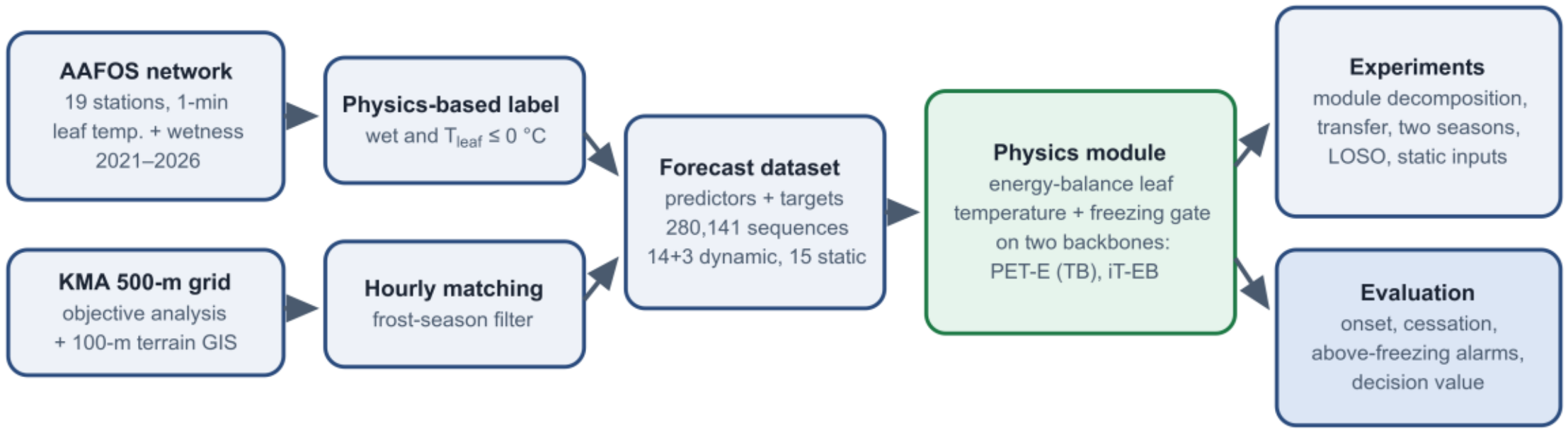

*Figure 1. Research flow. AAFOS station observations (19 stations, 1-min leaf temperature and wetness) and the KMA 500-m objective analysis grid with 100-m terrain GIS are matched hourly*

*and labeled with a physics-based freezing label to form the forecast dataset of predictors and targets. Leaf-sensor observations define the label in all conditions; as inputs, the three leaf-sensor features are added only in the sensor-equipped condition (14 dynamic features otherwise). The physics module (energy-balance leaf-temperature path and multiplicative freezing gate) is attached to a time-token transformer (PET-E) and to iTransformer (iT-EB). Decision thresholds are selected on the validation set.*

## 3.2 Freezing label

The minute-level freezing label was defined by combining leaf-surface moisture detection and the leaf temperature:

$$\mathrm{freezing}_{1\mathrm{min}} = (\mathrm{LWM}_{\mathrm{wet}} > 0 \ \vee \ \mathrm{LWM}_{\mathrm{con}} > 0) \ \wedge \ (T_{\mathrm{leaf}} \leq 0\ ^{\circ}\mathrm{C}). \quad (1)$$

Here $\mathrm{LWM}_{\mathrm{wet}}$ and $\mathrm{LWM}_{\mathrm{con}}$ are the wet-state and con-state (frozen) readings of either of the two leaf-wetness sensors in that minute, and $T_{\mathrm{leaf}}$ is the infrared leaf temperature. This label represents freezing conditions defined from surface moisture and temperature observations; it does not directly confirm ice formation, freezing of plant tissues, or crop damage. For each hour with at least 30 one-minute records, the hourly label was set to 1 if at least 5 min were judged freezing. Because Eq. (1) accepts either the wet or the con state as evidence of surface water ("water on the surface and the surface temperature at or below freezing"), the label does not depend on which of the two states a firmware version reports for frozen water, and it is consistent across past and current observations. The uncertainty of the label boundary due to supercooling was quantified by a sensitivity analysis varying the threshold over 0/−0.5/−1/−2 °C and the minimum number of freezing minutes per hour over 1/5/10, and by the agreement rate with a con-only definition. We analyzed the full-period minute-level records of all 19 stations. Lowering the threshold from 0 °C to −2 °C reduced the hourly incidence from 0.131 to 0.101, with an hourly agreement rate of 0.970 against the reference definition. The agreement with the con-only definition was 0.983 (Table S1). The label does not capture black frost, in which tissue freezes without visible ice under dry conditions (Snyder and de Melo-Abreu, 2005; §6).

## 3.3 Gridded weather and terrain

As weather predictors we used the KMA API Hub's high-resolution blended-weather grid (500-m resolution, produced every 5 min; we used the on-the-hour fields), extracting the nearest grid-cell value for each station at hourly resolution. This dataset is an analysis field produced by applying three-dimensional objective analysis to surface observations of the KMA and related agencies; it is not the forecast field of an NWP model. The input therefore contains no forecast information. We refer to this dataset hereafter as the objective analysis grid. The seven extracted variables are air temperature, relative humidity, dew point, wind speed, station pressure, visibility, and precipitation. Details of the extraction are given in Supplementary Note S6. The 15 static inputs derived from the 100-m terrain and GIS data comprise elevation, two terrain-gradient components, distances to the coast and inland water, topographic position index, seven land-cover fractions (urban, agricultural, forest, grassland, wetland, barren, and water), and farm and paddy fractions. These inputs were complete at all 19 stations.

## 3.4 Input construction

The input window comprises 24 hourly observations labeled t−23 to t, and the target hours are labeled t+1 to t+12. The targets at lead $\ell$ are the freezing label $y_\ell$ and the hourly minimum leaf temperature $T_\ell$ over the one-hour interval starting at t + $\ell$. Hourly values are labeled by the start of the hour. Thus, the input hour labeled t contains the 1-min sensor records from t to t + 59 min, whereas each gridded field is the on-the-hour value at its label time. The input window closes at t + 1 h: lead 1 covers the next hour, and lead 12 covers the interval 11–12 h after closure. These intervals describe the evaluation convention; operational data-delivery and computation delays must be accounted for separately. The 17 dynamic features consist of 7 gridded weather variables, 1 derived variable (air temperature minus dew point), 3 station sensor histories (minimum and mean leaf temperature and wetness fraction), and 6 astronomical and time encodings (sun elevation, night indicator, and sin/cos of the hour and the day of year); the 15 static features describe terrain (§3.3). The sensor-free condition uses the 14 dynamic features without the three station sensor histories. Because the radiation budget is not observed directly, sun elevation and the night indicator serve as proxies for its astronomical component. Only the frost seasons (October–April) were used.

Samples with any missing value in the input window or the leaf-temperature targets were excluded, because interpolating missing values could fill the end of the input window with values from after the issue time. The data were split in time order: train 2021-10-02 to 2025-02-18 (171,973 samples), validation 2025-02-18 to 2025-04-30 (23,474 samples), and test 2025-10-02 to 2026-04-18 (84,694 samples); the training–validation boundary falls within 2025-02-18. Dates refer to the first target hour (t + 1), and the test period ends where the objective analysis grid data end. Training samples whose target interval reached the start of validation were removed so that training and validation targets did not overlap. Of the 19 test stations, 5 began observation late and have no observations in the training and validation periods; these are the unseen stations, and §5.1 reports the trained and unseen stations separately. Standardization and clipping of the inputs are described in Supplementary Note S6.

## 3.5 Physics module and backbones

The backbone is a sequence encoder transformer (Vaswani et al., 2017). It consists of the following components: input embedding ($d = 64$) + sinusoidal positional encoding → a two-layer self-attention encoder (4 heads, feed-forward width 128, dropout 0.1; implemented in-house to extract the attention weights) → mean pooling → concatenation with the terrain-MLP representation → a frost head and a leaf-temperature head. In symbols,

$$h = \mathrm{concat}\left(\frac{1}{24}\sum_{j=1}^{24} z_j\,,\ f_s(x_s)\right), \qquad u = f_u(h), \qquad \hat{T} = f_T(h) \tag{2}$$

where $z_j$ is the output of the second encoder layer at time $j$ ($j = 1, \ldots, 24$), concat is vector concatenation, $x_s$ is the static terrain feature vector, $f_s$ is the terrain MLP (15→32→32), and $f_u$ and $f_T$ are MLP heads with one hidden layer (96 units, Gaussian error linear unit activation). $u, \hat{T} \in \mathbb{R}^{12}$ contain the data-path logits and predicted leaf temperatures for leads $\ell = 1, \ldots, 12$, and the two heads share the same representation $h$. The encoder layers apply layer normalization after each residual connection. Boldface in Eq. (2) and below denotes vectors. This backbone uses the 24 time steps of the time axis as tokens (hereafter the time-token backbone). We attach the physics module of (a) and (b) below to this backbone and evaluate the attention prior of (c) separately in ATT and FULL (Figure 2).

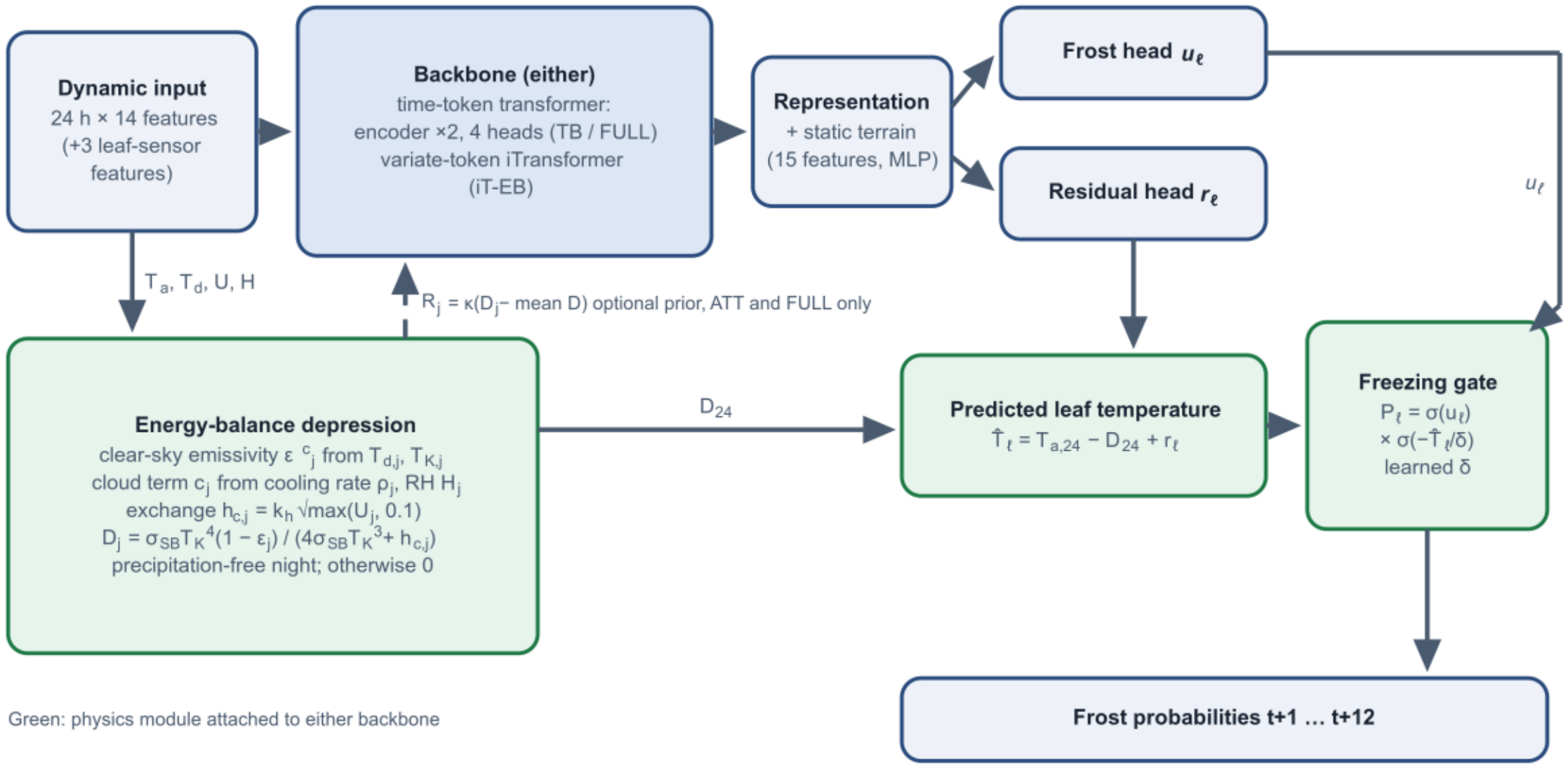

*Figure 2. Physics module and the two backbones. A 24-h dynamic sequence is encoded by the time-token transformer (PET-E) or by the variate-token iTransformer (iT-EB); the representation is concatenated with the output of a terrain MLP and passed to a frost head ($u_\ell$) and a residual head ($r_\ell$). The energy-balance and freezing-gate blocks form the physics module. The energy-balance block computes the leaf–air temperature depression $D_j$ from air temperature, dew point, wind speed, and relative humidity through the clear-sky emissivity, a cloud term, and leaf heat exchange (Eqs. (3)–(6)). The predicted leaf temperature $\hat{T}_\ell = T_{a,24} - D_{24} + r_\ell$ (Eq. (7)) enters the multiplicative freezing gate (Eq. (8)). The dashed arrow is the optional attention prior (Eq. (10)), included in FULL but not in the validation-selected PET-E (TB) or iT-EB.*

**(a) Energy-balance leaf-temperature path.** On clear nights with weak wind, sensible-heat exchange with the air cannot replace the heat that the leaf surface loses by longwave radiation, so the leaf temperature falls below the air temperature (§1.2). We compute this depression in °C from observed variables and use it to form the reference for the predicted leaf temperature. For each time step $j$ up to the issue time, we use the air temperature $T_{a,j}$ (°C; absolute temperature $T_{K,j} = T_{a,j} + 273.15$), dew point $T_{d,j}$, wind speed $U_j$, and relative humidity $H_j$ (%). The vapor pressure is obtained from the dew point (Eq. (10) of Bolton, 1980), and the clear-sky longwave emissivity of the atmosphere is computed with Eq. (11) of Brutsaert (1975):

$$e_j = 6.112\exp\left(\frac{17.67\,T_{d,j}}{T_{d,j}+243.5}\right), \qquad \varepsilon_j^c = \min\left\{1,\ 1.24\left(\frac{e_j}{T_{K,j}}\right)^{1/7}\right\}. \tag{3}$$

$e_j$ is in hPa. Clouds raise the sky emissivity toward 1 and reduce the radiative loss. We include the cloud fraction $c_j$ in the same form as Eq. (11) of Crawford and Duchon (1999):

$$\varepsilon_j = \varepsilon_j^{\mathrm{c}}\,(1 - c_j) + c_j, \qquad c_j = 1/2\left[\mathrm{clip}\left(1 - \frac{\rho_j}{\rho_0}\right) + \mathrm{clip}\left(\frac{H_j - H_0}{100 - H_0}\right)\right]. \tag{4}$$

The construction of the cloud term is explained in Supplementary Note S1. The two-sided sensible heat-exchange coefficient of the leaf is proportional to the square root of wind speed, following the Pohlhausen equation for laminar flow over a flat plate (Eq. (7) of Schuepp, 1993):

$$h_{c,j} = k_h \sqrt{\max(U_j,\, 0.1)}. \tag{5}$$

The lower wind-speed limit of 0.1 m $s^{-1}$ stands in for free convection under calm conditions. We set the leaf emissivity to 1 and assume that the upper leaf surface exchanges longwave radiation with the sky while the net longwave exchange of the lower surface is neglected; the heat-exchange coefficient $h_{c,j}$ is two-sided. If the net longwave loss per unit upper-surface area $\sigma_{\mathrm{SB}} T_K^4 (1 - \varepsilon)$ is taken up jointly by the sensible-heat exchange and the linearized emission term $4\sigma_{\mathrm{SB}} T_K^3$, the leaf–air temperature depression is

$$D_j = \frac{\sigma_{\mathrm{SB}}\, T_{K,j}^4\,(1 - \varepsilon_j)}{4\sigma_{\mathrm{SB}}\, T_{K,j}^3 + h_{c,j}} \quad \text{(precipitation-free nighttime conditions)}, \qquad D_j = 0 \quad \text{(otherwise)}. \tag{6}$$

$\sigma_{\mathrm{SB}}$ is the Stefan–Boltzmann constant. Nighttime denotes times with a sun elevation below 0°, and precipitation-free conditions denote zero precipitation in the corresponding hour. The predicted leaf temperature is constructed by subtracting the depression from the air temperature at the issue time ($j = 24$) and adding a learned residual $r_\ell$:

$$\hat{T}_\ell = T_{a,24} - D_{24} + r_\ell, \qquad r = f_T(h). \tag{7}$$

That is, in a model with (a) attached, $\hat{T} = f_T(h)$ of Eq. (2) is replaced by $r = f_T(h)$ and $\hat{T}_\ell$ is constructed by Eq. (7), so the leaf-temperature head learns the residual from the energy-balance reference. The calibration parameters $\rho_0 = \mathrm{softplus}(\xi_\rho)$, $H_0 = 50 + 45\,\sigma(\xi_H)$, and $k_h = \mathrm{softplus}(\xi_k)$ are learned subject to $\rho_0 > 0$, $50 \le H_0 \le 95$ %, and $k_h > 0$ ($\sigma$ is the logistic sigmoid; $\xi_\rho$, $\xi_H$, and $\xi_k$ are unconstrained learned parameters), starting from physical values: $\rho_0 = 1$ °C $h^{-1}$, $H_0 = 80$ %, and $k_h = 35.4$ W $m^{-2}$ $K^{-1}$ (m $s^{-1})^{-1/2}$. The last value agrees within 1 % with the two-sided Pohlhausen value for a leaf length of 0.05 m (35.2 with air properties at 0 °C). All inputs to this calculation are weather variables, not leaf sensors, so the same path operates in the condition without leaf-sensor inputs.

**(b) Multiplicative freezing gate.** The predicted leaf temperature $\hat{T}_\ell$ for lead $\ell$ (Eq. (2), or Eq. (7) in models with (a)) is passed through a learned soft threshold, and the probability is formed as a product:

$$P_\ell = \sigma(u_\ell)\,\sigma(-\hat{T}_\ell/\delta), \qquad \delta = (\mathrm{softplus}(\xi_\delta) + 10^{-3})^{-1}, \tag{8}$$

where $\sigma$ is the logistic sigmoid, $u_\ell$ is the data-path logit, and $\xi_\delta$ is an unconstrained learned parameter, a scalar shared by all leads. The additive variant (PET-A) used for comparison adds a physics term to the logit:

$$\text{logit}_\ell = u_\ell - \frac{w}{\tau}\,\hat{T}_\ell, \qquad P_\ell = \sigma(\text{logit}_\ell), \qquad w = \text{softplus}(\xi_w), \quad \tau = \text{softplus}(\xi_\tau) + 10^{-2}. \tag{9}$$

BASE, without physics mechanisms, is $P_\ell = \sigma(u_\ell)$. We initialized $\xi_\delta$ and $\xi_w$ to 0.55 and $\xi_\tau$ to 0. The corresponding initial scales are a multiplicative $\delta$ of about 0.99 °C and an additive $\tau/w$ of about 0.70 °C.

**(c) Time-axis attention prior.** In the time-token backbone, the same depression can be added to the attention scores as a prior on the key time step:

$$a_{ij} = \text{softmax}_j\left(\frac{q_i k_j^\top}{\sqrt{d_k}} + R_j\right). \tag{10}$$

$q_i$ and $k_j$ are the query and key vectors, and $d_k = d/4$ is the head dimension. $R$ is a key-direction bias shared by all queries and added identically to all heads of the two encoder layers. In the implementation, the input-window mean is subtracted from $R_j$, but by the shift invariance of softmax the attention weights do not change (Remark A.2). In ATT and FULL we set $R_j = \kappa\, D_j$ with $\kappa = \text{softplus}(\xi_\kappa) \geq 0$, which fixes the direction by construction so that keys at time steps with a larger depression receive a larger multiplicative prior factor $e^{\kappa D_j}$ (initial value of $\kappa$: 1). The final attention weights also depend on the query–key scores. For comparison, we also evaluated the learned proxy favorability, $R_j = g_\varphi(m_j)$, produced by a small MLP $g_\varphi$ (4→16→1) from the standardized sun elevation, wind speed, relative humidity, and night indicator $m_j$ (PET-M, §3.7).

**(d) Attachment and model names.** The physics module ((a) and (b)) connects to the output heads of the backbone, and we attached it to both backbones. The attention prior (c) connects to the time-token attention scores and is used only with the time-token backbone in this study. **PET-E** is the proposed model, with (a) and (b) attached to the time-token backbone (TB configuration); **FULL** additionally includes (c). **iT-EB** is the variate-token backbone iTransformer (Liu et al., 2024) with (a) and (b) attached. Because iTransformer uses variables, not time steps, as tokens, the time-axis prior (c) is not applied to it. The configuration of PET-E ((a) and (b)) was chosen by the validation CSI (§3.8). The attachment procedure and the details of the iTransformer configuration are given in §3.7 (ix).

**(e) Training.** The loss combines class-weighted binary cross-entropy and a mean squared error for leaf-temperature regression, with no separate physics penalty term. For training samples $n = 1, \ldots, N$,

$$\mathcal{L}_{\text{cls}} = -\frac{1}{12N}\sum_{n=1}^{N}\sum_{\ell=1}^{12} v_{n\ell}\,[\omega_+\, y_{n\ell}\log P_{n\ell} + (1 - y_{n\ell})\log(1 - P_{n\ell})], \tag{11}$$

$$\mathcal{L} = \mathcal{L}_{\text{cls}} + \eta\,\frac{1}{12N}\sum_{n=1}^{N}\sum_{\ell=1}^{12}\left(\hat{T}_{n\ell} - T_{n\ell}\right)^2 \tag{12}$$

Here $\omega_+ = (1-\pi)/\pi$ is the positive weight set by the positive fraction $\pi$ in the training set, and $\eta = 0.5$. Further training details and the training and decision procedure (Algorithm 1) are given in Supplementary Note S2.

## 3.6 Theoretical properties

Below, $\delta$ is the parameter of Eq. (8) in the multiplicative form and $\tau/w$ in the additive form (Eq. (9)); we denote the additive scale by $\delta_a$ only when treating both together. In the additive form we set $\psi = w/\tau = 1/\delta$.

**Proposition 1 (Sample-adaptive soft threshold).** For the additive gate, every input satisfies $P_\ell = \sigma\big((T_\ell^* - \hat{T}_\ell)/\delta\big)$, $T_\ell^* = \delta u_\ell$. That is, the additive model is equivalent to a logistic threshold model in the predicted leaf temperature, where the threshold center is set per sample by the data path and the transition scale $\delta$ is learned as a parameter common to all samples.

**Proposition 2 (Sign-constrained monotonicity).** For the additive gate with the data-path logit fixed, $\partial P_\ell / \partial \hat{T}_\ell = -\psi P_\ell (1 - P_\ell) < 0$. By the reparameterization of Eq. (9) ($w > 0, \tau > 0$), it is structurally impossible for the gate path to learn "warmer surface → higher frost probability."

**Proposition 3 (Nesting of baselines).** With the backbone and head parameters fixed, in the additive form the gate-free classifier of the same backbone (with or without the attention bias, as in the given model) is recovered pointwise in the limit $\psi \to 0^+$, and the "predict leaf temperature, then apply a 0 °C hard threshold" rule ($\mathbb{1}[\hat{T}_\ell < 0]$) is recovered pointwise, for all inputs with $\hat{T}_\ell \neq 0$, in the limit $\psi \to \infty$. The first limit also holds for inputs whose predicted temperature is 0. The two limits are approached within the parameterization of Eq. (9) as $\xi_w \to -\infty$ and $\xi_w \to +\infty$ but are not reached.

**Proposition 4 (Physics cap of the multiplicative gate).** For the multiplicative $P_\ell = \sigma(u_\ell)\,\sigma(-\hat{T}_\ell/\delta)$: (i) for all $u_\ell$, $P_\ell \leq \min\{\sigma(u_\ell), \sigma(-\hat{T}_\ell/\delta)\}$. The physics factor is a **uniform upper bound** that the data path cannot exceed (e.g., if $\hat{T} = +2\delta$ then $P \leq 0.12$). (ii) The additive $P_a = \sigma(u - \hat{T}/\delta_a)$ ($\delta_a = \tau/w$) has no such cap: for any fixed $\hat{T}$ and $p_0 \in (0,1)$ there exists a $u$ with $P_a > p_0$. That is, the data path can completely override the physics path. (iii) With the data-path logit fixed, monotonicity holds for the multiplicative form as well ($\partial P_\ell / \partial \hat{T}_\ell < 0$). In the multiplicative form $\delta$ enters as a single parameter, without the redundancy of the additive $(w, \tau)$ (Remark A.1).

**Corollary 1 (Cap and monotonicity under recalibration).** For $\tilde{P}_\ell = \min\{\mathcal{J}(P_\ell),\ \sigma(-\hat{T}_\ell/\delta)\}$ (Eq. (14)), obtained by recalibrating the multiplicative probability with a non-decreasing function $\mathcal{J}: [0,1] \to [0,1]$ and then clipping at the cap: (i) $\tilde{P}_\ell \leq \sigma(-\hat{T}_\ell/\delta)$ for every input, and (ii) with $u_\ell$ fixed, $\tilde{P}_\ell$ is non-increasing in $\hat{T}_\ell$.

The identifiability of the transition scale and its relation to the attention bias are discussed in Supplementary Note S3.

A bound on alarms follows from Proposition 4. The logarithmic bound below applies when the decision threshold is greater than 0 and less than 1. At the decision threshold $\theta_\ell^*$ (§3.8), a multiplicative alarm $P_\ell \geq \theta_\ell^*$ implies $\sigma(-\hat{T}_\ell/\delta) > P_\ell \geq \theta_\ell^*$, so it is possible only when $\hat{T}_\ell < \delta\ln\{(1-\theta_\ell^*)/\theta_\ell^*\}$. This bound is set by the learned $\delta$ and the lead-specific threshold; the additive form and BASE have no such restriction.

Combined with the energy-balance path (Eqs. (3)–(7)), this bound can be expressed in terms of observed variables.

**Proposition 5 (Energy-balance alarm bound).** For the depression $D$ of Eq. (6), the predicted leaf temperature of Eq. (7), and the multiplicative gate of Eq. (8), the following hold. (i) If an alarm $P_\ell \geq \theta$ is issued for lead $\ell$ at a decision threshold $\theta \in (0,1)$, then $T_{a,24} < D_{24} - r_\ell + \delta\ln\{(1-\theta)/\theta\}$. (ii) Regard $D$ under precipitation-free nighttime conditions as a function of $(T_K, T_d, U, H, \rho)$ through Eqs. (3)–(6), on the physical domain $T_K > 0$ and $T_d > -243.5$ °C. With the other arguments and the calibration parameters fixed, $D$ is non-increasing in wind speed $U$, dew point $T_d$, and relative humidity $H$, and non-decreasing in the air-temperature fall rate $\rho$. Moreover, $0 \leq D \leq \sigma_{\mathrm{SB}}T_K^4(1-\varepsilon^{\mathrm{c}})/(4\sigma_{\mathrm{SB}}T_K^3 + k_h\sqrt{0.1})$. (iii) Therefore, with the residual $r_\ell$, the issue-time air temperature, $\delta$, $\theta$, the calibration parameters, and the remaining arguments of (ii) fixed, the bound of (i), $D_{24} - r_\ell + \delta\ln\{(1-\theta)/\theta\}$, does not decrease as the wind weakens, as the dew point and relative humidity decrease, or as the air temperature falls faster.

Statement (iii) agrees in direction with the observation that leaves can freeze on clear nights with weak wind even when the air temperature is above 0 °C (§1.2). The proposition, however, holds only with $r_\ell$ fixed. The proof is in Appendix A of the Supplementary material.

## 3.7 Baselines and ablations

The comparators answer five questions. Items (i), (ii), and (iv) test the skill of the proposed model beyond the state at issue time and relative to tabular models; items (iii) and (v) test physics imposed other than through the structure, and item (viii) a training objective matched to state transitions; items (vi) and (vii) test what each part of the physics module adds; and item (ix) tests whether the same module transfers to another backbone. Under identical inputs, splits, and evaluation conventions we compared the following.

(i) **Persistence**: keeps the freezing state at issue time. Because the state at issue time is known only from leaf sensors, persistence is defined only in the sensor-equipped condition.

(ii) **Network climatology**: the month × hour freezing fraction of the training period, thresholded by the same validation rule as the other models. Because it does not distinguish stations, it can be applied to stations without leaf sensors.

(iii) **Two-stage**: the same backbone regresses only the leaf temperature, and the decision is made with a threshold chosen on the validation set. We also evaluated a classical rule that includes the wetness term of the label ($\hat{T} < 0 \wedge$ wet), combining the two-stage $\hat{T}$ with wetness at issue time (a wetness fraction above 0 in the issue hour).

(iv) **Flattened-feature baselines**: logistic regression (linear) and gradient boosting (LightGBM, nonlinear; Ke et al., 2017) on the same flattened features (24 h × dynamic features + terrain). Their settings are given in Supplementary Note S4.

(v) **Loss-embedding**: a learning-bias control in which BASE (without the gate or bias) was trained with Eq. (12) plus a penalty on the frost probability weighted by the predicted leaf temperature above 0 °C, $(\beta/12N)\sum_{n=1}^{N}\sum_{\ell=1}^{12}\sigma\,(u_{n\ell})\max(0,\hat{T}_{n\ell})$. It was trained with $\beta \in \{0.03, 0.1, 0.3, 1\}$; the tables show $\beta = 0.3$ as a representative value, and the CSI range across the four $\beta$ values is reported in §5.2.

(vi) **Physics-module decomposition**: in the time-token backbone, the parts of the module and the attention prior were added one at a time. We compared BASE without physics, the gate-only variant with only the multiplicative gate, the gate plus the energy-balance leaf-temperature path (PET-E, the TB configuration), the gate plus the energy-balance attention prior (ATT), and FULL with all three components. To see whether the content of the attention prior changes the results, we also evaluated variants using the proxy favorability: PET-M, with the learned $g_{\varphi}$ and the gate; the attention-bias-only variant, with $g_{\varphi}$ alone; and variants with $g_{\varphi}$ replaced by a direction-constrained form, with the gate (direction-constrained favorability) and alone (direction-constrained attention bias only). The direction-constrained form is specified in Supplementary Note S4. The gate×bias 2×2 ablation (the gate and the bias removed separately or together, with gate-only cells for both the multiplicative and additive forms) is included in this decomposition.

(vii) **Gate form**: the multiplicative gate (PET-M) was compared with the additive form (PET-A).

(viii) **Transition-weighted loss**: BASE and PET-A were trained with transition samples weighted by $1 + \gamma$ (the $v_{n\ell}$ of Eq. (11), $\gamma = 3$).

(ix) **iTransformer and module attachment**: to see whether the module transfers to another backbone, we used the official inverted embedding and encoder of the variate-token backbone iTransformer (Liu et al., 2024). Its configuration and the three compared forms are given in Supplementary Note S4.

The two-stage model and the variants in (v) and (viii) were trained only in the sensor-equipped condition, and the classical rule was also evaluated only in this condition. To assess performance at stations without leaf sensors, the sensor-free condition without the leaf-sensor inputs was applied to the transformer family, logistic regression, and GBM. Logistic regression, a deterministic fit from zero initialization, was run once; the transformer variants shared the architecture of §3.5 and the training hyperparameters of Supplementary Note S2.

## 3.8 Evaluation protocol and statistical analysis

The primary metric is the CSI. Freezing occurs in only a minority of test samples. Because the CSI excludes correct negatives, the abundant correctly forecast non-freezing samples do not dominate the score. Probability quality was evaluated with the AP, the Brier score, and the Brier skill score (BSS) relative to network climatology. With TP, FP, and FN the numbers of hits, false alarms, and misses of the decision $\mathbb{1}[P_\ell \geq \theta]$ at threshold $\theta$,

$$\mathrm{CSI} = \frac{\mathrm{TP}}{\mathrm{TP+FP+FN}}, \qquad \mathrm{POD} = \frac{\mathrm{TP}}{\mathrm{TP+FN}} \tag{13}$$

The CSI tables of §5 (Tables 3 and 4) and Supplementary Table S9 also report the F1 score, $2\,\mathrm{TP}/(2\,\mathrm{TP}+\mathrm{FP}+\mathrm{FN}) = 2\,\mathrm{CSI}/(1+\mathrm{CSI})$, computed per seed and lead and averaged in the same way as the CSI; no separate significance test was applied to it. The onset and cessation of freezing were measured separately. The onset POD is the fraction of onset samples forecast as freezing, and the cessation false-alarm rate is the fraction of cessation samples forecast as freezing. Discrimination of onset and cessation was measured with these two rates, their difference (the Peirce skill score), and the threshold-free transition AUC. The transition AUC is the probability that an onset sample receives a higher forecast probability than a cessation sample, with ties counted as one half. Because this Peirce skill score is computed on transition samples only, it differs from the Peirce skill score over all samples (POD minus the probability of false detection). Because persistence repeats the state at issue time, it detects no onset and issues a false alarm at every cessation.

The decision threshold for each lead, $\theta_\ell^*$, was chosen as the value maximizing the validation CSI among 120 quantiles (0.05–0.999) of the validation-set probabilities and was applied unchanged to the test set (for the two-stage model, the same rule was applied to the score $-\hat{T}_\ell$). Within each split, the test set was used for neither threshold selection nor early stopping, and the 2025–2026 test set was not used for configuration selection. TB, ATT, and FULL (the configuration with all three components) were compared using sensor-free validation CSI, averaged over leads and seeds, in the 2025–2026 split. This criterion selected TB in the final experiments. This validation period (2025-02-18 to 2025-04-30) lies within the 2024–2025 test season. Consequently, the choice of TB as the final configuration was not independent of that test season. We therefore interpret the 2024–2025 results as a retrospective robustness check rather than an independent test of the selected configuration. Throughout the paper, PET-E (§3.5 (d)) denotes this validation-selected TB configuration in temporal evaluation, LOSO, sensitivity analyses, and probability recalibration. FULL remains a comparator for the additional attention prior. To see differences that the overall average hides, we report trained-station and unseen-station strata and onset and cessation samples separately. The evaluation conditions are described in §4.

For differences between models, percentile confidence intervals and two-sided p values (twice the smaller of the two bootstrap tail probabilities of the difference) were obtained by a block bootstrap with one-day blocks defined by the date of the start of the first target interval (t+1 h, the end of the input window) for each test sample (date blocks; 2,000 replicates; Künsch, 1989), and the Holm correction was applied across the 12 leads of each comparison (Holm, 1979). All stations on the same date formed one block, and the two models of a comparison were resampled with the same block weights. Because the block bootstrap treats the trained models as fixed, it does not include variation due to training seeds. In each bootstrap replicate, the metric was recomputed for each seed from the resampled blocks and then averaged over seeds. Differences in the CSI and the Peirce skill score were judged significant only when the Holm-corrected $p < 0.05$ and the absolute difference exceeded twice the seed standard error. The seed standard error is the square root of the sum of the two models' seed variances, each divided by the number of seeds; seeds were not paired, and the seed variance of a single-run comparator (logistic regression, persistence, network climatology) was set to zero. The AP, Brier score, BSS, onset POD, cessation false-alarm rate, transition AUC, and relative economic value (REV; §3.9) are reported as point estimates. Because temporal autocorrelation may exceed one day, a difference was declared significant only when the criteria were met under both the date-block and the 7-day-block bootstraps (non-overlapping weeks counted from the first test date). Differences in the text were computed before rounding, so they may differ in the last digit from the difference of the two printed values.

We also measured how often alarms were issued when the observed leaf temperature was above freezing. Because the hourly freezing label is 1 only if at least 5 min of the hour have a leaf temperature of 0 °C or lower, an alarm at a time when the observed hourly minimum leaf temperature at the target time exceeds 0 °C is a false alarm by definition. Because this criterion uses the observed, not the predicted, leaf temperature, it is distinct from the structural cap and was applied to all models in the same way. As a stricter criterion that does not depend on small leaf-temperature errors near 0 °C, alarms at times when the observed hourly minimum leaf temperature was 2 °C or higher were counted separately. Each count was expressed as a fraction of all alarms. To identify which physics mechanism produced the differences, the gate-only variant was also evaluated in the same way in both test seasons and both sensor conditions, and the attention-bias-only variant in the 2025–2026 season. Alarms were defined with the same thresholds as for the CSI. Differences in the fractions were tested at each lead with the same block-bootstrap and seed-variability criteria as for the CSI, with Holm correction across the 12 leads; the fractions reported in the results (§5.3, §5.4, §5.8) are count fractions pooled over seeds and leads.

## 3.9 Probability recalibration and decision value

To compare probability quality and the operational value of warnings across methods, we fitted an isotonic recalibration on the validation set (Zadrozny and Elkan, 2002; hereafter recalibration) and computed the cost–loss decision value. To preserve the cap of Proposition 4 after recalibration, we clipped the recalibrated value at the cap:

$$\mathcal{J}_\ell = \operatorname{argmin}_{m \text{ non-decreasing}} \textstyle\sum_{n\in\text{val}} (m(P_{n\ell}) - y_{n\ell})^2, \qquad \tilde{P}_\ell = \min\{\mathcal{J}_\ell(P_\ell),\ \sigma(-\hat{T}_\ell/\delta)\}. \quad (14)$$

$\mathcal{J}_\ell$ was fitted on the validation set per method, seed, and lead by the pool-adjacent-violators algorithm (PAVA) and extended to [0, 1] by linear interpolation between the fitted points and constant values beyond them, which keeps it non-decreasing. The cap clipping was applied only to the multiplicative form, the only form with a cap (hereafter cap-preserving recalibration). By Corollary 1, $\tilde{P}_\ell$ respects the cap and, with $u_\ell$ fixed, is non-increasing in $\hat{T}_\ell$.

The value of a warning depends on the ratio $\alpha = C/L$ of the protection cost $C$ to the no-protection loss $L$. Reflecting this, we computed the relative economic value (REV) on the 2025–2026 five-seed predictions for 25 values of $\alpha$ from 0.02 to 0.50 in steps of 0.02 (Richardson, 2000). Writing expected cost in units of $L$, REV is

$$\mathrm{REV}(\alpha) = \frac{E_C - E_F}{E_C - E_P}, \qquad E_F = (\mathrm{tp} + \mathrm{fp})\,\alpha + \mathrm{fn}, \quad E_C = \min(\alpha, \bar{\mathrm{o}}), \quad E_P = \bar{\mathrm{o}}\,\alpha \tag{15}$$

Here tp, fp, and fn are the relative frequencies of hits, false alarms, and misses of the decision, and $\bar{\mathrm{o}} = \mathrm{tp} + \mathrm{fn}$ is the incidence. $E_C$ is the cost of choosing, using the incidence alone, the lower-expected-cost strategy between always-protect and always-no-protect, and $E_P$ is the expected cost of a perfect forecast, so REV is 1 for a perfect forecast and 0 for a forecast whose value equals that of the climatological strategy. If the probability $P$ is reliable, protection is optimal for a user with cost–loss ratio $\alpha$ when $P \geq \alpha$ (Perez-Lebel et al., 2025). We therefore used the rule "protect if $P \geq \alpha$" at each $\alpha$, which allows methods to be compared independently of the decision-threshold selection. Because the uncalibrated probabilities from class-weighted training cannot be assumed reliable, we also report REV after validation-set recalibration (with cap clipping for the multiplicative form). REV was computed per lead, then averaged over leads, and then averaged over seeds; the maximum REV is the largest of these values over the 25 values of $\alpha$.

---

## 4. Experiments

Using the input data of §3.4, we ran seven groups of experiments; Table 1 lists their training runs.

**Table 1.** Experimental conditions and training runs. The training-run counts include the repeat runs already listed in the experiment manifests. The six additional final reproducibility checks

described in §4 are excluded from these counts. Logistic regression, a deterministic fit, was run once.

| Setting | Data split | Methods | Seeds | Training runs |
|---|---|---|---|---|
| Season hold-out (2025–2026) | Train 2021-10 to 2025-02, validation 2025-02 to 2025-04, test 2025-10 to 2026-04 | PET-M, PET-A, BASE, 2×2 gate–bias ablation, GBM, logistic regression (both sensor conditions); two-stage, classical rule, loss-embedding (β ∈ {0.03, 0.1, 0.3, 1}), transition-weighted loss, δ initial values, training-data reduction (sensor-equipped); persistence, network climatology | 0–4 (logistic regression and no-learning baselines: one deterministic run) | 167 |
| Rolling origin (2024–2025) | Test 2024-10 to 2025-04; all later data excluded | PET-M, PET-A, BASE, GBM, logistic regression (both sensor conditions); two-stage (sensor-equipped) | 0–4 | 47 |
| LOSO, 15 terrain inputs | 19 folds; held-out station excluded from training and threshold selection; test 2025–2026 | PET-M, BASE, GBM (both sensor conditions) | 0–2 | 342 |
| LOSO, static-input candidates (sensor-free) | As above; no static input (S1) or elevation and distances to coast and inland water (S2) | PET-M, BASE, GBM | 0–2 | 342 |
| Season hold-out without static inputs | As in the season hold-out | PET-M, BASE, GBM (both sensor conditions) | 0–4 | 30 |
| LOSO without static inputs (sensor-equipped) | As in LOSO | PET-M, BASE, GBM | 0–2 | 171 |
| iTransformer backbone | Both temporal test seasons | iTransformer (both sensor conditions) | 0–4 (one repeat) | 21 |
| iTransformer + gate | Both temporal test seasons | iTransformer + multiplicative gate (sensor-free; sensor-equipped) | 0–4 (one repeat each) | 22 |
| Direction-constrained attention bias | Both temporal test seasons | PET-M with a direction-constrained bias; bias only (2025–2026) | 0–4 (one repeat) | 31 |
| Physics module (PET-E cells) | Both temporal test seasons | Gate + energy-balance leaf temperature (PET-E/TB), gate + attention (ATT), all three (FULL) | 0–4 (one repeat) | 61 |
| PET-E and FULL LOSO (sensor-free) | As in LOSO; S0 and S1 | PET-E (TB), FULL | 0–2 | 228 |
| Physics module on iTransformer | Both temporal test seasons | iT-EB (both sensor conditions) | 0–4 (one repeat) | 21 |
| LOSO and supplements for the module | As in LOSO; S0 and S1; 2024–2025 hold-out; training-data reduction; δ initial values | iT-EB, iTransformer (both conditions), PET-E (TB), FULL (sensor-equipped); gate only (2024–2025); PET-E and FULL since 1–3 seasons and δ0 = 0.5, 2.0 | 0–2 (LOSO), 0–4 (one LOSO repeat) | 745 |

**(i) Season hold-out.** In the 2025–2026 season we compared the decomposition of the physics module (BASE, gate only, PET-E (TB), ATT, FULL), the proxy-favorability variants (PET-M, attention bias only, direction-constrained favorability, direction-constrained attention bias only), the gate×bias 2×2 (including the gate-only cells for the additive and multiplicative forms), the multiplicative and additive forms, the three forms of iTransformer (§3.7 (ix)), the two-stage model, the classical rule, loss-embedding (four values of $\beta$), the transition-weighted loss, persistence, network climatology, logistic regression, and GBM. The transformer family, logistic regression, and GBM were trained in both the sensor-equipped and sensor-free conditions, and the remaining learned methods in the sensor-equipped condition, each with five seeds (logistic regression once). In the sensor-equipped condition we also ran the sensitivity to the initial value of $\delta$ (0.5 and 2.0 °C) and the training-data reduction (most recent 1, 2, and 3 seasons). The sensitivity to the initial value of $\delta$ covered PET-M, PET-E, and FULL, and the training-data reduction covered PET-E, FULL, PET-M, BASE, and PET-A. The season hold-out run count in Table 1 includes five additional sensor-equipped PET-M runs configured to record the learning trajectory of the transition scale, one per seed.

**(ii) Retrospective rolling-origin evaluation.** For the rolling-origin repetition, the test period was moved to 2024-10-02 to 2025-04-30, a separate input set excluding all later data was built with the same rules, and training, validation, and threshold selection were repeated (training 2021-10-02 to 2024-03-05, 113,099 samples; validation 2024-03-05 to 2024-04-30, 15,446; test 66,935). PET-E and its decomposition variants (gate only, ATT, FULL), PET-M, PET-M with the direction-constrained favorability, BASE, PET-A, the three forms of iTransformer, GBM, and logistic regression were compared in both sensor conditions, the two-stage model and the classical rule were run in the sensor-equipped condition, and network climatology was computed from this training period. Model fitting, early stopping, and decision-threshold selection were repeated using the earlier training and validation periods specified above. However, PET-E retained the TB configuration selected as described in §3.8, using a validation period that overlaps the 2024–2025 test season. This experiment therefore provides a retrospective robustness check rather than an independent test of configuration selection.

**(iii) LOSO.** In the 19-fold LOSO, one station was excluded in each fold; models were trained on the training-period data of the remaining stations, with their validation-period data used for early stopping and threshold selection, and then used to predict the 2025–2026 test season at the held-out station. Standardization statistics and the positive-class weight were recomputed in each fold from the training data of the remaining stations, and test metrics were computed per seed from the pooled predictions of the 19 folds. The network climatology used as the BSS reference was computed from the training period of the whole network. Within each fold, the held-out station was excluded from model fitting, early stopping, and decision-threshold selection. Separately, a common static-input configuration for each model and sensor condition was selected using pooled out-of-fold validation predictions and labels from the 14 trained stations (§4 (iv); Supplementary Note S5). Thus, station exclusion applies to fold-specific model fitting and threshold selection, whereas static-input configuration selection used validation information pooled across stations. PET-E, FULL, PET-M, BASE, iT-EB, iTransformer, and GBM were trained in both sensor conditions with seeds 0–2.

**(iv) Static inputs in LOSO.** We ran LOSO with different configurations of the static terrain inputs. The candidate sets, the selection rule, and the station-level resampling are given in Supplementary Note S5.

**(v) Static inputs in the season hold-out.** To check whether removing the static terrain inputs harms prediction at trained stations, only the static inputs were removed (S1) in the 2025–2026 test setting, and PET-M, BASE, and GBM were retrained in both sensor conditions with seeds 0–4. The static inputs of the main model were selected based on the validation CSI. The rule, fixed before the analysis was run, was to keep S0 if the S1 validation CSI (lead and seed mean) of PET-M was lower than that of S0 by more than twice the seed standard error in at least one sensor condition, and otherwise to use S1.

**(vi) Analyses of saved predictions.** From the saved predictions and model weights we computed the false alarms at above-freezing observed leaf temperatures (§3.8), probability recalibration and REV (§3.9), the learned calibration parameters and the energy-balance depression $D$ in the test season, and, for PET-M, the partial dependence of $g_{\varphi}$ and the attention weights by input time step. To distinguish contemporaneous depression diagnostics from future leaf-temperature prediction, we also computed lead-specific MAE and RMSE from saved test predictions on identical samples for PET-E versus gate-only and iT-EB versus iT+gate. Errors were computed separately for each seed and lead before averaging over five seeds; lead means are arithmetic means of these seed-averaged errors (Supplementary Table S5). These error comparisons are descriptive, without significance testing.

**(vii) External backbone.** The three forms of iTransformer were trained for the two temporal test seasons in both sensor conditions with seeds 0–4, and in LOSO the plain form and iT-EB were trained as in (iii). The temporal comparisons retained all 15 static inputs, and LOSO evaluated both S0 and S1 as in (iii). The data splits, validation-based threshold selection, and statistical tests followed the same protocol as for the time-token backbone.

The final experiments were trained on one NVIDIA RTX 4090 GPU using Python 3.9.12, PyTorch 2.8.0 (CUDA 12.8), LightGBM 4.6.0, and NumPy 1.26.3. Three representative runs from the original experiments and three from the additional TB experiments were repeated on the same device. These covered temporal and LOSO settings and yielded bit-identical saved prediction arrays in all six comparisons.

# 5. Results

## 5.1 Lead-time performance and skill over persistence

In the 2025–2026 sensor-equipped setting, PET-E achieved CSI values of 0.799, 0.592, and 0.510 at t+1, t+6, and t+12, respectively. Persistence achieved 0.804, 0.370, and 0.211 at the same leads. Lead-specific results are given in Table S6 and Figure 3, and AP, Brier score, and onset–cessation metrics in Table S2.

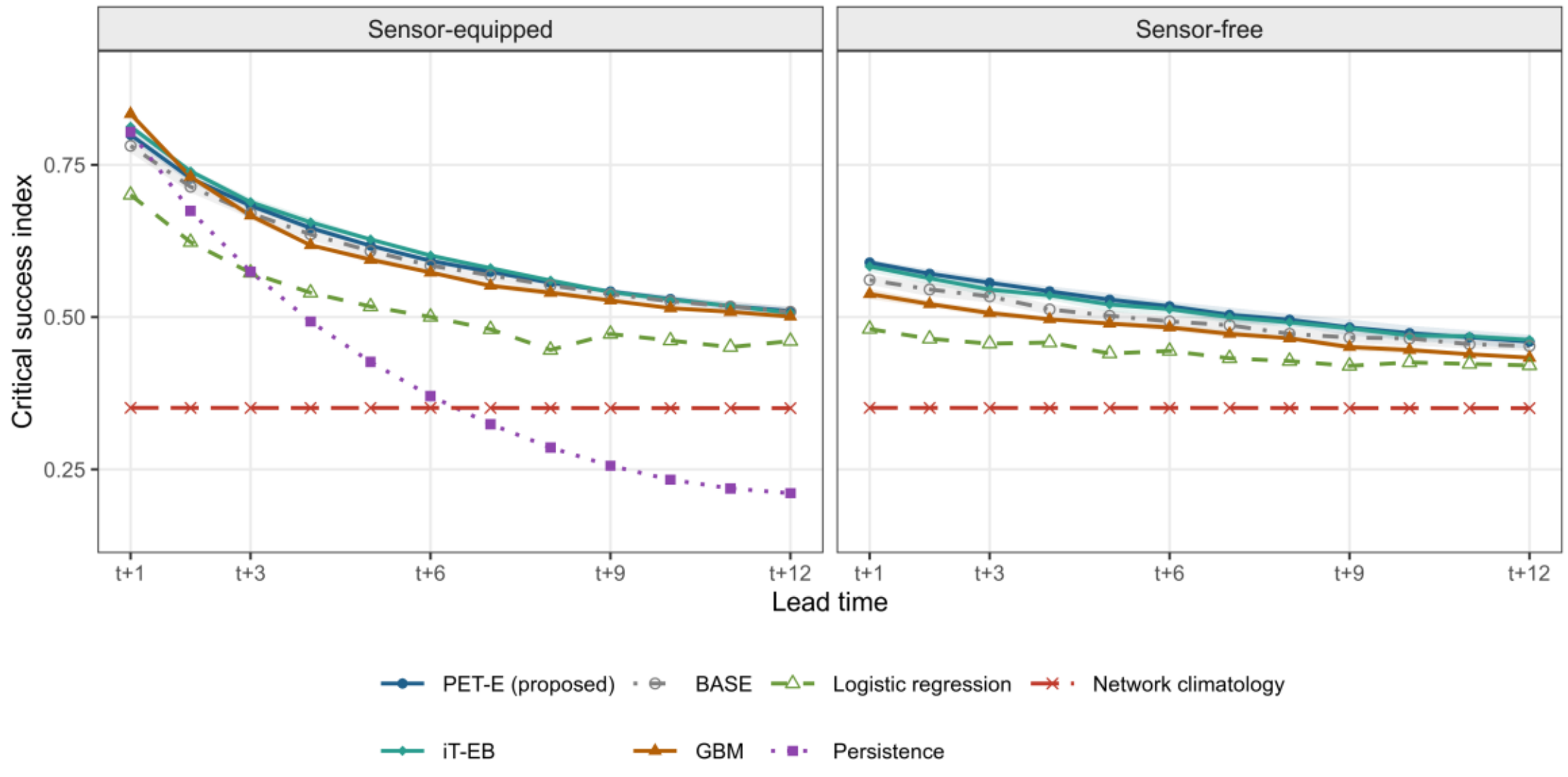

*Figure 3. Critical success index by lead time in the 2025–2026 test season, with leaf-sensor inputs (left) and without (right). Lines are seed means; shading, one seed standard deviation. Persistence is defined only with leaf-sensor inputs. Lead ℓ covers the interval from ℓ−1 h (inclusive) to ℓ h (exclusive) after the end of the input window; lead 1 is the immediately following 0–1 h interval. The nominal label t+ℓ identifies the start of the target interval.*

The lead-mean CSI of PET-E was 0.608 with leaf-sensor inputs and 0.516 without them. In the sensor-free condition, its CSI was 0.590 at t+1 and 0.460 at t+12. CSI was lower at unseen stations than at trained stations: 0.459 versus 0.639 with leaf-sensor inputs, and 0.360 versus 0.549 without them (Table 2).

**Table 2.** Lead-mean CSI by station stratum (a; 2025–2026 season) and under training-data reduction (b; 2025–2026 season, sensor-equipped). In this and the following tables, a blank cell in a leading column repeats the entry above it.

### (a) Station strata, 2025–2026 season

| Condition | Method | All stations | Trained stations | Unseen stations |
|---|---|---|---|---|
| sensor-equipped | PET-E | 0.608 | 0.639 | 0.459 |
| | iT-EB | 0.613 | 0.640 | 0.489 |
| | PET-M | 0.596 | 0.627 | 0.470 |
| | BASE | 0.600 | 0.628 | 0.487 |
| | GBM | 0.597 | 0.621 | 0.487 |
| sensor-free | PET-E | 0.516 | 0.549 | 0.360 |
| | iT-EB | 0.511 | 0.543 | 0.377 |
| | PET-M | 0.505 | 0.540 | 0.374 |
| | BASE | 0.496 | 0.534 | 0.367 |

| Condition | Method | All stations | Trained stations | Unseen stations |
|---|---|---|---|---|
| | GBM | 0.478 | 0.509 | 0.362 |

**(b) Training-data reduction, 2025–2026 season, sensor-equipped**

| Training seasons | PET-E | PET-M | BASE | PET-A |
|---|---|---|---|---|
| Most recent 1 | 0.578 | 0.572 | 0.568 | 0.562 |
| Most recent 2 | 0.592 | 0.589 | 0.593 | 0.591 |
| Most recent 3 | 0.601 | 0.600 | 0.600 | 0.596 |
| All (2021–2025) | 0.608 | 0.596 | 0.600 | 0.604 |

## 5.2 Two-stage, flattened, and loss-embedding baselines

With leaf-sensor inputs, GBM had a higher CSI than PET-E at t+1 (0.834 versus 0.799). At t+6, the values were 0.592 for PET-E and 0.573 for GBM; at t+12, they were 0.510 and 0.501. Their lead-mean CSI values were 0.608 and 0.597, a difference of +0.011. Under both block criteria, PET-E was significantly higher at 6 leads and GBM at 1 lead.

The four loss-embedding variants achieved CSI values of 0.602, 0.605, 0.607, and 0.603 for β = 0.03, 0.1, 0.3, and 1, respectively, compared with 0.608 for PET-E. Lead-specific CSI and onset–cessation metrics for the two-stage model, classical rule, logistic regression, and transition-weighted loss variants are reported alongside these results in Table S6 and Table S2.

## 5.3 Decomposition of the physics module

In the 2025–2026 sensor-free setting, attaching the physics module to BASE increased lead-mean CSI from 0.496 to 0.516, reduced the Brier score from 0.1289 to 0.1053, and lowered the fraction of alarms at observed leaf temperatures ≥ 2 °C from 0.087 to 0.051 (Table 3, Figure 4). Lead-mean CSI was 0.506 for the gate-only variant and 0.512 for both ATT and FULL. Validation CSI was 0.62427 for TB, 0.62254 for FULL, and 0.61049 for ATT; the prespecified criterion selected TB as PET-E.

**Table 3.** Decomposition of the physics module in the time-token backbone, with GBM for reference. Gate only, the multiplicative gate alone; ATT, the gate with the energy-balance attention prior; FULL, the gate, the energy-balance temperature path, and the attention prior. CSI, F1 score, and Brier score are lead and seed means; an alarm fraction is the number of alarms at the stated observed leaf temperature, summed over seeds and leads, divided by all alarms. The additive-gate, proxy-favorability, and attention-bias variants are in Table S9.

| Season | Condition | Method | CSI | F1 score | Brier score | Alarms at observed leaf temp. > 0 °C | Alarms at observed leaf temp. ≥ 2 °C |
|---|---|---|---|---|---|---|---|
| 2025–2026 | sensor-free | BASE | 0.496 | 0.662 | 0.1289 | 0.189 | 0.087 |
| | | Gate only | 0.506 | 0.671 | 0.1112 | 0.151 | 0.060 |
| | | PET-E | 0.516 | 0.680 | 0.1053 | 0.137 | 0.051 |
| | | ATT | 0.512 | 0.676 | 0.1105 | 0.159 | 0.064 |

| Season | Condition | Method | CSI | F1 score | Brier score | Alarms at observed leaf temp. > 0 °C | Alarms at observed leaf temp. ≥ 2 °C |
|---|---|---|---|---|---|---|---|
| | | FULL | 0.512 | 0.676 | 0.1056 | 0.141 | 0.053 |
| | | GBM | 0.478 | 0.647 | 0.1108 | 0.235 | 0.122 |
| | sensor-equipped | BASE | 0.600 | 0.747 | 0.1002 | 0.154 | 0.062 |
| | | Gate only | 0.593 | 0.741 | 0.0935 | 0.128 | 0.046 |
| | | PET-E | 0.608 | 0.753 | 0.0847 | 0.115 | 0.041 |
| | | ATT | 0.599 | 0.746 | 0.0894 | 0.128 | 0.048 |
| | | FULL | 0.610 | 0.754 | 0.0848 | 0.116 | 0.042 |
| | | GBM | 0.597 | 0.743 | 0.0859 | 0.164 | 0.072 |
| 2024–2025 | sensor-free | BASE | 0.566 | 0.722 | 0.1192 | 0.139 | 0.062 |
| | | Gate only | 0.589 | 0.741 | 0.1134 | 0.107 | 0.039 |
| | | PET-E | 0.592 | 0.743 | 0.1103 | 0.096 | 0.033 |
| | | ATT | 0.585 | 0.738 | 0.1148 | 0.102 | 0.036 |
| | | FULL | 0.591 | 0.742 | 0.1114 | 0.098 | 0.033 |
| | | GBM | 0.542 | 0.702 | 0.1287 | 0.173 | 0.088 |
| | sensor-equipped | BASE | 0.656 | 0.791 | 0.0934 | 0.109 | 0.041 |
| | | Gate only | 0.656 | 0.791 | 0.0898 | 0.092 | 0.030 |
| | | PET-E | 0.663 | 0.795 | 0.0903 | 0.085 | 0.028 |
| | | ATT | 0.663 | 0.796 | 0.0902 | 0.081 | 0.026 |
| | | FULL | 0.663 | 0.796 | 0.0903 | 0.084 | 0.027 |
| | | GBM | 0.647 | 0.783 | 0.0909 | 0.119 | 0.051 |

Adding the energy-balance temperature path to the gate, forming PET-E, lowered the fraction of alarms issued at observed leaf temperature ≥ 2 °C in all four settings. The numbers of leads with significant reductions were 2, 1, 5, and 1 in the 2025–2026 sensor-free and sensor-equipped settings, followed by the 2024–2025 sensor-free and sensor-equipped settings. In that order, CSI increased significantly at 4, 11, 1, and 5 leads, with no significant decreases.

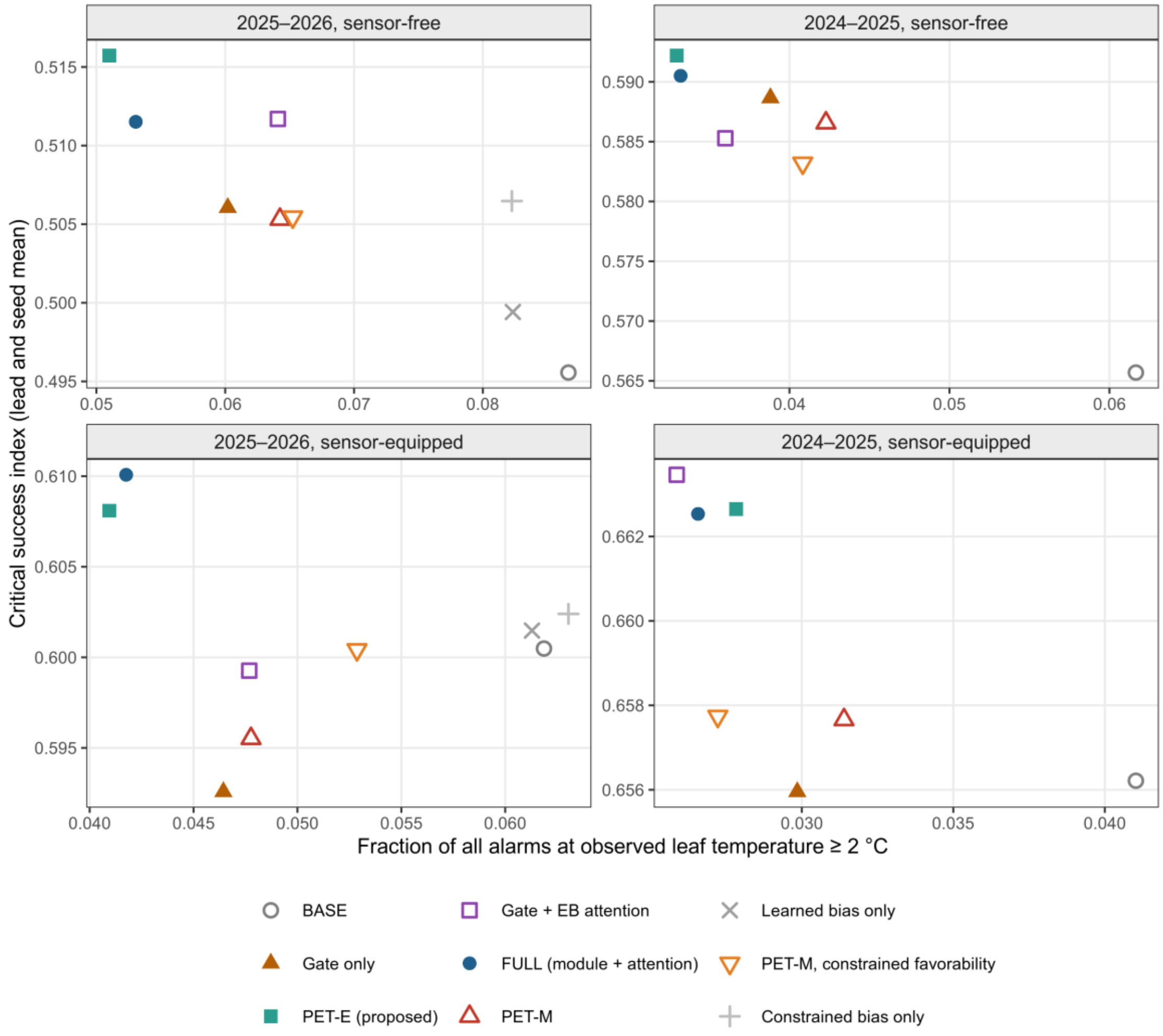

*Figure 4. Decomposition of the physics module. Fraction of all alarms issued at observed leaf temperature ≥ 2 °C (horizontal axis; alarms pooled over seeds and leads) against lead-mean CSI (vertical axis; seed mean) in four settings (two test seasons × with and without leaf-sensor inputs). PET-E is the gate plus energy-balance leaf-temperature path (TB), FULL additionally includes the attention prior, and Gate + EB attention is ATT; in the legend, Gate only is the multiplicative gate only, and 'bias' denotes the attention bias of Table S9. The proxy-favorability variants (PET-M; learned and direction-constrained favorability) are also shown; the two attention-bias-only variants (learned and direction-constrained) were trained only in the 2025–2026 season. Axis ranges differ between panels.*

Adding the same physics to the gate-only variant as an attention prior, forming ATT, significantly raised CSI in three of the four settings. In the same order, significant CSI increases occurred at 1, 5, 0, and 9 leads. Significant reductions in the alarm fraction occurred at 5 leads in the 2024–2025 sensor-equipped setting and at no lead in the other three settings. FULL, which combines both placements, had lead-mean CSI close to PET-E: its CSI differed significantly only at one of the 12 leads in each sensor-free season (lower for FULL) and at no lead in either sensor-equipped setting. In no setting did FULL have a significantly higher CSI or Peirce score, or a significantly lower alarm fraction, than PET-E at any lead.

Relative to BASE, PET-E reduced the alarm fraction significantly at 12, 12, 12, and 11 leads in the order above. CSI increased significantly at 9, 2, 12, and 2 leads, with no significant decreases. Results for the additive gate, the proxy-favorability module, learned and direction-constrained attention, and reduced training data are given in Table S9, Table S2, and Table 2.

## 5.4 Transfer of the physics module to iTransformer

Attaching the module to iTransformer reduced both Brier score and the ≥ 2 °C alarm fraction in all four temporal settings (Table 4, Figure 5). The alarm fraction was significantly lower at all 12 leads in every setting, with no significant CSI decrease. In the 2025–2026 sensor-free and sensor-equipped settings, followed by the 2024–2025 sensor-free and sensor-equipped settings, iT-EB achieved CSI values of 0.511, 0.613, 0.589, and 0.669, compared with 0.509, 0.607, 0.573, and 0.654 for the plain backbone. Significant CSI increases occurred at 0, 5, 11, and 12 leads, respectively. These changes were obtained by adding only four trainable parameters to iTransformer, with the remaining initial weights held identical between the module-equipped and unmodified models for each seed.

**Table 4.** Physics module attached to iTransformer, with the corresponding configurations of the time-token backbone (four settings). The alarm fraction is defined as in Table 3.

| Season | Condition | Method | CSI | F1 score | Brier score | Peirce skill score | Alarms at observed leaf temp. ≥ 2 °C |
|---|---|---|---|---|---|---|---|
| 2025–2026 | sensor-free | BASE | 0.496 | 0.662 | 0.1289 | 0.449 | 0.087 |
| | | Gate only | 0.506 | 0.671 | 0.1112 | 0.448 | 0.060 |
| | | PET-E | 0.516 | 0.680 | 0.1053 | 0.397 | 0.051 |
| | | iTransformer | 0.509 | 0.674 | 0.1120 | 0.400 | 0.081 |
| | | iTransformer + gate | 0.509 | 0.674 | 0.1092 | 0.438 | 0.060 |
| | | iT-EB | 0.511 | 0.676 | 0.1078 | 0.454 | 0.063 |
| | sensor-equipped | BASE | 0.600 | 0.747 | 0.1002 | 0.287 | 0.062 |
| | | Gate only | 0.593 | 0.741 | 0.0935 | 0.332 | 0.046 |
| | | PET-E | 0.608 | 0.753 | 0.0847 | 0.317 | 0.041 |
| | | iTransformer | 0.607 | 0.752 | 0.0913 | 0.308 | 0.064 |
| | | iTransformer + gate | 0.615 | 0.757 | 0.0839 | 0.359 | 0.046 |
| | | iT-EB | 0.613 | 0.757 | 0.0842 | 0.353 | 0.045 |
| 2024–2025 | sensor-free | BASE | 0.566 | 0.722 | 0.1192 | 0.362 | 0.062 |
| | | Gate only | 0.589 | 0.741 | 0.1134 | 0.367 | 0.039 |
| | | PET-E | 0.592 | 0.743 | 0.1103 | 0.362 | 0.033 |

| Season | Condition | Method | CSI | F1 score | Brier score | Peirce skill score | Alarms at observed leaf temp. ≥ 2 °C |
|---|---|---|---|---|---|---|---|
| | | iTransformer | 0.573 | 0.728 | 0.1177 | 0.339 | 0.066 |
| | | iTransformer + gate | 0.586 | 0.739 | 0.1138 | 0.347 | 0.044 |
| | | iT-EB | 0.589 | 0.741 | 0.1153 | 0.395 | 0.045 |
| | sensor-equipped | BASE | 0.656 | 0.791 | 0.0934 | 0.254 | 0.041 |
| | | Gate only | 0.656 | 0.791 | 0.0898 | 0.276 | 0.030 |
| | | PET-E | 0.663 | 0.795 | 0.0903 | 0.284 | 0.028 |
| | | iTransformer | 0.654 | 0.788 | 0.0936 | 0.284 | 0.046 |
| | | iTransformer + gate | 0.668 | 0.799 | 0.0871 | 0.312 | 0.029 |
| | | iT-EB | 0.669 | 0.800 | 0.0861 | 0.341 | 0.029 |

No significant CSI difference was found between iT-EB and iT+gate in any of the four settings. The alarm fraction of iT-EB was significantly higher at two leads and lower at one lead in the 2025–2026 sensor-equipped setting, and lower at one lead in the 2024–2025 sensor-equipped setting. Neither sensor-free setting showed a significant difference in this fraction.

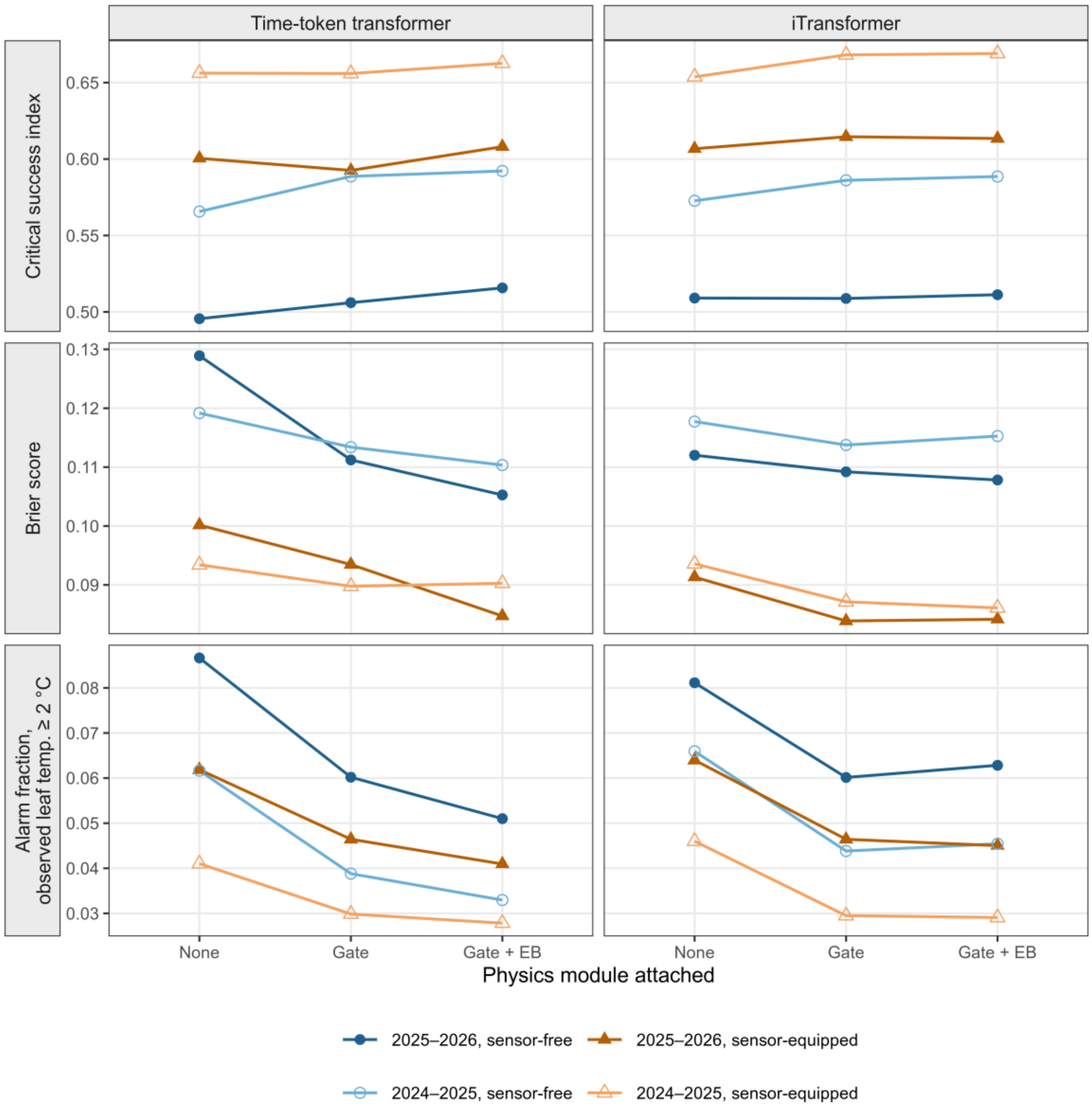

*Figure 5. Stepwise attachment of the physics module to the two backbones. Left, time-token transformer (None = BASE, Gate = gate only, Gate + EB = PET-E (TB), without the attention prior); right, iTransformer (None = plain iTransformer, Gate = iT+gate, Gate + EB = iT-EB). From top: CSI and Brier score (lead and seed means) and the fraction of alarms at observed leaf temperature ≥ 2 °C (alarms pooled over seeds and leads).*

## 5.5 Onset–cessation discrimination and the second test season

In the 2025–2026 sensor-free setting, PET-E had an onset POD of 0.691 and a cessation false-alarm rate of 0.294, compared with 0.766 and 0.443 for GBM (Table S7, Figure 6). For BASE, the corresponding onset POD and cessation false-alarm rate were 0.795 and 0.346, respectively. The Peirce skill score difference (PET-E − GBM) was +0.074 and was significant at 10 leads.

Transition AUC was 0.764 for PET-E and 0.710 for GBM. Network climatology had a higher Peirce skill score than PET-E (0.449 versus 0.397).

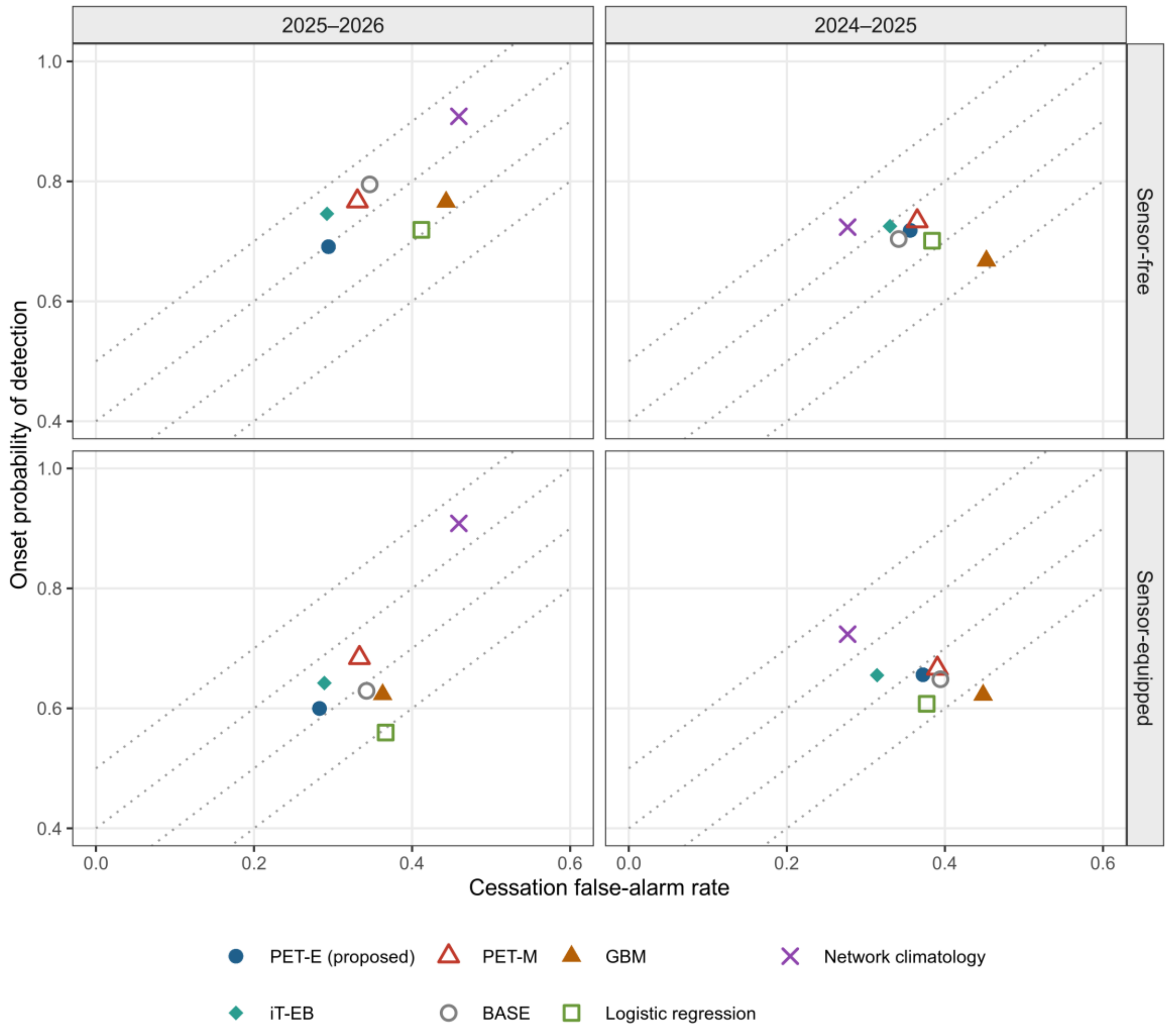

*Figure 6. Onset POD against cessation false-alarm rate in the two test seasons (columns) without and with leaf-sensor inputs (rows); lead and seed means. The Peirce skill score is onset POD − cessation false-alarm rate; dotted lines join points of equal Peirce skill score (0.2–0.5).*

In the 2024–2025 sensor-free setting, CSI was 0.592 for PET-E, 0.542 for GBM, 0.573 for iTransformer, 0.589 for iT-EB, and 0.587 for PET-M. PET-E was significantly higher than GBM at 12 leads and iTransformer at 10 leads, and PET-M at 3 leads. Its transition AUC was 0.740 and cessation false-alarm rate was 0.356.

## 5.6 Deployment without leaf sensors and at held-out stations

In the 2025–2026 sensor-free setting, PET-E and GBM achieved lead-mean CSI values of 0.516 and 0.478; PET-E was significantly higher at all 12 leads. iTransformer and PET-M achieved 0.509 and 0.505, respectively, with no significant CSI difference between PET-E and iTransformer; PET-E exceeded PET-M significantly at 4 leads (Table S6).

In sensor-free LOSO with 15 static inputs (S0), CSI was 0.446 for PET-E, 0.463 for iT-EB, 0.456 for iTransformer, and 0.434 for GBM (Table 5). PET-E did not differ significantly from GBM in CSI, whereas iT-EB was significantly higher at all 12 leads. No significant CSI difference was found between iT-EB and iTransformer. BSS was 0.095 for PET-E, 0.124 for iT-EB, 0.082 for iTransformer, and 0.129 for GBM.

**Table 5.** LOSO performance in both sensor conditions with 15 terrain inputs (S0) and without static inputs (S1). Metrics are means over leads and seeds 0–2; the alarm fraction is the number of alarms at observed leaf temperature ≥ 2 °C, summed over seeds and leads, divided by all alarms.

| Condition | Static input | Method | CSI | AP | BSS | Peirce skill score | Transition AUC | Alarms at observed leaf temp. ≥ 2 °C |
|---|---|---|---|---|---|---|---|---|
| sensor-free | 15 terrain inputs (S0) | PET-E | 0.446 | 0.660 | 0.095 | 0.357 | 0.745 | 0.058 |
| | | FULL | 0.444 | 0.652 | 0.085 | 0.356 | 0.747 | 0.055 |
| | | iT-EB | 0.463 | 0.661 | 0.124 | 0.404 | 0.773 | 0.063 |
| | | iTransformer | 0.456 | 0.656 | 0.082 | 0.390 | 0.755 | 0.086 |
| | | PET-M | 0.448 | 0.656 | 0.092 | 0.381 | 0.760 | 0.067 |
| | | BASE | 0.430 | 0.637 | −0.002 | 0.385 | 0.756 | 0.084 |
| | | GBM | 0.434 | 0.650 | 0.129 | 0.315 | 0.703 | 0.124 |
| | No static input (S1) | PET-E | 0.468 | 0.680 | 0.121 | 0.429 | 0.783 | 0.057 |
| | | FULL | 0.464 | 0.675 | 0.105 | 0.430 | 0.785 | 0.058 |
| | | iT-EB | 0.467 | 0.677 | 0.112 | 0.442 | 0.789 | 0.060 |
| | | iTransformer | 0.470 | 0.680 | 0.071 | 0.428 | 0.771 | 0.085 |
| | | PET-M | 0.470 | 0.676 | 0.106 | 0.435 | 0.785 | 0.063 |
| | | BASE | 0.471 | 0.682 | 0.075 | 0.436 | 0.773 | 0.081 |
| | | GBM | 0.437 | 0.649 | 0.108 | 0.325 | 0.707 | 0.117 |
| sensor-equipped | 15 terrain inputs (S0) | PET-E | 0.576 | 0.798 | 0.330 | 0.271 | 0.695 | 0.049 |
| | | FULL | 0.578 | 0.796 | 0.335 | 0.269 | 0.693 | 0.049 |
| | | iT-EB | 0.588 | 0.807 | 0.356 | 0.319 | 0.720 | 0.051 |
| | | iTransformer | 0.589 | 0.808 | 0.304 | 0.277 | 0.693 | 0.066 |
| | | PET-M | 0.576 | 0.800 | 0.319 | 0.325 | 0.710 | 0.052 |
| | | BASE | 0.579 | 0.799 | 0.256 | 0.262 | 0.680 | 0.065 |
| | | GBM | 0.590 | 0.804 | 0.384 | 0.243 | 0.650 | 0.070 |
| | No static input (S1) | PET-E | 0.591 | 0.811 | 0.346 | 0.335 | 0.724 | 0.048 |
| | | FULL | 0.593 | 0.814 | 0.348 | 0.335 | 0.721 | 0.047 |

| Condition | Static input | Method | CSI | AP | BSS | Peirce skill score | Transition AUC | Alarms at observed leaf temp. ≥ 2 °C |
|---|---|---|---|---|---|---|---|---|
| | | iT-EB | 0.602 | 0.819 | 0.360 | 0.365 | 0.744 | 0.050 |
| | | iTransformer | 0.595 | 0.813 | 0.286 | 0.304 | 0.706 | 0.065 |
| | | PET-M | 0.591 | 0.813 | 0.339 | 0.312 | 0.707 | 0.054 |
| | | BASE | 0.590 | 0.806 | 0.287 | 0.258 | 0.682 | 0.063 |
| | | GBM | 0.586 | 0.802 | 0.362 | 0.248 | 0.656 | 0.073 |

In sensor-equipped LOSO with S1, CSI was 0.591 for PET-E, 0.602 for iT-EB, and 0.586 for GBM. Relative to GBM, PET-E was significantly higher at 3 leads and iT-EB at 10 leads; each was lower at one lead. Both had significantly higher Peirce skill scores than GBM at all 12 leads. Station-level resampling gave a CSI difference of +0.016 for iT-EB minus GBM (95% interval 0.010–0.021) and +0.005 for PET-E minus GBM (−0.003–0.012). In sensor-equipped S0, iT-EB had a significantly higher CSI than the plain iTransformer at two leads and a lower CSI at one lead.

## 5.7 Static terrain inputs and transfer to new stations

Static-input configuration selection, based on pooled out-of-fold validation predictions and labels from the 14 trained stations, selected S1 for every model in the sensor-free condition. In the sensor-equipped condition, BASE and PET-M retained S0, while the other models selected S1 (Table S8). In the sensor-free condition, validation CSI with S0 and S1 was 0.550 and 0.575 for PET-E, 0.553 and 0.571 for iT-EB, and 0.512 and 0.518 for GBM. The GBM improvement with S1, +0.006, exceeded twice the seed standard error (0.005). Validation CSI with S2 was 0.550 for PET-M, 0.547 for BASE, and 0.508 for GBM; S1 was selected.

In sensor-free LOSO with S1, CSI was 0.468 for PET-E, 0.467 for iT-EB, 0.470 for iTransformer, and 0.437 for GBM. PET-E and iT-EB had significantly higher CSI and Peirce skill scores than GBM at all 12 leads. The CSI increase from S0 to S1 was +0.022 for PET-E and +0.004 for iT-EB, significant at 12 and 3 leads, respectively (Figure 7). With S1, iT-EB and iTransformer did not differ significantly in CSI. Comparing the validation-selected S1 configurations, station-level resampling gave a PET-E − GBM CSI difference of +0.032 (95% interval 0.014–0.048) and an iT-EB − GBM difference of +0.031 (0.015–0.048).

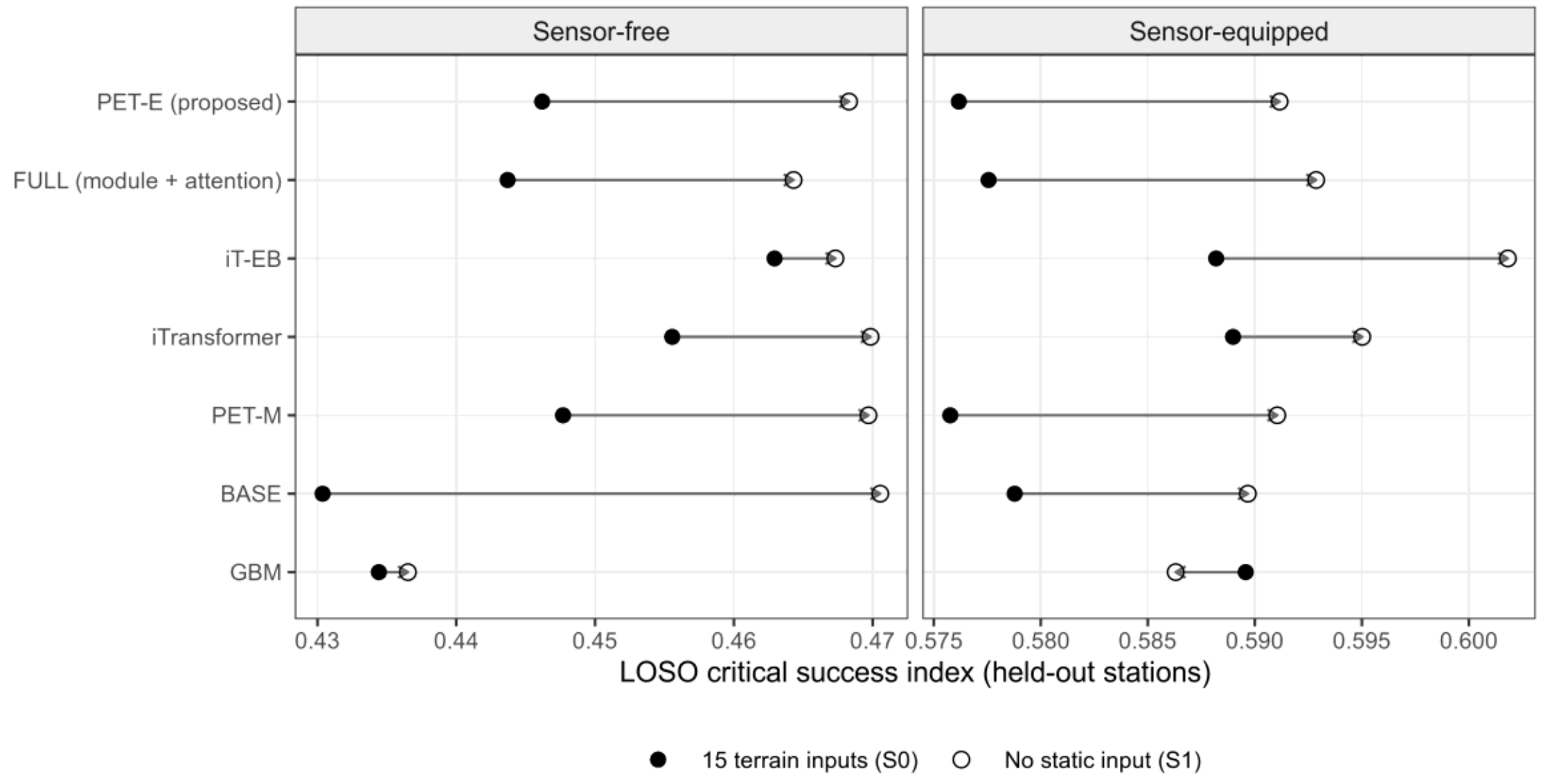

*Figure 7. Effect of the static terrain inputs in LOSO. Lead-mean CSI at held-out stations (mean over seeds 0–2) with 15 terrain inputs (S0, filled) and without static inputs (S1, open) in both sensor conditions; arrows point from S0 to S1. Horizontal-axis ranges differ between panels.*

In the sensor-free condition, the correlation between station freezing rate and mean predicted probability was 0.59 for PET-E, 0.57 for iT-EB, and 0.66 for GBM with S0, and 0.62, 0.59, and 0.63 with S1. For temporal extrapolation, sensor-free PET-M validation CSI decreased from 0.607 with S0 to 0.593 with S1. The magnitude of this difference (−0.014) exceeded twice the seed standard error (0.007), so S0 was retained for temporal evaluation under the rule in §4 (v).

## 5.8 False alarms at above-freezing leaf temperatures

In the 2025–2026 sensor-free setting, the fraction of alarms issued at observed leaf temperatures above 0 °C was 0.137 for PET-E, 0.189 for BASE, and 0.235 for GBM. Under the ≥ 2 °C criterion, the fractions were 0.051, 0.087, and 0.122, respectively; PET-E was significantly lower than BASE and GBM at all 12 leads (Table 3). In LOSO, this fraction was 0.058 for PET-E versus 0.124 for GBM with sensor-free S0, and 0.048 versus 0.073 with sensor-equipped S1 (Table 5).

## 5.9 Learned calibration parameters and transition scale

For PET-E, the learned parameters ranged from 1.04 to 1.56 °C h$^{-1}$ for $\rho_0$, 77.0 to 79.4 % for $H_0$, 35.5 to 36.2 W m$^{-2}$ K$^{-1}$ (m s$^{-1}$)$^{-1/2}$ for $k_h$ (Table S3). The attention coefficient $\kappa$ in FULL ranged from 0.69 to 0.92. The corresponding iT-EB ranges were 1.14–1.64 °C h$^{-1}$, 74.5–78.7 %, and 35.6–36.2 W m$^{-2}$ K$^{-1}$ (m s$^{-1}$)$^{-1/2}$. PET-E's learned $\delta$ ranged from 1.04 to 1.31 °C. Initializing it at 0.5 or 2.0 °C yielded ranges of 1.03–1.10 °C and 1.17–2.01 °C, respectively.

Across 44,285 precipitation-free nighttime samples in the 2025–2026 test season, the correlation between $D$ and the observed air-minus-leaf temperature difference was 0.478 with the initial parameter values and 0.492 with the learned parameters (Figure 8). The observed mean difference was 2.59 °C; mean $D$ was 1.96 °C initially and 1.78 °C after training. In sensor-equipped PET-M, learned favorability increased with wind speed in 4 seeds and with the night indicator in 2 seeds. Partial-dependence curves and attention weights are shown in Figure S1.

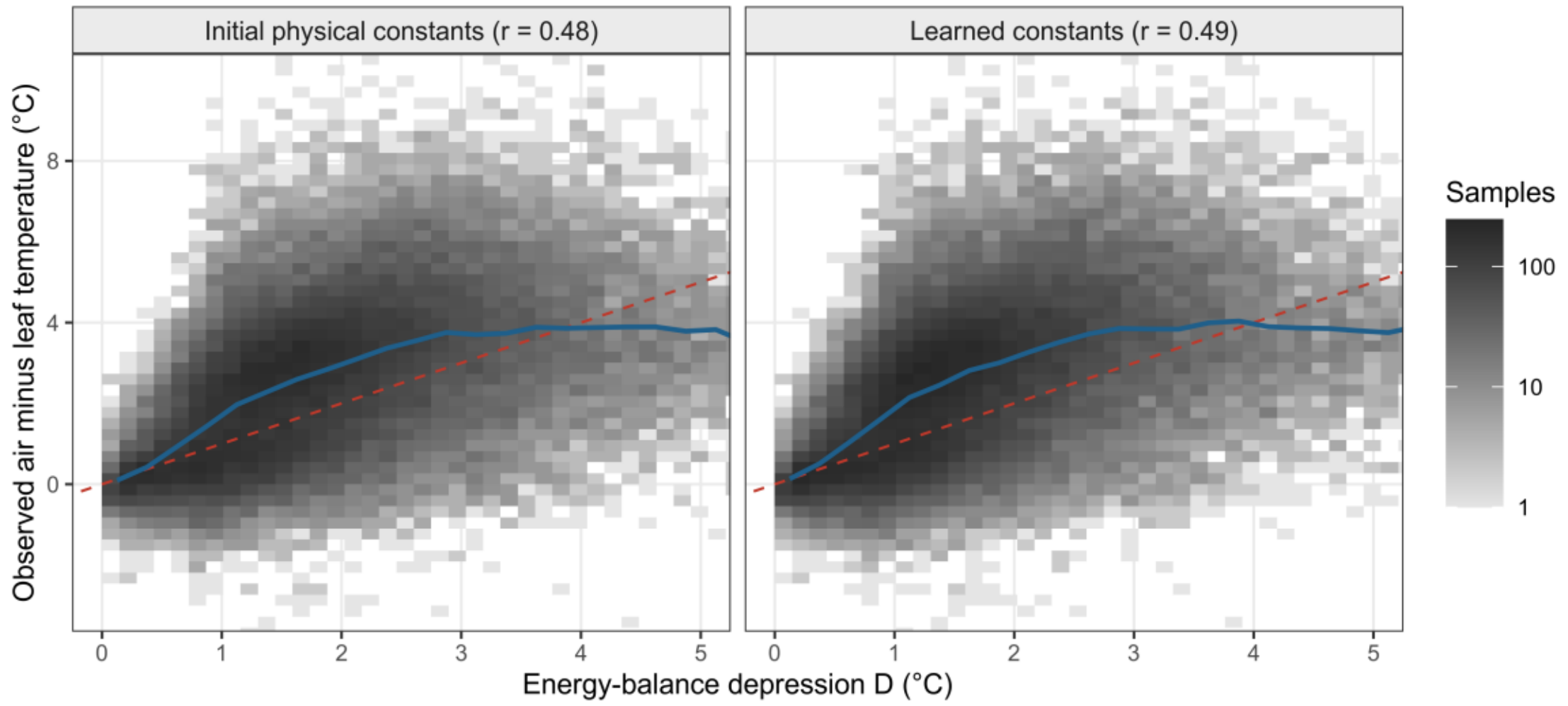

*Figure 8. Energy-balance depression D against the observed air-minus-leaf temperature difference (air temperature − leaf temperature) in the 2025–2026 test season at issue times under precipitation-free nighttime conditions. Left, D computed with the initial parameter values; right, mean of D computed with the learned parameters of the five sensor-free PET-E seeds. Gray cells, number of samples (log scale); solid line, mean observed difference per 0.25 °C bin of D (bins with at least 50 samples); dashed line, 1:1; r, Pearson correlation.*

For future leaf-temperature prediction, PET-E had lower lead-mean MAE than gate-only in all four temporal settings (Supplementary Table S5). In the 2025–2026 sensor-free and sensor-equipped settings, MAE was 2.127 versus 2.215 °C and 1.978 versus 2.034 °C, respectively; in 2024–2025, it was 2.300 versus 2.304 °C and 2.101 versus 2.111 °C. Lead-mean RMSE was lower for PET-E in three settings, with a small increase in the 2024–2025 sensor-free setting (2.943 versus 2.941 °C). For iTransformer, both errors were lower with the temperature path in three settings but higher in the 2024–2025 sensor-equipped setting. These are point estimates.

## 5.10 Probability recalibration and decision value

In the 2025–2026 sensor-free setting, maximum REV before recalibration was 0.634 for PET-E, 0.635 for iT-EB, and 0.605 for GBM (Table S4, Figure S2). After cap-preserving isotonic recalibration, PET-E and iT-EB achieved 0.638 and 0.640; GBM achieved 0.605 after isotonic recalibration. PET-E's Brier score decreased from 0.1053 to 0.1040 after cap-preserving recalibration. In the sensor-equipped setting, recalibrated maximum REV was 0.722 for PET-E, 0.729 for iT-EB, and 0.710 for GBM.

# 6. Discussion

This study combines a mathematically explicit temperature-dependent probability bound with controlled comparisons of where energy-balance information enters a transformer. The results distinguish the structural constraint supplied by the output pathway from the predictive contribution of the attention prior. Across both evaluated backbones, the complete module reduced Brier scores and warm-surface alarm fractions in the temporal evaluations. Table 6 summarizes these findings and their scope. The comparisons address three questions: whether an energy-balance reference can impose a useful constraint on frost probability, whether that module transfers between token designs, and how its effect changes when leaf sensors or station-specific terrain inputs are unavailable.

**Table 6.** Evaluation settings, comparisons, main results, and the scope of the supported claims.

| Setting | Comparison | Main result | Scope of the supported claim |
|---|---|---|---|
| Skill without leaf-sensor inputs (2025–2026, all 19 test stations) | PET-E vs GBM, iTransformer, PET-M, BASE | CSI 0.516 vs GBM 0.478 and iTransformer 0.509; higher than GBM at 12 of 12 leads | Simulated by removing leaf-sensor inputs; supported within one network (19 stations pooled) |
| Another season (2024–2025, sensor-free) | PET-E vs GBM, iTransformer | CSI 0.592 vs GBM 0.542 and iTransformer 0.573; higher than GBM at 12 and than iTransformer at 10 of 12 leads | Simulated sensor-free condition; supported for two seasons of one network |
| Decomposition of the physics module (sensor-free, 2025–2026) | Gate only, PET-E (TB), ATT, FULL | Fraction of alarms at observed leaf temp. ≥ 2 °C: gate only 0.060, PET-E 0.051 (lower at 2 leads); ATT CSI higher/lower than gate only at 1/0 leads | Gate alone lowered the fraction; the energy-balance path lowered it further in all four settings of the time-token backbone (significant at 2, 1, 5, and 1 leads); the incremental effects of the attention prior with the gate retained are reported separately for CSI and the alarm fraction |
| Transfer to iTransformer (two seasons × two sensor conditions) | iT-EB vs iTransformer | Fraction of alarms at observed leaf temp. ≥ 2 °C 0.063 vs 0.081 (sensor-free, 2025–2026); CSI higher at 11 and 12 leads in 2024–2025 | Lower fraction and Brier score in all four settings; no statistically significant CSI reduction at any lead |
| Onset and cessation (sensor-free, 2025–2026) | PET-E vs GBM | Cessation false-alarm rate 0.294 vs 0.443; onset POD 0.691 vs 0.766; Peirce skill score higher at 10 of 12 leads | Fewer cessation false alarms at a lower onset POD; higher Peirce skill score than GBM |
| New stations, sensor-free LOSO | PET-E, iT-EB vs GBM | 15 terrain inputs: CSI 0.446 vs 0.434; no static input: 0.468 (PET-E), 0.467 (iT-EB) vs 0.437, higher at 12 of 12 leads | Terrain and no-static-input configurations evaluated separately in held-out-station folds |
| New stations, sensor-equipped LOSO | iT-EB, PET-E vs GBM | No static input: CSI 0.602 (iT-EB), 0.591 (PET-E) vs 0.586; Peirce skill score higher at 12 (iT-EB) and 12 (PET-E) of 12 leads | Onset/cessation discrimination and overall CSI are evaluated separately |
| Leaf sensors available (2025–2026) | PET-E vs GBM, persistence | Lead-mean CSI 0.608 vs GBM 0.597; t+1 GBM and persistence higher | Overall lead-mean CSI and first-lead performance are reported separately |
| Decision value (sensor-free, 2025–2026) | PET-E vs GBM after recalibration | Maximum REV 0.638 vs 0.605 | Higher maximum relative economic value |

**Placement of physics.** The temperature path and attention prior made different contributions. In the time-token backbone, adding the energy-balance temperature path to the gate reduced the ≥ 2 °C alarm fraction in all four temporal settings and improved CSI at some leads in every setting (§5.3). Adding the energy-balance prior to attention while retaining the gate improved CSI in three settings, demonstrating predictive gains from a physically directed key-time prior. Its reduction of the alarm fraction, however, was significant in only one setting. The output path supplies a structural probability cap through the predicted temperature, whereas attention changes the weights assigned to inputs. This distinction connects the mathematical guarantee to the experimental comparison. No statistically significant additional gains in CSI, Peirce skill score, or either alarm fraction were detected from adding the attention prior on top of the temperature pathway under the evaluated settings (§5.3). A single output-side placement therefore delivers the predictive gain together with the structural bound. The prespecified validation criterion selected this more parsimonious configuration (TB) as PET-E, and the temporal and held-out-station evaluations use it.

**Interpretation and structural guarantees.** With the data-path logit fixed, frost probability cannot increase as predicted leaf temperature rises (Propositions 2 and 4). The multiplicative cap excludes alarms above the predicted-temperature bound (§3.6). Coupling the energy-balance reference to this gate raises the air-temperature bound for an alarm by the depression, with the residual and other arguments fixed; the depression is non-decreasing as wind speed, dew-point temperature, and relative humidity decrease and the air-temperature fall rate increases (Proposition 5). This direction is consistent with radiative frost on calm nights even above 0 °C air temperature (Snyder and de Melo-Abreu, 2005). The temperature reference retains units of °C and an explicit physical interpretation. After calibration, the estimated depression D was more strongly correlated with the observed air–leaf temperature difference, although its mean moved farther below the observed mean (§5.9). The residual can account for differences between the reference and observations; the reference alone is not a complete leaf-temperature model. The dependence of learned $\delta$ on its initialization also argues for interpreting it as an effective model scale rather than an identified supercooling width. Like structural constraints in climate and lake-temperature models (Beucler et al., 2021; Daw et al., 2020), the module enforces a specific relation rather than all governing physical processes.

**Cross-backbone applicability.** Across both temporal test seasons and both sensor conditions, the module-equipped PET-E and iT-EB reduced Brier score and the ≥ 2 °C alarm fraction relative to their plain backbones, with no significant CSI decrease (§5.3–§5.4). The energy-balance path gave additional reductions beyond the gate more consistently on the time-token backbone than on iTransformer. A possible explanation is that variate-token attention already represents relations among the weather variables used to construct the reference, but the experiments do not establish that mechanism. At held-out stations, the effect is more conditional: iT-EB improved BSS across the four sensor/static-input cases, but had a significant CSI decrease at one lead in sensor-equipped S0 (§5.6). Temporal evidence of no significant CSI loss should therefore not be extended to every spatial test.

**False alarms and state transitions.** An alarm issued for an observed leaf temperature above 0 °C is a false alarm under this study's label. Reducing such alarms can matter for unnecessary frost-protection activation even when overall CSI changes little. The observed reduction is an empirical result; the mathematical cap applies to predicted leaf temperature, not directly to the observed temperature. Probability recalibration alone does not enforce the temperature-dependent cap, whereas cap-preserving recalibration retains it (§3.9). Separately, onset and cessation metrics distinguish state-change discrimination from persistence. In the 2025–2026 sensor-free setting, PET-E improved overall CSI and reduced the fraction of alarms at observed leaf temperatures ≥ 2 °C relative to BASE, while onset POD was lower at the evaluated decision thresholds (Table 3, Table S7). Relative to GBM, PET-E also reduced cessation false alarms but had a lower onset POD (§5.5). Operational use should therefore consider onset sensitivity alongside overall skill and false alarms. Network climatology achieved a higher Peirce skill score than PET-E in that setting, showing that a high all-sample CSI does not imply the best transition discrimination. Its month–hour frequencies encode the timing of transitions within the seasonal and diurnal cycles. These metrics should be considered together when deciding when to start or stop protection (Snyder and de Melo-Abreu, 2005).

**Sensor-free deployment and terrain.** PET-E exceeded GBM in CSI at every lead in both sensor-free seasons. The input condition was simulated at observation-network stations, so it tests removal of leaf-sensor history rather than deployment at independently observed agricultural sites. The physics module nevertheless provides a surface-temperature reference without requiring a leaf sensor. This complements approaches that exclude unavailable surface-temperature predictors (Kim et al., 2024a). At held-out stations without leaf-sensor inputs, validation selected no static inputs for every evaluated model, including GBM. PET-E and iT-EB then exceeded GBM at all leads, and station-resampling intervals for both CSI differences were positive (§5.7). With static inputs, iT-EB already exceeded GBM, while PET-E did not differ significantly. Thus, the results do not support a blanket claim that terrain prevents useful spatial transfer. They show that static-input choice and backbone both matter. Nineteen stations provide limited coverage of a 15-variable terrain representation; station-specific associations may not transfer to a new site. This is an interpretation, not proof that the model memorized station identities. In temporal extrapolation, the validation comparison instead retained static inputs (§5.7).

**Comparison with physics-based loss penalties and operational implications.** The tested physics-penalty variants did not exceed PET-E in CSI (§5.2). This supports the evaluated structural formulation under the tested settings, rather than a general ranking of structural and loss-based physics incorporation (Karniadakis et al., 2021). Training-data reduction results are reported in Table 2; they do not establish a statistically significant advantage under every reduced-data condition. With leaf-sensor inputs, persistence and GBM remain strong first-lead baselines. Their advantage need not persist after removing the sensor history. Findings for tabular learning (Grinsztajn et al., 2022) provide context, but do not replace the matched time-series comparisons here. Model use should also account for calibration and the user's cost–loss ratio (Richardson, 2000; Bouttier and Marchal, 2024). In the sensor-free setting, cap-preserving recalibration gave PET-E a maximum REV of 0.638, compared with 0.605 for recalibrated GBM. These are retrospective decision-value estimates, not measured farm benefits.

**Limitations.** (1) Because the label depends on wetness detection, it does not capture black frost under dry conditions. (2) The prediction target is the surface freezing state, and the predictions were not validated against actual ice formation or crop-damage events. (3) The data come from a single network (19 stations) in a single country, and both test seasons come from the same network. (4) The energy-balance path estimates cloud from the air-temperature fall rate and relative humidity without observing it, fixes leaf size and emissivity, and neglects the latent heat of freezing and condensation and the heat storage of the leaf. The depression also uses only its value at issue time and therefore does not reflect future changes in cloud and wind. (5) The experiments assumed that the objective analysis grid was available at issue time and did not account for operational production latency. (6) GBM and iTransformer were trained with fixed settings without tuning. The GBM comparison concerns a single model trained jointly across leads and does not establish superiority over separately trained lead-specific or tuned models. iTransformer was the only external backbone evaluated; TFT (Lim et al., 2021), PatchTST (Nie et al., 2023), and DLinear (Zeng et al., 2023) were not evaluated. (7) The sensor-free condition simulates deployment by removing leaf-sensor inputs at AAFOS stations and was not validated at agricultural sites that actually lack leaf sensors. (8) In LOSO, neighboring stations observe the synoptic events of the same period as the held-out station, so the estimate of spatial generalization may be optimistic. (9) In the sensor-free condition, the onset–cessation discrimination was below that of network climatology (§5.5); coupling with short-range forecast fields and a two-gate factorization that adds a wetness gate are directions for improvement. (10) Performance was evaluated over October–April rather than separately for crop-specific flowering periods. Applicability to flowering-season frost protection therefore requires targeted evaluation.

## 7. Conclusion

We establish a transformer-compatible module that couples an energy-balance surface-temperature reference to a bounded freezing probability. Its conditional monotonicity and alarm boundary are explicit: with the residual and other arguments fixed, the air-temperature bound rises by the physically computed depression. The structure can be attached to both time-token and variate-token backbones.

In the 2025–2026 hold-out season without leaf-sensor inputs, PET-E achieved higher CSI than the fixed-configuration, joint-lead GBM model at all 12 leads, with lead-mean CSI values of 0.516 and 0.478, respectively. In the retrospective 2024–2025 evaluation, the corresponding values were 0.592 and 0.542; this evaluation was not independent of final configuration selection. Across these two seasonal evaluations and both sensor conditions, PET-E and iT-EB had lower Brier scores and lower fractions of alarms at observed leaf temperatures ≥ 2 °C than their respective unmodified backbones, with no statistically significant CSI decrease at any lead. In the time-token backbone, adding the energy-balance temperature path reduced this alarm fraction relative to the gate-only variant in all four settings; this incremental reduction was not consistent in iTransformer. Adding the energy-balance prior to attention while retaining the gate improved CSI in three settings but significantly reduced this alarm fraction in only one, and adding the attention prior on top of the temperature path yielded no statistically significant additional gains under the evaluated settings. These results distinguish the predictive contribution of attention from the temperature-dependent bound supplied by the output path.

At held-out stations without leaf-sensor inputs, validation favored excluding static inputs for every evaluated model. Under that configuration, sensor-free PET-E and iT-EB exceeded GBM at all leads, and both also improved onset–cessation Peirce skill in the sensor-equipped condition. Spatial results remain conditional on backbone and input configuration; in particular, the absence of significant CSI loss in temporal tests did not hold in every spatial comparison. The next steps are independent validation at agricultural sites without leaf sensors and in other networks, and refinement of the temperature reference using cloud observations and short-range forecast fields.

## CRediT authorship contribution statement

Juyeon Kim and Chaeyeon Yi contributed equally to this work. **Juyeon Kim:** Methodology, Software, Formal analysis, Investigation, Data curation, Validation, Visualization, Writing – original draft. **Chaeyeon Yi:** Conceptualization, Methodology, Data curation, Investigation, Resources, Writing – review & editing. **Yun Am Seo:** Conceptualization, Supervision, Writing – review & editing, Project administration, Funding acquisition.

## Declaration of competing interest

The authors declare that they have no known competing financial interests or personal relationships that could have appeared to influence the work reported in this paper.

## Declaration of generative AI and AI-assisted technologies in the writing process

During the preparation of this work the authors used Claude (Anthropic) and ChatGPT (OpenAI) to improve the readability and language of the manuscript. After using these tools, the authors reviewed and edited the content as needed and take full responsibility for the content of the published article.

## Data and code availability

The AAFOS station observations were provided by the National Institute of Meteorological Sciences of the KMA and cannot be redistributed by the authors; they can be requested from that institute. The KMA high-resolution blended-weather grid (500-m objective analysis grid) used as weather predictors is available from the KMA API Hub (apihub.kma.go.kr). The 100-m terrain and GIS variables were derived from products of their originating agencies. The code for preprocessing, modeling, and evaluation is available from the corresponding author on reasonable request.

## Funding

This research was supported by Global - Learning & Academic research institution for Master's·PhD students, and Postdocs (G-LAMP) Program of the National Research Foundation of Korea (NRF) grant funded by the Ministry of Education (No. RS-2026-25610465). This work was also supported by the Disaster Safety Ministry Collaboration R&D Program [RS-2025-02311088] of the Ministry of the Interior and Safety. The funding sources had no role in the study design; in the collection, analysis, and interpretation of data; in the writing of the report; or in the decision to submit the article for publication.

## Acknowledgments

The authors thank the National Institute of Meteorological Sciences of the Korea Meteorological Administration for providing the AAFOS observations.

## References

Alsafadi, K., Alatrach, B., Sammen, S.Sh., Cao, W., 2024. Prediction of daily leaf wetness duration using multi-step machine learning. Comput. Electron. Agric. 224, 109131. https://doi.org/10.1016/j.compag.2024.109131

Barooni, M., Ziarati, K., Barooni, A., 2023. Frost prediction using machine learning methods in Fars province. In: 28th International Computer Conference, Computer Society of Iran (CSICC). https://doi.org/10.1109/CSICC58665.2023.10105391

Beucler, T., Pritchard, M., Rasp, S., Ott, J., Baldi, P., Gentine, P., 2021. Enforcing analytic constraints in neural networks emulating physical systems. Phys. Rev. Lett. 126, 098302. https://doi.org/10.1103/PhysRevLett.126.098302

Bolton, D., 1980. The computation of equivalent potential temperature. Mon. Weather Rev. 108, 1046–1053. https://doi.org/10.1175/1520-0493(1980)108<1046:TCOEPT>2.0.CO;2

Bouaziz, S., Hafiane, A., Canals, R., Nedjai, R., 2026. SPyCer: semi-supervised physics-guided contextual attention for near-surface air temperature estimation from satellite imagery. arXiv preprint arXiv:2603.05219. https://doi.org/10.48550/arXiv.2603.05219

Bouttier, F., Marchal, H., 2024. Probabilistic short-range forecasts of high-precipitation events: optimal decision thresholds and predictability limits. Nat. Hazards Earth Syst. Sci. 24, 2793–2816. https://doi.org/10.5194/nhess-24-2793-2024

Brutsaert, W., 1975. On a derivable formula for long-wave radiation from clear skies. Water Resour. Res. 11, 742–744. https://doi.org/10.1029/WR011i005p00742

Chun, J.A., Lee, H.-J., Im, S.-H., Kim, D., Baek, S.-S., 2021. Comparative assessment of frost event prediction models using logistic regression, random forest, and LSTM networks (in Korean). J. Korea Water Resour. Assoc. 54, 667–680. https://doi.org/10.3741/JKWRA.2021.54.9.667

Crawford, T.M., Duchon, C.E., 1999. An improved parameterization for estimating effective atmospheric emissivity for use in calculating daytime downwelling longwave radiation. J. Appl. Meteorol. 38, 474–480. https://doi.org/10.1175/1520-0450(1999)038<0474:AIPFEE>2.0.CO;2

Das, P., Posch, A., Barber, N., Hicks, M., Vandal, T.J., Duffy, K., Singh, D., van Werkhoven, K., Ganguly, A.R., 2024. Hybrid physics-AI outperforms numerical weather prediction for extreme precipitation nowcasting. npj Clim. Atmos. Sci. 7, 282. https://doi.org/10.1038/s41612-024-00834-8

Daw, A., Thomas, R.Q., Carey, C.C., Read, J.S., Appling, A.P., Karpatne, A., 2020. Physics-guided architecture (PGA) of neural networks for quantifying uncertainty in lake temperature modeling. In: Proceedings of the 2020 SIAM International Conference on Data Mining (SDM), pp. 532–540. https://doi.org/10.1137/1.9781611976236.60

de Almeida, V.A., Anochi, J.A., Rozante, J.R., de Campos Velho, H.F., 2025. Machine learning with voting committee for frost prediction. Meteorology 4, 6. https://doi.org/10.3390/meteorology4010006

Diedrichs, A.L., Bromberg, F., Dujovne, D., Brun-Laguna, K., Watteyne, T., 2018. Prediction of frost events using machine learning and IoT sensing devices. IEEE Internet Things J. 5, 4589–4597. https://doi.org/10.1109/JIOT.2018.2867333

Ding, L., Noborio, K., Shibuya, K., 2019. Frost forecast using machine learning—from association to causality. Procedia Comput. Sci. 159, 1001–1010. https://doi.org/10.1016/j.procs.2019.09.267

Drepper, B., Bamps, B., Gobin, A., Van Orshoven, J., 2022. Strategies for managing spring frost risks in orchards: effectiveness and conditionality—a systematic review. Environ. Evid. 11, 29. https://doi.org/10.1186/s13750-022-00281-z

Grinsztajn, L., Oyallon, E., Varoquaux, G., 2022. Why do tree-based models still outperform deep learning on typical tabular data? In: Advances in Neural Information Processing Systems 35 (Datasets and Benchmarks Track). https://neurips.cc/virtual/2022/poster/55627 (accessed 15 September 2026)

Holm, S., 1979. A simple sequentially rejective multiple test procedure. Scand. J. Stat. 6, 65–70. http://www.jstor.org/stable/4615733 (accessed 10 November 2026)

Hua, W., Heinemann, P., He, L., 2025. Frost management in agriculture with advanced sensing, modeling, and artificial intelligent technologies: a review. Comput. Electron. Agric. 231, 110027. https://doi.org/10.1016/j.compag.2025.110027

Jordan, D.N., Smith, W.K., 1994. Energy balance analysis of nighttime leaf temperatures and frost formation in a subalpine environment. Agric. For. Meteorol. 71, 359–372. https://doi.org/10.1016/0168-1923(94)90020-5

Kang, S., Seong, S.-K., Sohn, E., Kim, J., Shim, J., 2025. Frost forecasting via weakly-supervised semantic segmentation of satellite imagery. GIScience Remote Sens. 62, 2496013. https://doi.org/10.1080/15481603.2025.2496013

Karniadakis, G.E., Kevrekidis, I.G., Lu, L., Perdikaris, P., Wang, S., Yang, L., 2021. Physics-informed machine learning. Nat. Rev. Phys. 3, 422–440. https://doi.org/10.1038/s42254-021-00314-5

Ke, G., Meng, Q., Finley, T., Wang, T., Chen, W., Ma, W., Ye, Q., Liu, T.-Y., 2017. LightGBM: A highly efficient gradient boosting decision tree. In: Advances in Neural Information Processing Systems 30 (NIPS 2017), pp. 3146–3154. https://papers.nips.cc/paper/6907-lightgbm-a-highly-efficient-gradient-boosting-decision-tree (accessed 10 November 2026)

Kim, H., Kim, S., 2022. A study on frost prediction model using machine learning (in Korean). Korean J. Appl. Stat. 35, 543–552. https://doi.org/10.5351/KJAS.2022.35.4.543

Kim, K.R., Jo, E.-S., Ko, M.-S., Kang, J.-H., Hwang, Y.-J., Lee, Y.-H., 2024b. Implementation of an automated agricultural frost observation system (AAFOS) (in Korean). Korean J. Agric. For. Meteorol. 26, 63–74. https://doi.org/10.5532/KJAFM.2024.26.1.63

Kim, T.-H., Chae, H.-K., 2024. Responses of agriculture and rural areas to the climate crisis: (4) Spring frost damage (in Korean). Issue+ No. 17. Korea Rural Economic Institute (KREI). https://repository.krei.re.kr/handle/2018.oak/30904 (accessed 7 September 2026)

Kim, Y., Heo, J., Kim, E.-S., Shim, K.-M., Jo, S., Kang, M., 2024a. Estimation of frost occurrence using multi-input deep learning (in Korean). Korean J. Agric. For. Meteorol. 26, 53–62. https://doi.org/10.5532/KJAFM.2024.26.1.53

Kim, Y., Shim, K.-M., Jung, M.-P., Choi, I.-T., 2017. Study on the estimation of frost occurrence classification using machine learning methods (in Korean). Korean J. Agric. For. Meteorol. 19, 86–92. https://doi.org/10.5532/KJAFM.2017.19.3.86

Künsch, H.R., 1989. The jackknife and the bootstrap for general stationary observations. Ann. Stat. 17, 1217–1241. https://doi.org/10.1214/aos/1176347265

Lim, B., Arık, S.Ö., Loeff, N., Pfister, T., 2021. Temporal fusion transformers for interpretable multi-horizon time series forecasting. Int. J. Forecast. 37, 1748–1764. https://doi.org/10.1016/j.ijforecast.2021.03.012

Lira, H., Martí, L., Sanchez-Pi, N., 2022. A graph neural network with spatio-temporal attention for multi-sources time series data: an application to frost forecast. Sensors 22, 1486. https://doi.org/10.3390/s22041486

Liu, Y., Hu, T., Zhang, H., Wu, H., Wang, S., Ma, L., Long, M., 2024. iTransformer: inverted transformers are effective for time series forecasting. In: International Conference on Learning

Representations (ICLR). https://openreview.net/forum?id=JePfAI8fah (accessed 10 November 2026)

Lyu, T., Zhang, W., Liu, H., 2026. Physics-informed teleconnection-aware transformer for global subseasonal-to-seasonal forecasting. In: Proceedings of the 32nd ACM SIGKDD Conference on Knowledge Discovery and Data Mining (KDD). https://doi.org/10.1145/3770854.3780198

Nie, Y., Nguyen, N.H., Sinthong, P., Kalagnanam, J., 2023. A time series is worth 64 words: long-term forecasting with transformers. In: International Conference on Learning Representations (ICLR). https://openreview.net/forum?id=Jbdc0vTOcol (accessed 10 November 2026)

Pérez Tárraga, J., Castillo-Cara, M., Arias-Antúnez, E., Dujovne, D., 2025. Frost forecasting through machine learning algorithms. Earth Sci. Inform. 18, 183. https://doi.org/10.1007/s12145-025-01710-6

Perez-Lebel, A., Varoquaux, G., Koyejo, S., Doutreligne, M., Le Morvan, M., 2025. Decision from suboptimal classifiers: Excess risk pre- and post-calibration. In: Proceedings of the 28th International Conference on Artificial Intelligence and Statistics, Proceedings of Machine Learning Research 258, pp. 2395–2403. https://proceedings.mlr.press/v258/perez-lebel25a.html (accessed 10 November 2026)

Prabha, T., Hoogenboom, G., 2008. Evaluation of the Weather Research and Forecasting model for two frost events. Comput. Electron. Agric. 64, 234–247. https://doi.org/10.1016/j.compag.2008.05.019

Raissi, M., Perdikaris, P., Karniadakis, G.E., 2019. Physics-informed neural networks: a deep learning framework for solving forward and inverse problems involving nonlinear partial differential equations. J. Comput. Phys. 378, 686–707. https://doi.org/10.1016/j.jcp.2018.10.045

Richardson, D.S., 2000. Skill and relative economic value of the ECMWF ensemble prediction system. Q. J. R. Meteorol. Soc. 126, 649–667. https://doi.org/10.1002/qj.49712656313

Rogers, R.R., Yau, M.K., 1989. A Short Course in Cloud Physics. 3rd ed. Pergamon Press.

Rozante, J.R., Ramirez, E., Ramirez, D., Rozante, G., 2023. Improved frost forecast using machine learning methods. Artif. Intell. Geosci. 4, 164–181. https://doi.org/10.1016/j.aiig.2023.10.001

Schaefer, J.T., 1990. The critical success index as an indicator of warning skill. Weather Forecast. 5, 570–575. https://doi.org/10.1175/1520-0434(1990)005<0570:TCSIAA>2.0.CO;2

Schuepp, P.H., 1993. Tansley Review No. 59: Leaf boundary layers. New Phytol. 125, 477–507. https://doi.org/10.1111/j.1469-8137.1993.tb03898.x

Shaw, P., Uszkoreit, J., Vaswani, A., 2018. Self-attention with relative position representations. In: Proceedings of NAACL-HLT, pp. 464–468. https://doi.org/10.18653/v1/N18-2074

Snyder, R.L., de Melo-Abreu, J.P., 2005. Frost Protection: Fundamentals, Practice and Economics. Food and Agriculture Organization of the United Nations, Rome. https://www.fao.org/4/y7223e/y7223e00.htm (accessed 7 September 2026)

Solís, M., Rojas-Herrera, V., 2021. Approaches for the prediction of leaf wetness duration with machine learning. Biomimetics 6, 29. https://doi.org/10.3390/biomimetics6020029

Su, Y., Zhao, L., Guo, J., Huang, W., Li, X., Li, H., Chen, J., 2026. GrasshopperML: A mechanism-guided machine learning framework with combinatorial optimization for grasshopper risk warning. Artif. Intell. Agric. 16, 1504–1518. https://doi.org/10.1016/j.aiia.2026.05.006

Talsma, C.J., Solander, K.C., Mudunuru, M.K., Crawford, B., Powell, M.R., 2023. Frost prediction using machine learning and deep neural network models. Front. Artif. Intell. 5, 963781. https://doi.org/10.3389/frai.2022.963781

Vaswani, A., Shazeer, N., Parmar, N., Uszkoreit, J., Jones, L., Gomez, A.N., Kaiser, Ł., Polosukhin, I., 2017. Attention is all you need. In: Advances in Neural Information Processing Systems 30. https://papers.nips.cc/paper/7181-attention-is-all-you-need (accessed 10 November 2026)

Verma, Y., Heinonen, M., Garg, V., 2024. ClimODE: climate and weather forecasting with physics-informed neural ODEs. In: International Conference on Learning Representations (ICLR). https://openreview.net/forum?id=xuY33XhEGR (accessed 10 November 2026)

Zadrozny, B., Elkan, C., 2002. Transforming classifier scores into accurate multiclass probability estimates. In: Proceedings of the Eighth ACM SIGKDD International Conference on Knowledge Discovery and Data Mining, pp. 694–699. https://doi.org/10.1145/775047.775151

Zeng, A., Chen, M., Zhang, L., Xu, Q., 2023. Are transformers effective for time series forecasting? In: Proceedings of the AAAI Conference on Artificial Intelligence 37, pp. 11121–11128. https://doi.org/10.1609/aaai.v37i9.26317

Zhang, Y., Long, M., Chen, K., Xing, L., Jin, R., Jordan, M.I., Wang, J., 2023. Skilful nowcasting of extreme precipitation with NowcastNet. Nature 619, 526–532. https://doi.org/10.1038/s41586-023-06184-4

Zhou, J., Ru, X., He, Z., Gong, K., Wang, J., Wang, Z., Li, M., Feng, H., Yu, Q., He, J., 2026. Integrating explainable machine learning models with geospatial features to enhance the prediction of late-spring frost risks in apple production. Artif. Intell. Agric. https://doi.org/10.1016/j.aiia.2026.08.017

**Supplementary material**

**A plug-in energy-balance freezing module for transformer frost nowcasting without leaf-sensor inputs**

Juyeon Kim, Chaeyeon Yi, Yun Am Seo

Contents: Tables S1–S10, Figures S1–S2, Supplementary Notes S1–S6 (method details), and Appendix A (proofs of Propositions 1–5 and Corollary 1). Table, figure, equation, and proposition numbers follow the main text.

# Supplementary tables

**Table S1.** Label sensitivity to the leaf-temperature threshold and the label definition; k, minimum number of freezing minutes per hour; reference definition, 0 °C with k = 5.

| Leaf-temperature threshold | Minutes per hour (k) | Hourly incidence | Agreement with reference |
|---|---|---|---|
| −2.0 °C | 1 | 0.106 | 0.972 |
| −2.0 °C | 5 | 0.101 | 0.970 |
| −2.0 °C | 10 | 0.096 | 0.966 |
| −1.0 °C | 1 | 0.120 | 0.985 |
| −1.0 °C | 5 | 0.114 | 0.984 |
| −1.0 °C | 10 | 0.110 | 0.980 |
| −0.5 °C | 1 | 0.128 | 0.991 |
| −0.5 °C | 5 | 0.121 | 0.991 |
| −0.5 °C | 10 | 0.117 | 0.986 |
| 0.0 °C | 1 | 0.137 | 0.994 |
| 0.0 °C | 5 | 0.131 | 1.000 |
| 0.0 °C | 10 | 0.126 | 0.995 |
| Freezing state (con) only | 5 | 0.114 | 0.983 |

**Table S2.** All metrics by method in the 2025–2026 test season (lead and seed means; ± for CSI, seed standard deviation). In this and the following tables, a blank cell in a leading column repeats the entry above it.

| Condition | Method | CSI | AP | Brier score | Onset POD | Cessation false-alarm rate | Peirce skill score | Transition AUC | Unseen-station CSI |
|---|---|---|---|---|---|---|---|---|---|
| sensor-equipped | PET-E | 0.608 ± 0.003 | 0.825 | 0.0847 | 0.600 | 0.283 | 0.317 | 0.721 | 0.459 |
| | iT-EB | 0.613 ± 0.002 | 0.832 | 0.0842 | 0.642 | 0.289 | 0.353 | 0.737 | 0.489 |
| | iTransformer | 0.607 ± 0.003 | 0.825 | 0.0913 | 0.624 | 0.316 | 0.308 | 0.705 | 0.489 |
| | PET-M | 0.596 ± 0.008 | 0.822 | 0.0915 | 0.684 | 0.333 | 0.351 | 0.721 | 0.470 |
| | PET-A | 0.604 ± 0.006 | 0.822 | 0.1041 | 0.619 | 0.322 | 0.297 | 0.702 | 0.487 |

| Condition | Method | CSI | AP | Brier score | Onset POD | Cessation false-alarm rate | Peirce skill score | Transition AUC | Unseen-station CSI |
|---|---|---|---|---|---|---|---|---|---|
| | Additive gate only | 0.604 ± 0.007 | 0.822 | 0.1066 | 0.623 | 0.319 | 0.304 | 0.707 | 0.485 |
| | BASE | 0.600 ± 0.008 | 0.821 | 0.1002 | 0.629 | 0.342 | 0.287 | 0.692 | 0.487 |
| | GBM | 0.597 ± 0.001 | 0.816 | 0.0859 | 0.623 | 0.363 | 0.260 | 0.660 | 0.487 |
| | Logistic regression | 0.519 | 0.741 | 0.1144 | 0.560 | 0.367 | 0.193 | 0.627 | 0.395 |
| | Two-stage ($\hat{T}$ threshold) | 0.429 ± 0.002 | 0.472 | — | 0.833 | 0.387 | 0.446 | 0.779 | 0.278 |
| | Classical rule ($\hat{T} < 0 \wedge$ wet) | 0.499 ± 0.001 | — | — | 0.149 | 0.393 | −0.244 | — | 0.382 |
| | Loss-embedding (β = 0.3) | 0.607 ± 0.007 | 0.824 | 0.0962 | 0.624 | 0.312 | 0.313 | 0.714 | 0.494 |
| | Transition-weighted PET-A | 0.577 ± 0.004 | 0.784 | 0.1275 | 0.724 | 0.159 | 0.565 | 0.860 | 0.450 |
| | Transition-weighted BASE | 0.568 ± 0.009 | 0.773 | 0.1394 | 0.746 | 0.159 | 0.588 | 0.866 | 0.448 |
| | Persistence | 0.406 | — | — | 0.000 | 1.000 | −1.000 | — | 0.311 |
| | Network climatology | 0.351 | 0.459 | 0.1416 | 0.908 | 0.459 | 0.449 | 0.776 | 0.226 |
| sensor-free | PET-E | 0.516 ± 0.009 | 0.737 | 0.1053 | 0.691 | 0.294 | 0.397 | 0.764 | 0.360 |
| | iT-EB | 0.511 ± 0.006 | 0.741 | 0.1078 | 0.746 | 0.292 | 0.454 | 0.792 | 0.377 |
| | iTransformer | 0.509 ± 0.005 | 0.729 | 0.1120 | 0.745 | 0.345 | 0.400 | 0.756 | 0.377 |
| | PET-M | 0.505 ± 0.006 | 0.736 | 0.1118 | 0.767 | 0.331 | 0.436 | 0.778 | 0.374 |
| | PET-A | 0.501 ± 0.007 | 0.724 | 0.1256 | 0.772 | 0.350 | 0.422 | 0.766 | 0.376 |
| | Additive gate only | 0.494 ± 0.017 | 0.713 | 0.1352 | 0.795 | 0.357 | 0.439 | 0.771 | 0.361 |
| | BASE | 0.496 ± 0.007 | 0.710 | 0.1289 | 0.795 | 0.346 | 0.449 | 0.774 | 0.367 |
| | GBM | 0.478 ± 0.003 | 0.707 | 0.1108 | 0.766 | 0.443 | 0.323 | 0.710 | 0.362 |
| | Logistic regression | 0.441 | 0.636 | 0.1333 | 0.719 | 0.412 | 0.307 | 0.707 | 0.311 |
| | Network climatology | 0.351 | 0.459 | 0.1416 | 0.908 | 0.459 | 0.449 | 0.776 | 0.226 |

**Table S3.** Learned calibration parameters and transition scale δ. Calibration parameters and δ are given as the minimum–maximum over five seeds (δ is a scalar shared by all leads). $k_h$ is in W $m^{-2}$ $K^{-1}$ $(m\ s^{-1})^{-1/2}$. Test CSI is the lead and seed mean; the initial-value row gives values before

training. The last five rows give the sensitivity to the initial value of δ (2025–2026 season, sensor-equipped; PET-E with 0.5 and 2, PET-M with the default, 0.5, and 2).

| Model | Season | Condition | $\rho_0$ (°C h$^{-1}$) | $H_0$ (%) | $k_h$ | κ | δ (°C) | Test CSI |
|---|---|---|---|---|---|---|---|---|
| PET-E | 2025–2026 | sensor-free | 1.11–1.50 | 77.1–79.0 | 35.5–36.1 | — | 1.13–1.31 | 0.516 |
| | 2025–2026 | sensor-equipped | 1.15–1.56 | 77.0–78.9 | 35.6–36.2 | — | 1.14–1.21 | 0.608 |
| | 2024–2025 | sensor-free | 1.04–1.08 | 79.4–79.4 | 35.5–35.5 | — | 1.04–1.07 | 0.592 |
| | 2024–2025 | sensor-equipped | 1.06–1.26 | 78.2–79.3 | 35.5–35.8 | — | 1.06–1.18 | 0.663 |
| FULL | 2025–2026 | sensor-free | 1.18–1.30 | 76.5–77.8 | 35.8–36.1 | 0.69–0.76 | 1.29–1.31 | 0.512 |
| | 2025–2026 | sensor-equipped | 1.15–1.42 | 75.5–78.6 | 35.7–36.1 | 0.71–0.80 | 1.21–1.23 | 0.610 |
| | 2024–2025 | sensor-free | 1.04–1.07 | 79.0–79.3 | 35.5–35.5 | 0.87–0.92 | 1.04–1.09 | 0.591 |
| | 2024–2025 | sensor-equipped | 1.07–1.23 | 78.4–79.4 | 35.6–35.8 | 0.78–0.88 | 1.07–1.20 | 0.663 |
| iT-EB | 2025–2026 | sensor-free | 1.31–1.46 | 76.0–77.2 | 35.8–36.0 | — | 1.41–1.44 | 0.511 |
| | 2025–2026 | sensor-equipped | 1.37–1.64 | 74.5–77.2 | 35.9–36.2 | — | 1.27–1.35 | 0.613 |
| | 2024–2025 | sensor-free | 1.14–1.24 | 77.6–78.6 | 35.6–35.7 | — | 1.16–1.32 | 0.589 |
| | 2024–2025 | sensor-equipped | 1.14–1.37 | 77.1–78.7 | 35.6–35.8 | — | 1.13–1.33 | 0.669 |
| Initial value | — | — | 1.00 | 80.0 | 35.4 | 1.00 | 0.99 | — |
| PET-E, δ0 = 0.5 | 2025–2026 | sensor-equipped | — | — | — | — | 1.03–1.10 | 0.608 |
| PET-E, δ0 = 2 | 2025–2026 | sensor-equipped | — | — | — | — | 1.17–2.01 | 0.603 |
| PET-M, δ0 = 0.99 (default) | 2025–2026 | sensor-equipped | — | — | — | — | 1.31–1.38 | 0.596 |
| PET-M, δ0 = 0.5 | 2025–2026 | sensor-equipped | — | — | — | — | 1.01–1.22 | 0.604 |
| PET-M, δ0 = 2 | 2025–2026 | sensor-equipped | — | — | — | — | 1.31–1.65 | 0.598 |

**Table S4.** Probability recalibration and relative economic value (2025–2026 test season; lead and seed means). AUC, area under the ROC curve over all test samples; reliability, the reliability term of the Brier score computed over 10 equal-width probability bins (Murphy, 1973). Cap-preserving recalibration was applied only to models with a multiplicative gate.

| Condition | Method | Calibration | Brier score | Reliability | AUC | Maximum REV | Cost–loss ratios with REV > 0 (of 25) |
|---|---|---|---|---|---|---|---|
| sensor-free | PET-E | Uncalibrated | 0.1053 | 0.0065 | 0.907 | 0.634 | 25 |
| | | Isotonic | 0.1043 | 0.0052 | 0.906 | 0.638 | 25 |
| | | Cap-preserving isotonic | 0.1040 | 0.0049 | 0.907 | 0.638 | 25 |
| | FULL | Uncalibrated | 0.1056 | 0.0064 | 0.905 | 0.632 | 25 |
| | | Isotonic | 0.1048 | 0.0054 | 0.905 | 0.634 | 25 |

| Condition | Method | Calibration | Brier score | Reliability | AUC | Maximum REV | Cost–loss ratios with REV > 0 (of 25) |
|---|---|---|---|---|---|---|---|
| | | Cap-preserving isotonic | 0.1045 | 0.0051 | 0.906 | 0.634 | 25 |
| | iT-EB | Uncalibrated | 0.1078 | 0.0092 | 0.907 | 0.635 | 25 |
| | | Isotonic | 0.1044 | 0.0056 | 0.907 | 0.639 | 25 |
| | | Cap-preserving isotonic | 0.1041 | 0.0053 | 0.907 | 0.640 | 25 |
| | iTransformer | Uncalibrated | 0.1120 | 0.0122 | 0.907 | 0.626 | 25 |
| | | Isotonic | 0.1051 | 0.0050 | 0.907 | 0.637 | 25 |
| | PET-M | Uncalibrated | 0.1118 | 0.0131 | 0.908 | 0.619 | 25 |
| | | Isotonic | 0.1079 | 0.0088 | 0.908 | 0.631 | 25 |
| | | Cap-preserving isotonic | 0.1076 | 0.0086 | 0.908 | 0.631 | 25 |
| | BASE | Uncalibrated | 0.1289 | 0.0259 | 0.902 | 0.575 | 25 |
| | | Isotonic | 0.1126 | 0.0094 | 0.901 | 0.621 | 25 |
| | GBM | Uncalibrated | 0.1108 | 0.0056 | 0.893 | 0.605 | 25 |
| | | Isotonic | 0.1115 | 0.0062 | 0.893 | 0.605 | 25 |
| sensor-equipped | PET-E | Uncalibrated | 0.0847 | 0.0070 | 0.938 | 0.716 | 25 |
| | | Isotonic | 0.0786 | 0.0009 | 0.938 | 0.722 | 25 |
| | | Cap-preserving isotonic | 0.0786 | 0.0009 | 0.938 | 0.722 | 25 |
| | FULL | Uncalibrated | 0.0848 | 0.0074 | 0.939 | 0.716 | 25 |
| | | Isotonic | 0.0784 | 0.0010 | 0.939 | 0.725 | 25 |
| | | Cap-preserving isotonic | 0.0784 | 0.0010 | 0.939 | 0.725 | 25 |
| | iT-EB | Uncalibrated | 0.0842 | 0.0085 | 0.941 | 0.720 | 25 |
| | | Isotonic | 0.0771 | 0.0015 | 0.941 | 0.729 | 25 |
| | | Cap-preserving isotonic | 0.0771 | 0.0015 | 0.941 | 0.729 | 25 |
| | iTransformer | Uncalibrated | 0.0913 | 0.0139 | 0.940 | 0.698 | 25 |
| | | Isotonic | 0.0792 | 0.0019 | 0.940 | 0.723 | 25 |
| | PET-M | Uncalibrated | 0.0915 | 0.0130 | 0.938 | 0.693 | 25 |
| | | Isotonic | 0.0818 | 0.0034 | 0.938 | 0.711 | 25 |
| | | Cap-preserving isotonic | 0.0818 | 0.0033 | 0.938 | 0.711 | 25 |
| | BASE | Uncalibrated | 0.1002 | 0.0214 | 0.939 | 0.672 | 25 |
| | | Isotonic | 0.0809 | 0.0024 | 0.939 | 0.718 | 25 |
| | GBM | Uncalibrated | 0.0859 | 0.0067 | 0.935 | 0.703 | 25 |

| Condition | Method | Calibration | Brier score | Reliability | AUC | Maximum REV | Cost–loss ratios with REV > 0 (of 25) |
|---|---|---|---|---|---|---|---|
| | | Isotonic | 0.0815 | 0.0021 | 0.934 | 0.710 | 25 |

**Table S5.** Lead-specific leaf-temperature prediction errors.

MAE and RMSE are in °C. Each cell reports MAE / RMSE, averaged over five seeds after computing each error separately by seed and lead on identical test samples. The Mean row is the arithmetic mean across leads, not a pooled RMSE. Differences are descriptive; no significance test was performed. PET-E is the validation-selected TB configuration.

### 2025–2026, without leaf-sensor inputs

| Lead | gate-only | PET-E | iT+gate | iT-EB |
|---|---|---|---|---|
| 1 | 1.624 / 2.053 | 1.370 / 1.773 | 1.481 / 1.890 | 1.397 / 1.799 |
| 2 | 1.743 / 2.212 | 1.547 / 2.002 | 1.647 / 2.103 | 1.579 / 2.039 |
| 3 | 1.877 / 2.393 | 1.720 / 2.224 | 1.804 / 2.311 | 1.741 / 2.251 |
| 4 | 2.002 / 2.561 | 1.874 / 2.423 | 1.935 / 2.481 | 1.883 / 2.438 |
| 5 | 2.107 / 2.703 | 2.012 / 2.600 | 2.046 / 2.624 | 2.005 / 2.599 |
| 6 | 2.200 / 2.826 | 2.133 / 2.753 | 2.140 / 2.746 | 2.123 / 2.753 |
| 7 | 2.298 / 2.947 | 2.249 / 2.897 | 2.232 / 2.861 | 2.225 / 2.879 |
| 8 | 2.389 / 3.059 | 2.354 / 3.025 | 2.322 / 2.970 | 2.320 / 3.000 |
| 9 | 2.474 / 3.159 | 2.447 / 3.136 | 2.396 / 3.058 | 2.390 / 3.083 |
| 10 | 2.546 / 3.249 | 2.528 / 3.230 | 2.460 / 3.136 | 2.466 / 3.173 |
| 11 | 2.618 / 3.333 | 2.606 / 3.323 | 2.533 / 3.223 | 2.531 / 3.249 |
| 12 | 2.707 / 3.437 | 2.685 / 3.419 | 2.607 / 3.313 | 2.598 / 3.328 |
| Mean | 2.215 / 2.828 | 2.127 / 2.734 | 2.134 / 2.726 | 2.105 / 2.716 |

### 2025–2026, with leaf-sensor inputs

| Lead | gate-only | PET-E | iT+gate | iT-EB |
|---|---|---|---|---|
| 1 | 1.113 / 1.445 | 0.986 / 1.289 | 1.036 / 1.342 | 1.023 / 1.331 |
| 2 | 1.410 / 1.825 | 1.269 / 1.664 | 1.320 / 1.708 | 1.300 / 1.688 |
| 3 | 1.636 / 2.110 | 1.504 / 1.967 | 1.543 / 1.995 | 1.520 / 1.971 |
| 4 | 1.799 / 2.317 | 1.708 / 2.225 | 1.739 / 2.244 | 1.703 / 2.206 |
| 5 | 1.943 / 2.500 | 1.875 / 2.436 | 1.880 / 2.421 | 1.862 / 2.406 |
| 6 | 2.066 / 2.653 | 2.022 / 2.622 | 2.020 / 2.596 | 1.995 / 2.575 |
| 7 | 2.179 / 2.793 | 2.153 / 2.783 | 2.135 / 2.733 | 2.115 / 2.726 |
| 8 | 2.279 / 2.914 | 2.265 / 2.920 | 2.235 / 2.854 | 2.223 / 2.858 |
| 9 | 2.369 / 3.020 | 2.366 / 3.041 | 2.314 / 2.945 | 2.316 / 2.965 |
| 10 | 2.453 / 3.124 | 2.455 / 3.149 | 2.394 / 3.040 | 2.395 / 3.061 |
| 11 | 2.536 / 3.233 | 2.531 / 3.239 | 2.478 / 3.145 | 2.478 / 3.159 |
| 12 | 2.629 / 3.340 | 2.606 / 3.330 | 2.563 / 3.252 | 2.561 / 3.255 |
| Mean | 2.034 / 2.606 | 1.978 / 2.555 | 1.972 / 2.523 | 1.958 / 2.517 |

### 2024–2025, without leaf-sensor inputs

| Lead | gate-only | PET-E | iT+gate | iT-EB |
|---|---|---|---|---|
| 1 | 1.710 / 2.175 | 1.563 / 2.006 | 1.704 / 2.190 | 1.469 / 1.897 |
| 2 | 1.835 / 2.344 | 1.765 / 2.261 | 1.836 / 2.364 | 1.670 / 2.149 |
| 3 | 1.958 / 2.508 | 1.932 / 2.484 | 1.967 / 2.537 | 1.847 / 2.372 |
| 4 | 2.077 / 2.663 | 2.084 / 2.681 | 2.089 / 2.691 | 2.008 / 2.572 |
| 5 | 2.188 / 2.808 | 2.212 / 2.848 | 2.198 / 2.823 | 2.146 / 2.749 |
| 6 | 2.301 / 2.950 | 2.322 / 2.984 | 2.305 / 2.951 | 2.272 / 2.910 |
| 7 | 2.398 / 3.069 | 2.424 / 3.106 | 2.398 / 3.063 | 2.374 / 3.032 |
| 8 | 2.506 / 3.195 | 2.519 / 3.218 | 2.467 / 3.145 | 2.457 / 3.136 |
| 9 | 2.593 / 3.302 | 2.596 / 3.312 | 2.539 / 3.235 | 2.536 / 3.242 |
| 10 | 2.636 / 3.356 | 2.654 / 3.386 | 2.605 / 3.316 | 2.613 / 3.336 |
| 11 | 2.678 / 3.410 | 2.716 / 3.462 | 2.662 / 3.389 | 2.680 / 3.421 |
| 12 | 2.769 / 3.514 | 2.808 / 3.572 | 2.715 / 3.454 | 2.737 / 3.494 |
| Mean | 2.304 / 2.941 | 2.300 / 2.943 | 2.290 / 2.930 | 2.234 / 2.859 |

## 2024–2025, with leaf-sensor inputs

| Lead | gate-only | PET-E | iT+gate | iT-EB |
|---|---|---|---|---|
| 1 | 1.140 / 1.521 | 1.133 / 1.494 | 1.162 / 1.527 | 1.181 / 1.553 |
| 2 | 1.427 / 1.888 | 1.424 / 1.880 | 1.453 / 1.910 | 1.458 / 1.907 |
| 3 | 1.668 / 2.182 | 1.662 / 2.186 | 1.674 / 2.193 | 1.692 / 2.206 |
| 4 | 1.857 / 2.410 | 1.843 / 2.414 | 1.853 / 2.418 | 1.876 / 2.437 |
| 5 | 2.020 / 2.611 | 1.992 / 2.596 | 2.012 / 2.611 | 2.028 / 2.624 |
| 6 | 2.154 / 2.774 | 2.135 / 2.769 | 2.160 / 2.787 | 2.165 / 2.790 |
| 7 | 2.277 / 2.922 | 2.258 / 2.916 | 2.281 / 2.932 | 2.288 / 2.938 |
| 8 | 2.388 / 3.051 | 2.371 / 3.051 | 2.390 / 3.064 | 2.389 / 3.059 |
| 9 | 2.491 / 3.172 | 2.480 / 3.178 | 2.469 / 3.159 | 2.485 / 3.172 |
| 10 | 2.567 / 3.267 | 2.561 / 3.275 | 2.533 / 3.241 | 2.564 / 3.270 |
| 11 | 2.630 / 3.345 | 2.635 / 3.365 | 2.596 / 3.317 | 2.629 / 3.350 |
| 12 | 2.714 / 3.446 | 2.717 / 3.460 | 2.664 / 3.398 | 2.690 / 3.427 |
| Mean | 2.111 / 2.716 | 2.101 / 2.715 | 2.104 / 2.713 | 2.120 / 2.728 |

**Table S6.** CSI by lead time in the 2025–2026 test season (seed means; lead mean ± seed standard deviation). Seven of the 12 leads are shown; the lead mean is over all 12 leads. Decision thresholds were selected on the validation set and frozen.

| Condition | Method | t+1 | t+2 | t+3 | t+4 | t+6 | t+9 | t+12 | Lead mean |
|---|---|---|---|---|---|---|---|---|---|
| sensor-equipped | PET-E | 0.799 | 0.728 | 0.683 | 0.646 | 0.592 | 0.542 | 0.510 | 0.608 ± 0.003 |
| | iT-EB | 0.813 | 0.740 | 0.689 | 0.656 | 0.601 | 0.541 | 0.507 | 0.613 ± 0.002 |
| | iTransformer | 0.801 | 0.721 | 0.674 | 0.642 | 0.589 | 0.544 | 0.513 | 0.607 ± 0.003 |
| | PET-M | 0.784 | 0.715 | 0.668 | 0.633 | 0.582 | 0.529 | 0.501 | 0.596 ± 0.008 |
| | BASE | 0.781 | 0.714 | 0.671 | 0.636 | 0.585 | 0.538 | 0.509 | 0.600 ± 0.008 |
| | GBM | 0.834 | 0.730 | 0.667 | 0.618 | 0.573 | 0.527 | 0.501 | 0.597 ± 0.001 |
| | Logistic regression | 0.701 | 0.623 | 0.573 | 0.540 | 0.500 | 0.472 | 0.461 | 0.519 |

| Condition | Method | t+1 | t+2 | t+3 | t+4 | t+6 | t+9 | t+12 | Lead mean |
|---|---|---|---|---|---|---|---|---|---|
| | Two-stage ($\hat{T}$ threshold) | 0.475 | 0.460 | 0.449 | 0.441 | 0.428 | 0.412 | 0.404 | 0.429 ± 0.002 |
| | Classical rule ($\hat{T} < 0 \wedge$ wet) | 0.823 | 0.729 | 0.654 | 0.589 | 0.483 | 0.367 | 0.314 | 0.499 ± 0.001 |
| | Loss-embedding ($\beta = 0.3$) | 0.786 | 0.720 | 0.673 | 0.640 | 0.591 | 0.544 | 0.516 | 0.607 ± 0.007 |
| | Transition-weighted PET-A | 0.759 | 0.694 | 0.646 | 0.612 | 0.559 | 0.514 | 0.484 | 0.577 ± 0.004 |
| | Transition-weighted BASE | 0.738 | 0.675 | 0.637 | 0.601 | 0.553 | 0.511 | 0.480 | 0.568 ± 0.009 |
| | Persistence | 0.804 | 0.674 | 0.575 | 0.493 | 0.370 | 0.256 | 0.211 | 0.406 |
| | Network climatology | 0.351 | 0.351 | 0.351 | 0.351 | 0.351 | 0.351 | 0.350 | 0.351 |
| sensor-free | PET-E | 0.590 | 0.571 | 0.556 | 0.542 | 0.518 | 0.483 | 0.460 | 0.516 ± 0.009 |
| | iT-EB | 0.583 | 0.564 | 0.545 | 0.536 | 0.513 | 0.482 | 0.463 | 0.511 ± 0.006 |
| | iTransformer | 0.572 | 0.561 | 0.544 | 0.531 | 0.509 | 0.485 | 0.462 | 0.509 ± 0.005 |
| | PET-M | 0.574 | 0.559 | 0.545 | 0.527 | 0.504 | 0.477 | 0.458 | 0.505 ± 0.006 |
| | BASE | 0.561 | 0.545 | 0.534 | 0.513 | 0.493 | 0.466 | 0.453 | 0.496 ± 0.007 |
| | GBM | 0.538 | 0.521 | 0.507 | 0.496 | 0.483 | 0.451 | 0.433 | 0.478 ± 0.003 |
| | Logistic regression | 0.480 | 0.464 | 0.456 | 0.458 | 0.445 | 0.420 | 0.421 | 0.441 |
| | Network climatology | 0.351 | 0.351 | 0.351 | 0.351 | 0.351 | 0.351 | 0.350 | 0.351 |

**Table S7.** Discrimination of freezing onset and cessation in two test seasons (lead and seed means).

| Season | Condition | Method | Onset POD | Cessation false-alarm rate | Peirce skill score | Transition AUC |
|---|---|---|---|---|---|---|
| 2025–2026 | sensor-free | PET-E | 0.691 | 0.294 | 0.397 | 0.764 |
| | | iT-EB | 0.746 | 0.292 | 0.454 | 0.792 |
| | | PET-M | 0.767 | 0.331 | 0.436 | 0.778 |
| | | BASE | 0.795 | 0.346 | 0.449 | 0.774 |
| | | GBM | 0.766 | 0.443 | 0.323 | 0.710 |
| | | Logistic regression | 0.719 | 0.412 | 0.307 | 0.707 |
| | | Network climatology | 0.908 | 0.459 | 0.449 | 0.776 |
| | sensor-equipped | PET-E | 0.600 | 0.283 | 0.317 | 0.721 |
| | | iT-EB | 0.642 | 0.289 | 0.353 | 0.737 |
| | | PET-M | 0.684 | 0.333 | 0.351 | 0.721 |
| | | BASE | 0.629 | 0.342 | 0.287 | 0.692 |
| | | GBM | 0.623 | 0.363 | 0.260 | 0.660 |

| Season | Condition | Method | Onset POD | Cessation false-alarm rate | Peirce skill score | Transition AUC |
|---|---|---|---|---|---|---|
| | | Logistic regression | 0.560 | 0.367 | 0.193 | 0.627 |
| | | Network climatology | 0.908 | 0.459 | 0.449 | 0.776 |
| 2024–2025 | sensor-free | PET-E | 0.718 | 0.356 | 0.362 | 0.740 |
| | | iT-EB | 0.725 | 0.330 | 0.395 | 0.748 |
| | | PET-M | 0.734 | 0.365 | 0.369 | 0.736 |
| | | BASE | 0.704 | 0.341 | 0.362 | 0.729 |
| | | GBM | 0.668 | 0.452 | 0.215 | 0.641 |
| | | Logistic regression | 0.701 | 0.384 | 0.317 | 0.696 |
| | | Network climatology | 0.724 | 0.277 | 0.447 | 0.765 |
| | sensor-equipped | PET-E | 0.656 | 0.372 | 0.284 | 0.681 |
| | | iT-EB | 0.655 | 0.314 | 0.341 | 0.712 |
| | | PET-M | 0.666 | 0.391 | 0.276 | 0.667 |
| | | BASE | 0.648 | 0.394 | 0.254 | 0.657 |
| | | GBM | 0.622 | 0.448 | 0.174 | 0.596 |
| | | Logistic regression | 0.608 | 0.377 | 0.231 | 0.635 |
| | | Network climatology | 0.724 | 0.277 | 0.447 | 0.765 |

**Table S8.** Static-input selection in LOSO. Configurations were selected by the validation-period CSI of the 14 trained stations, each predicted by the fold model that excluded it. The station-level r is the Pearson correlation over 19 stations between the station freezing rate and the station mean predicted probability (seed means); parentheses give the 95% station-resampling interval.

| Condition | Method | Static input | Validation CSI (14 trained stations) | LOSO CSI | LOSO BSS | Station-level r (95% CI) |
|---|---|---|---|---|---|---|
| sensor-free | PET-E | S0 | 0.550 | 0.446 | 0.095 | 0.59 (0.39, 0.76) |
| | FULL | S0 | 0.547 | 0.444 | 0.085 | 0.56 (0.36, 0.76) |
| | iT-EB | S0 | 0.553 | 0.463 | 0.124 | 0.57 (0.35, 0.79) |
| | iTransformer | S0 | 0.541 | 0.456 | 0.082 | 0.59 (0.40, 0.78) |
| | PET-M | S0 | 0.526 | 0.448 | 0.092 | 0.59 (0.34, 0.79) |
| | BASE | S0 | 0.511 | 0.430 | −0.002 | 0.47 (0.14, 0.70) |
| | GBM | S0 | 0.512 | 0.434 | 0.129 | 0.66 (0.45, 0.86) |
| | PET-E | S1 | 0.575 | 0.468 | 0.121 | 0.62 (0.42, 0.81) |
| | FULL | S1 | 0.573 | 0.464 | 0.105 | 0.59 (0.35, 0.80) |
| | iT-EB | S1 | 0.571 | 0.467 | 0.112 | 0.59 (0.35, 0.80) |

| Condition | Method | Static input | Validation CSI (14 trained stations) | LOSO CSI | LOSO BSS | Station-level r (95% CI) |
|---|---|---|---|---|---|---|
| | iTransformer | S1 | 0.562 | 0.470 | 0.071 | 0.62 (0.39, 0.82) |
| | PET-M | S1 | 0.569 | 0.470 | 0.106 | 0.63 (0.40, 0.82) |
| | BASE | S1 | 0.563 | 0.471 | 0.075 | 0.66 (0.45, 0.85) |
| | GBM | S1 | 0.518 | 0.437 | 0.108 | 0.63 (0.41, 0.82) |
| sensor-equipped | PET-E | S0 | 0.603 | 0.576 | 0.330 | 0.90 (0.80, 0.96) |
| | FULL | S0 | 0.603 | 0.578 | 0.335 | 0.90 (0.79, 0.96) |
| | iT-EB | S0 | 0.614 | 0.588 | 0.356 | 0.91 (0.81, 0.97) |
| | iTransformer | S0 | 0.616 | 0.589 | 0.304 | 0.92 (0.83, 0.97) |
| | PET-M | S0 | 0.611 | 0.576 | 0.319 | 0.91 (0.80, 0.96) |
| | BASE | S0 | 0.609 | 0.579 | 0.256 | 0.91 (0.81, 0.96) |
| | GBM | S0 | 0.602 | 0.590 | 0.384 | 0.97 (0.93, 0.99) |
| | PET-E | S1 | 0.619 | 0.591 | 0.346 | 0.95 (0.88, 0.98) |
| | FULL | S1 | 0.619 | 0.593 | 0.348 | 0.95 (0.87, 0.98) |
| | iT-EB | S1 | 0.631 | 0.602 | 0.360 | 0.95 (0.88, 0.98) |
| | iTransformer | S1 | 0.624 | 0.595 | 0.286 | 0.95 (0.87, 0.98) |
| | PET-M | S1 | 0.620 | 0.591 | 0.339 | 0.96 (0.90, 0.98) |
| | BASE | S1 | 0.619 | 0.590 | 0.287 | 0.95 (0.88, 0.98) |
| | GBM | S1 | 0.609 | 0.586 | 0.362 | 0.95 (0.88, 0.98) |

**Table S9.** Additive-gate PET-A and the proxy-favorability and attention-bias variants of the physics module, in the same settings and with the same metrics as Table 3. PET-M, the proxy-favorability module; PET-A, the additive gate; Bias only, the learned attention bias alone; DC, the direction-constrained form. The two attention-bias-only variants (learned and direction-constrained) were trained only in the 2025–2026 season.

| Season | Condition | Method | CSI | F1 score | Brier score | Alarms at observed leaf temp. > 0 °C | Alarms at observed leaf temp. ≥ 2 °C |
|---|---|---|---|---|---|---|---|
| 2025–2026 | sensor-free | PET-M | 0.505 | 0.671 | 0.1118 | 0.159 | 0.064 |
| | | PET-A | 0.501 | 0.667 | 0.1256 | 0.187 | 0.083 |
| | | Bias only | 0.499 | 0.665 | 0.1260 | 0.183 | 0.082 |
| | | PET-M (DC) | 0.505 | 0.671 | 0.1130 | 0.162 | 0.065 |
| | | Bias only (DC) | 0.506 | 0.672 | 0.1180 | 0.186 | 0.082 |
| | sensor-equipped | PET-M | 0.596 | 0.743 | 0.0915 | 0.129 | 0.048 |
| | | PET-A | 0.604 | 0.750 | 0.1041 | 0.149 | 0.059 |
| | | Bias only | 0.601 | 0.748 | 0.0994 | 0.153 | 0.061 |

| Season | Condition | Method | CSI | F1 score | Brier score | Alarms at observed leaf temp. > 0 °C | Alarms at observed leaf temp. ≥ 2 °C |
|---|---|---|---|---|---|---|---|
| | | PET-M (DC) | 0.600 | 0.747 | 0.0916 | 0.140 | 0.053 |
| | | Bias only (DC) | 0.602 | 0.749 | 0.0953 | 0.157 | 0.063 |
| 2024–2025 | sensor-free | PET-M | 0.587 | 0.739 | 0.1137 | 0.112 | 0.042 |
| | | PET-A | 0.554 | 0.713 | 0.1251 | 0.151 | 0.068 |
| | | PET-M (DC) | 0.583 | 0.736 | 0.1145 | 0.108 | 0.041 |
| | sensor-equipped | PET-M | 0.658 | 0.792 | 0.0909 | 0.094 | 0.031 |
| | | PET-A | 0.654 | 0.789 | 0.1004 | 0.109 | 0.041 |
| | | PET-M (DC) | 0.658 | 0.792 | 0.0913 | 0.086 | 0.027 |

**Table S10.** Station characteristics and data availability for the 19 AAFOS stations used in this study

| ID | Station | Latitude (°N) | Longitude (°E) | Elevation (m) | First target date | Last target date | Samples | Freezing (%) |
|---|---|---|---|---|---|---|---|---|
| AAFOS01 | Yeongok | 37.8516 | 128.8192 | 8 | 2021-10-02 | 2026-04-18 | 24,117 | 11.8 |
| AAFOS02 | Daegwallyeong | 37.6771 | 128.7183 | 772 | 2021-10-02 | 2026-04-18 | 22,744 | 35.7 |
| AAFOS03 | Jinbu | 37.6479 | 128.5645 | 542 | 2021-11-26 | 2026-04-18 | 23,036 | 21.0 |
| AAFOS04 | Bongpyeong | 37.6123 | 128.3772 | 552 | 2021-11-27 | 2026-04-18 | 23,053 | 24.3 |
| AAFOS05 | Myeonon | 37.5620 | 128.3776 | 533 | 2021-11-27 | 2026-04-18 | 22,741 | 34.5 |
| AAFOS06 | Pyeongchang | 37.3775 | 128.3947 | 303 | 2021-11-27 | 2026-04-18 | 23,111 | 31.2 |
| AAFOS07 | Sinbuk | 37.9546 | 127.7763 | 140 | 2023-01-03 | 2026-04-18 | 17,208 | 22.5 |
| AAFOS08 | Daehwa | 37.5455 | 128.4411 | 562 | 2022-04-06 | 2026-04-18 | 19,736 | 19.0 |
| AAFOS09 | Daehwa 2 | 37.5366 | 128.4361 | 604 | 2022-04-06 | 2026-01-19 | 17,076 | 25.1 |
| AAFOS10 | Bukgangneung | 37.8046 | 128.8553 | 75 | 2023-01-04 | 2026-04-18 | 15,971 | 7.8 |
| AAFOS11 | Gangneung | 37.7515 | 128.8910 | 27 | 2023-01-04 | 2026-04-18 | 17,144 | 6.4 |
| AAFOS12 | Bukchuncheon | 37.9474 | 127.7544 | 96 | 2023-02-04 | 2026-04-18 | 14,963 | 25.8 |
| AAFOS13 | Hwadong | 36.3721 | 127.8949 | 360 | 2024-10-25 | 2026-04-18 | 9,017 | 13.4 |
| AAFOS14 | Hyeonseo | 36.2794 | 128.8960 | 316 | 2024-10-26 | 2026-04-18 | 8,597 | 19.9 |
| AAFOS15 | Sabeolguk | 36.4734 | 128.2246 | 84 | 2025-10-03 | 2026-04-18 | 4,705 | 14.0 |
| AAFOS16 | Andong | 36.5729 | 128.7073 | 141 | 2025-10-02 | 2026-04-18 | 4,746 | 13.4 |
| AAFOS17 | Seoul | 37.5714 | 126.9658 | 86 | 2025-10-02 | 2026-04-18 | 4,746 | 7.1 |
| AAFOS18 | Suwon | 37.2575 | 126.9830 | 40 | 2025-10-02 | 2026-04-18 | 4,746 | 14.7 |
| AAFOS21 | Gangneung-Wangsan | 37.6104 | 128.7730 | 687 | 2025-12-24 | 2026-04-18 | 2,684 | 27.2 |

Samples are the final forecast samples with a complete 24-h input window and all 12 target intervals (§3.4), totalling 280,141 over the five frost seasons. Freezing (%) is the percentage of these samples whose first target interval is labeled as surface freezing (§3.2). First and last target dates refer to the start of the first target interval (t + 1 h) in the earliest and latest retained samples, respectively; observation records begin earlier.

# Supplementary figures

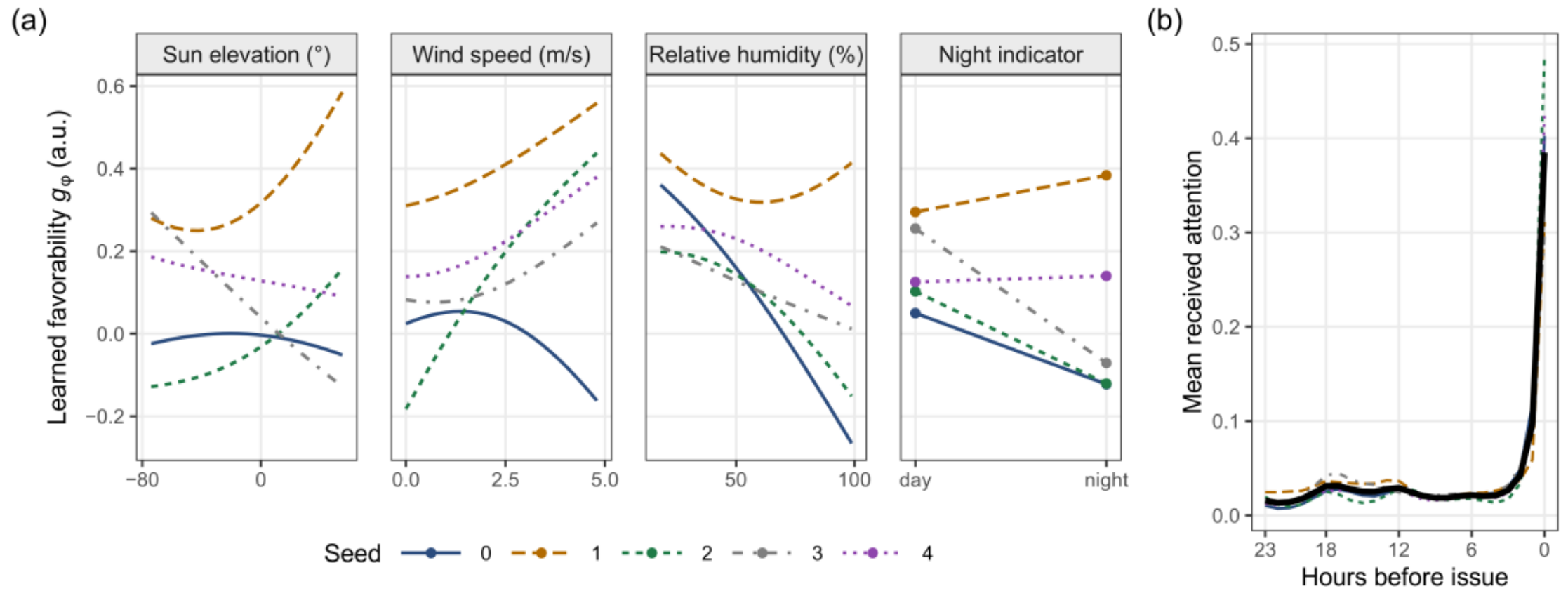

*Figure S1. Learned favorability and received attention in PET-M (sensor-equipped, five seeds, all test windows). (a) Partial dependence of the favorability network $g_\varphi$ on each of its four inputs, with the other inputs kept at their sampled values; one line per seed. (b) Mean attention received by each input hour, averaged over layers, heads, and test windows; one line per seed and the across-seed mean in black. On the horizontal axis, 0 is the issue time (the last hour of the input window).*

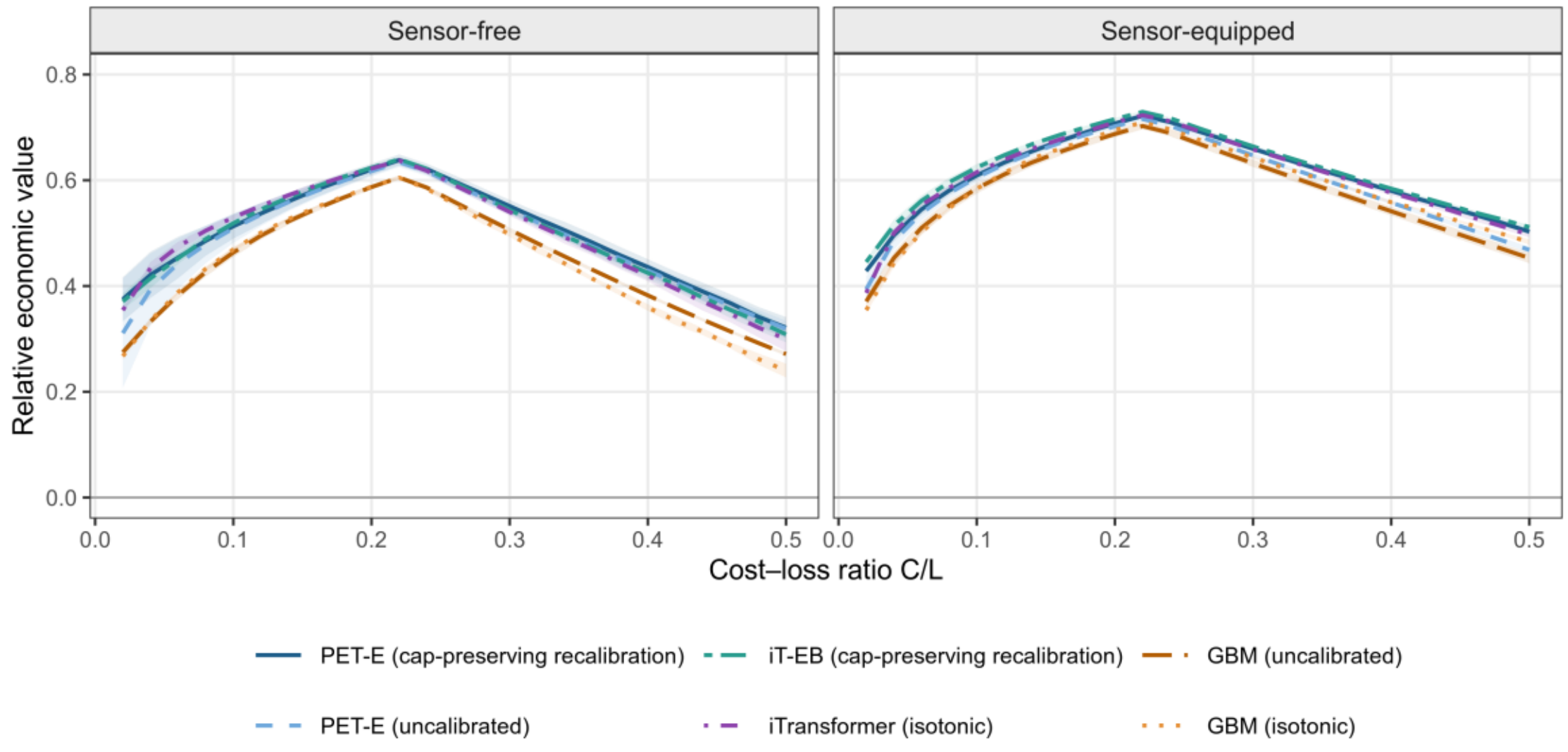

*Figure S2. Relative economic value in the 2025–2026 test season (seed mean; shading, seed standard deviation). PET-E and iT-EB are shown with cap-preserving recalibration and iTransformer and GBM with isotonic recalibration; uncalibrated curves are added for PET-E and GBM.*

# Supplementary notes

**Note S1. Cloud term of the energy-balance path**

Crawford and Duchon (1999) estimated the cloud fraction in the daytime from the ratio of observed to clear-sky solar radiation. Because solar radiation is unavailable at night, we use an effective cloudiness proxy formed as the mean of two observed signals. The first is the air-temperature fall rate $\rho_j = (T_{a,j-\Delta_j} - T_{a,j})/\Delta_j$ (°C h$^{-1}$; for $j \geq 2$, an end-point difference over $\Delta_j = \min(3, j-1)$ hours, with $\rho_1 = 0$ at the first time step of the input window and negative values under warming), which uses the faster cooling of clearer nights. The second is the relative humidity: near saturation, cloud or fog is taken to be present. The function clip truncates values to [0, 1], and $\rho_0$ and $H_0$ are calibration parameters.

**Note S2. Training details and the training and decision procedure**

**Training details.** The sample weight $v_{n\ell}$ is 1 in all models except the transition-weighted variants (§3.7), where $v_{n\ell} = 1 + \gamma\,\mathbb{1}[y_{n\ell} \neq s_n]$ with $\gamma = 3$; $\mathbb{1}[\cdot]$ is the indicator function and $s_n$ the freezing state at issue time. Because Eq. (11) is a function of $P_{n\ell}$, the multiplicative form (Eq. (8)) and the additive form (Eq. (9)) share the same objective, and the difference between the two forms is confined to the gate structure. For the multiplicative form, $P_{n\ell}$ was clipped to $[10^{-6}, 1-10^{-6}]$ for numerical stability when computing Eq. (11). We optimized with AdamW (Loshchilov and Hutter, 2019). We used a learning rate of $10^{-3}$, weight decay of $10^{-4}$ (applied to all parameters, including the gate and calibration parameters), a batch size of 256, and at most 30 epochs, and clipped the gradient norm to 1.0. For classification models, early stopping used the positive-class-weighted validation binary cross-entropy, excluding the auxiliary regression term, the transition weights, and the physics penalty. If this criterion did not improve for 6 consecutive epochs, training stopped and the parameters at its minimum were restored; the two-stage regression exception is specified below Algorithm 1 in this note. The learned models were trained with seeds 0–4 (logistic regression once; §3.7), and LOSO models with seeds 0–2 for each fold (§4). The procedure from training to decision is summarized in Algorithm 1 below.

**Algorithm 1.** Training and decision procedure for the transformer family (per condition and seed)

1. Standardize the input features with training-period statistics and clip the standardized values at an absolute value of 8. Set $\omega_+ = (1-\pi)/\pi$ from the training positive fraction $\pi$.
2. Initialize the parameters with the seed and set up AdamW (lr $10^{-3}$, weight decay $10^{-4}$).
3. For each epoch 1–30:
    - For each random minibatch of size 256, compute $u_\ell$, $\hat{T}_\ell$, and $P_\ell$ by Eqs. (2)–(10) (for iTransformer, the architecture of §3.7 (ix)), and update the parameters after clipping the gradient norm of the loss in Eq. (12) to 1.0.
    - Compute the validation loss (positive-class-weighted binary cross-entropy; §3.5 (e)); save the parameters if it is a new minimum, and stop if there is no improvement for 6 consecutive epochs.
4. Restore the saved parameters and compute the validation and test probabilities $P_\ell$.

5. For each lead, choose $\theta_\ell^*$ maximizing the validation CSI, and compute the metrics of Eq. (13) and the onset–cessation metrics with the test decision $\mathbb{1}[P_\ell \geq \theta_\ell^*]$.
6. (§3.9) For each lead, fit $\mathcal{I}_\ell$ of Eq. (14) on the validation set and apply it to the test probabilities; at each $\alpha$, apply the rule "protect if the probability is at least $\alpha$" to the uncalibrated $P_\ell$ and to the recalibrated probability ($\mathcal{I}_\ell(P_\ell)$, or $\tilde{P}_\ell$ for the multiplicative form) and compute REV (Eq. (15)).

The two-stage model has no gate or attention bias; in step 3 it uses only the leaf-temperature regression error for training and early stopping, and in step 5 it thresholds the score $-\hat{T}_\ell$. The loss-embedding and transition-weighted variants change only the training loss of step 3 as in §3.7.

**Note S3. Transition scale and attention bias**

Because the additive probability depends on $(w, \tau)$ only through the ratio $\psi$, the only reportable transition scale is $\delta$. The regression term of Eq. (12) anchors $\hat{T}$ to the observed leaf temperature in °C. In models that use Eq. (7), the reference $T_{a,24} - D_{24}$, expressed in °C, also breaks the symmetry under joint rescaling of the temperature head and $\delta$ described in Remark A.1; this does not identify $\delta$ separately from the data path (Remark A.1). $\delta$ is an effective transition scale conditioned on the data path, and its learned value can depend on the initial value (§5.9). Applying the attention bias is equivalent to multiplying the bias-free attention weight $a_{ij}^0$ by the exponential of the key prior, $e^{R_j}$, and then renormalizing. This relation can be written as $a_{ij} \propto a_{ij}^0 e^{R_j}$ (Remark A.2). Proofs are in Appendix A. Both gate forms impose their constraint in the architecture, which Beucler et al. (2021) call a hard constraint. Their content differs: the additive form is a **soft prior** that the data path can override (Proposition 4(ii)), and the multiplicative form is a **bounded modulation** with a cap. The per-sample threshold $T^*$ of the additive form is not fixed at 0 °C. The physics factor $\sigma(-\hat{T}/\delta)$ of the multiplicative form is 1/2 at $\hat{T} = 0$. Total-derivative monotonicity with respect to the raw input is outside the scope of these propositions.

**Note S4. Baseline and backbone details**

**Flattened-feature baselines (iv).** LightGBM was trained as a single model for all leads, stacking sample × lead as rows and adding lead one-hot variables. Its settings were fixed in advance (num_leaves 63, learning rate 0.05, at least 50 samples per leaf, feature and bagging fractions of 0.8 with bagging at every iteration, positive-class weight $(1 - \pi)/\pi$, and at most 600 trees with early stopping after 50 rounds without improvement in the validation log loss) and not tuned. Logistic regression predicted all leads with one weight matrix and was fitted by full-batch Adam (learning rate 0.05, 300 iterations, L2 penalty $10^{-4}$, positive-class weight $(1 - \pi)/\pi$, with $\pi$ taken as the training positive fraction at lead 1 for this model).

**iTransformer (ix).** Each variable's 24-h history formed one token. The encoder had two layers, a representation dimension of 64, four attention heads, a feed-forward dimension of 128, and dropout of 0.1, followed by a final layer normalization. The token outputs, kept in variable order, were flattened and concatenated with a 32-dimensional representation of the 15 static inputs. Two output heads, each with 96 hidden units, produced 12 frost logits and 12 leaf-temperature predictions. The sensor-free and sensor-equipped versions of the plain form had 250,936 and 287,800 trainable parameters, respectively. To preserve the absolute temperature level, the training-statistic standardization and clipping of §3.4 were applied, and per-window mean removal was not used. The objective was the same as for BASE (Eq. (12), $\eta = 0.5$), and the optimizer, batch size, maximum number of epochs, and early-stopping rule were the same as in Supplementary Note S2. The settings were fixed before inspecting the test scores of this model. Three forms of the same backbone were compared: the plain form without physics (iTransformer), the form with only the multiplicative gate attached (iT+gate), and the form with the gate and the energy-balance leaf-temperature path attached (iT-EB). The module adds only four parameters ($\xi_\delta$, $\xi_\rho$, $\xi_H$, $\xi_k$), and with the same seed the remaining initial weights are identical to those of the plain form. The time-axis attention prior was not included because it does not apply to a variate-token backbone.

**Gate-parameter check.** To check whether the gate parameter has a physical meaning, PET-M and PET-E were trained in the sensor-equipped condition with the initial value of $\delta$ set to the default (0.99 °C), 0.5, and 2.0, and the learned $\delta$ and test CSI were compared. The calibration parameters of the energy-balance path ($\rho_0$, $H_0$, $k_h$) and, for FULL, the attention coefficient $\kappa$ are reported before and after training as ranges across seeds. To check whether the learned parameters bring the depression calculation closer to observations, we used the issue-time samples under precipitation-free nighttime conditions in the 2025–2026 test season. On these samples, $D$ computed with the initial parameter values and the mean of the $D$ values computed with the learned parameters of each of the five seeds of sensor-free PET-E were compared with the observed air-minus-leaf temperature difference by Pearson correlation and by mean. To check whether the learned favorability follows the intended radiative-cooling response, the response direction of $g_\varphi$ of PET-M to each input was also compared across seeds. To check whether the gain from the physics mechanisms grows when training data are scarce, PET-E, PET-M, BASE, and PET-A were retrained with the training data reduced to the most recent 1, 2, or 3 seasons.

**Direction-constrained favorability.** The direction-constrained form gives the night indicator a positive coefficient and wind speed and sun elevation negative coefficients (magnitudes made positive by softplus), so that it increases at night and decreases in wind speed and sun elevation, and it represents relative humidity by an unconstrained small MLP (one hidden layer of 8 units). The three coefficient magnitudes are initialized at softplus(0) = 0.693.

**Note S5. Static-input candidates and selection in LOSO**

**Static-input candidates in LOSO (Experiment (iv)).** The candidates were all 15 variables (S0) and no static input (S1); for PET-M, BASE, and GBM in the sensor-free condition, a three-variable set (S2: elevation, distance to coast, and distance to inland water) was also included. S0 and S1 were run for all models of (iii) in both sensor conditions. The test season was not used for candidate selection. For each of the 14 trained stations, the fold model that excluded that station predicted the station's validation period. The lead-specific decision thresholds were set on the validation data of the other stations in that fold. For each seed, the predictions of the 14 stations were pooled, the CSI was computed from the pooled hit, false-alarm, and miss counts at each lead, and the result was averaged over leads. Candidate selection used the seed mean of this score. A candidate replaced S0 only if its improvement over S0 exceeded twice the seed standard error; if both candidates met this criterion, the one with the higher mean CSI was selected. Selection was made separately for each model, and the rule was fixed before the sensor-free analysis was run. The selected static-input configuration was then applied across all 19 LOSO folds for that model and sensor condition. Configuration selection therefore used validation labels from the 14 trained stations, although each station's validation predictions were generated by a fold model that excluded it. This procedure did not use test-season labels, but it was not a nested selection procedure that excluded the target station from every stage of configuration selection. Because there are only 19 stations, the main CSI differences were also checked by station-level resampling (2,000 replicates; the hits, false alarms, and misses of the resampled stations were summed and the CSI recomputed). The extent to which between-station differences in freezing carried into the predictions was measured by the Pearson correlation over the 19 stations between station freezing rate and station mean predicted probability (seed means), with an interval from the same station resampling. The same validation-based selection rule (S0, S1) was applied to sensor-equipped LOSO, and its test-season results were compared in the same way.

**Note S6. Data provenance, weather-grid extraction, and input standardization**

**Raw sensor files (§3.1).** The raw data combine an archive of 2021–2024 logger files with current logger files in time order; values agreed exactly where the two sources overlapped.

**Weather-grid extraction (§3.3).** Intervals missing from the primary extraction (before March 2023 and in parts of later winters) were taken from a second extraction of the same 5-min fields at the same grid coordinates. In a test in which a date covered by the primary extraction was re-extracted with the second method, all extracted variables agreed exactly with the primary extraction. Wind direction was not included in the second extraction and was therefore excluded from the features for full-period consistency.

**Input standardization (§3.4).** Inputs were standardized with training-period statistics, and the standardized values were clipped at an absolute value of 8. The physical variables of the energy-balance path (§3.5; air temperature, dew point, wind speed, relative humidity, sun elevation, and precipitation) are passed as a separate input standardized and clipped in the same way and are converted back to physical units inside the model with the training-period statistics before Eqs. (3)–(6) are applied. In the test season, precipitation was the only physical variable with clipped values, and only whether it is zero enters the calculation, so the clipping does not affect the depression. In the training period, wind speed was also clipped in some strong-wind hours, where $h_{c,j}$ is large and the depression is close to zero.

## Appendix A. Proofs

**Proof of Proposition 1.** Setting $\psi = w/\tau > 0$, $\delta = \tau/w = 1/\psi$ in Eq. (9),

$$\mathrm{logit}_\ell = u_\ell - \psi \hat{T}_\ell = \frac{\delta u_\ell - \hat{T}_\ell}{\delta} = \frac{T_\ell^* - \hat{T}_\ell}{\delta}, \qquad T_\ell^* = \delta u_\ell \tag{A.1}$$

so $P_\ell = \sigma\big((T_\ell^* - \hat{T}_\ell)/\delta\big)$ for every input. Because $\sigma$ is strictly increasing and $\sigma(0) = 1/2$, $P_\ell = 1/2$ holds only at $\hat{T}_\ell = T_\ell^*$. With $u_\ell$ (hence $T_\ell^*$) fixed, for the same reason $\left|\hat{T}_\ell - T_\ell^*\right| \le \delta$ is equivalent to $P_\ell \in [\sigma(-1), \sigma(1)] \approx [0.27, 0.73]$, so the width of the $\hat{T}_\ell$ interval over which the probability changes between approximately 0.27 and 0.73 is $2\delta$. ▫

**Proof of Proposition 2.** With $u_\ell$ fixed, from Eq. (A.1) $\partial\, \mathrm{logit}_\ell / \partial \hat{T}_\ell = -\psi$, and since $\sigma'(x) = \sigma(x)\{1 - \sigma(x)\}$, by the chain rule

$$\frac{\partial P_\ell}{\partial \hat{T}_\ell} = \sigma'(\mathrm{logit}_\ell)\, \frac{\partial\, \mathrm{logit}_\ell}{\partial \hat{T}_\ell} = -\psi\, P_\ell(1 - P_\ell) \tag{A.2}$$

In the parameterization of Eq. (9), $w > 0$ and $\tau > 10^{-2}$ for all $\xi_w$, $\xi_\tau$, so $\psi > 0$, and since $P_\ell \in (0,1)$, Eq. (A.2) is negative at any parameter value during training. This is the partial derivative with respect to the gate path. Because $u_\ell$ and $\hat{T}_\ell$ are functions of the same representation $\boldsymbol{h}$ (Eq. (2)), the sign of the total derivative with respect to the raw input does not follow from this result. ▫

**Proof of Proposition 3.** With the backbone and the two heads' parameters fixed, $u_\ell$ and $\hat{T}_\ell$ are finite for each input, and from Eq. (A.1) $\mathrm{logit}_\ell = u_\ell - \psi \hat{T}_\ell$. (i) If $\psi \to 0^+$, then $\mathrm{logit}_\ell \to u_\ell$ and, since $\sigma$ is continuous, $P_\ell \to \sigma(u_\ell)$. This is the probability of the gate-free classifier of the same backbone, corresponding to BASE if there is no attention bias and to the attention-bias-only variant if there is. (ii) If $\psi \to \infty$, then $\mathrm{logit}_\ell = u_\ell + \psi \left|\hat{T}_\ell\right| \to +\infty$ when $\hat{T}_\ell < 0$ and $\mathrm{logit}_\ell \to -\infty$ when $\hat{T}_\ell > 0$, so

$$\lim_{\psi \to \infty} P_\ell = \mathbb{1}[\hat{T}_\ell < 0] \qquad (\hat{T}_\ell \ne 0) \tag{A.3}$$

and if $\hat{T}_\ell = 0$ then $P_\ell = \sigma(u_\ell)$. Eq. (A.3) is the rule that applies a 0 °C hard threshold to the predicted leaf temperature; unlike the classical rule of §3.7 (iii), it has no wetness term. Fixing $\tau$ in Eq. (9) and sending $\xi_w \to -\infty$ or $+\infty$ makes $w = \mathrm{softplus}(\xi_w)$ go to $0^+$ or $\infty$, so the two limits are approached within the implementation's parameterization but not reached. Also, these limits concern the functional form and do not mean that the separately trained two-stage baseline (without gate or attention bias) shares the same training procedure. ▫

**Proof of Proposition 4.** (i) Since $0 < \sigma(x) < 1$, $P_\ell = \sigma(u_\ell)\, \sigma(-\hat{T}_\ell/\delta)$ is smaller than each of its two factors:

$$P_\ell \le \min\{\sigma(u_\ell),\ \sigma(-\hat{T}_\ell/\delta)\} \qquad \text{for all } u_\ell \in \mathbb{R}. \tag{A.4}$$

The second term on the right is independent of $u_\ell$ and is therefore a uniform upper bound. If $\hat{T}_\ell = 2\delta$, the bound is $\sigma(-2) \approx 0.119$. (ii) Fix $\hat{T}$ and $p_0 \in (0,1)$ and let $u^* = \hat{T}/\delta_a + \mathrm{logit}(p_0)$.

Since $\sigma$ is strictly increasing, for all $u > u^*$, $P_a = \sigma(u - \hat{T}/\delta_a) > \sigma(\text{logit}(p_0)) = p_0$. Hence $\sup_u P_a = 1$, and there is no upper bound below 1 determined by $\hat{T}$. (iii) With $u_\ell$ fixed,

$$\frac{\partial P_\ell}{\partial \hat{T}_\ell} = -\frac{1}{\delta}\,\sigma(u_\ell)\,\sigma(-\hat{T}_\ell/\delta)\,\{1 - \sigma(-\hat{T}_\ell/\delta)\} < 0 \tag{A.5}$$

because in Eq. (8) $\delta > 0$ and the three factors are all positive. ▫

**Proof of Corollary 1.** (i) The first property follows immediately from the definition of the minimum. (ii) Fix $u_\ell$ and let $\hat{T}_\ell < \hat{T}'_\ell$. By Eq. (A.5), $P_\ell(\hat{T}_\ell) > P_\ell(\hat{T}'_\ell)$, and since $\mathcal{J}$ is non-decreasing, $\mathcal{J}(P_\ell(\hat{T}_\ell)) \geq \mathcal{J}(P_\ell(\hat{T}'_\ell))$. Also $\sigma(-\hat{T}_\ell/\delta) > \sigma(-\hat{T}'_\ell/\delta)$. Since $A \geq A'$ and $B \geq B'$ imply $\min\{A, B\} \geq \min\{A', B'\}$,

$$\tilde{P}_\ell(\hat{T}_\ell) \geq \tilde{P}_\ell(\hat{T}'_\ell) \tag{A.6}$$

Because the $\mathcal{J}$ fitted by PAVA can have flat segments, strict decrease is not guaranteed. ▫

**Proof of Proposition 5.** (i) Since $0 < \sigma(x) < 1$, from Eq. (8) $P_\ell = \sigma(u_\ell)\,\sigma(-\hat{T}_\ell/\delta) < \sigma(-\hat{T}_\ell/\delta)$. If $P_\ell \geq \theta$, then $\sigma(-\hat{T}_\ell/\delta) > \theta$, and since $\sigma$ is strictly increasing and $\sigma^{-1}(\theta) = \ln\{\theta/(1-\theta)\}$, $-\hat{T}_\ell/\delta > \ln\{\theta/(1-\theta)\}$. Since $\delta > 0$, $\hat{T}_\ell < \delta\ln\{(1-\theta)/\theta\}$, and substituting $\hat{T}_\ell = T_{a,24} - D_{24} + r_\ell$ from Eq. (7) gives the claim.

(ii) Under precipitation-free nighttime conditions, the denominator $4\sigma_{\text{SB}}T_K^3 + h_c$ of Eq. (6) is positive, and rearranging Eq. (4) gives $1 - \varepsilon = (1 - c)(1 - \varepsilon^{\text{c}})$. Since $c \in [0,1]$ and $\varepsilon^{\text{c}} \leq 1$, $1 - \varepsilon \geq 0$, and hence $D \geq 0$. Regarding $D$ as a function of $(T_K, T_d, U, H, \rho)$, we show the result for each argument in turn, with the others held fixed. (a) $U$: $h_c = k_h\sqrt{\max(U, 0.1)}$ in Eq. (5) is non-decreasing in $U$ ($k_h > 0$), and the numerator of $D$ does not depend on $U$ and is non-negative, so $D$ is non-increasing in $U$. (b) $T_d$: in $e = 6.112\exp\{17.67T_d/(T_d + 243.5)\}$ of Eq. (3), $\frac{d}{dT_d}\frac{17.67T_d}{T_d+243.5} = \frac{17.67\times243.5}{(T_d+243.5)^2} > 0$ ($T_d > -243.5$), so $e$ is strictly increasing in $T_d$. Since $x \mapsto \min\{1, 1.24(x/T_K)^{1/7}\}$ is non-decreasing for $x > 0$, $\varepsilon^{\text{c}}$ is non-decreasing in $T_d$. In $1 - \varepsilon = (1 - c)(1 - \varepsilon^{\text{c}})$, $1 - c \geq 0$, so $1 - \varepsilon$ is non-increasing in $T_d$, and since the denominator does not depend on $T_d$, $D$ is also non-increasing. (c) $H$: since $H_0 \in [50,95]$, $(H - H_0)/(100 - H_0)$ is strictly increasing in $H$, and since the clip is non-decreasing, $c$ in Eq. (4) is non-decreasing in $H$. In $1 - \varepsilon = (1 - c)(1 - \varepsilon^{\text{c}})$, $1 - \varepsilon^{\text{c}} \geq 0$, so $1 - \varepsilon$ is non-increasing in $c$, and hence $D$ is non-increasing in $H$. (d) $\rho$: since $\rho_0 > 0$, $1 - \rho/\rho_0$ is decreasing in $\rho$, and $c$ is non-increasing in $\rho$. By the same argument as in (c), $D$ is non-decreasing in $\rho$. The upper bound follows from $1 - \varepsilon \leq 1 - \varepsilon^{\text{c}}$ and $h_c \geq k_h\sqrt{0.1}$. This argument uses only compositions of monotone functions, so it holds also at the points where the clip and the min are not differentiable.

(iii) With $r_\ell$ fixed, the bound $D_{24} - r_\ell + \delta\ln\{(1-\theta)/\theta\}$ in (i) is increasing in $D_{24}$, so each monotonicity in (ii) carries over to the bound in the same direction. ▫

**Remark A.1 (Scale of $\delta$).** (i) The probability of Eq. (9) depends on $(w, \tau)$ only through $\psi = w/\tau$. For $\zeta > 0$, $(w, \tau) \mapsto (\zeta w, \zeta\tau)$ leaves all logits unchanged (within the implementation range $\tau > 10^{-2}$, for $\zeta$ with $\zeta\tau > 10^{-2}$), so $w$ and $\tau$ are not separately determined and the only reportable additive transition scale is $\delta = 1/\psi$. (ii) In both gate forms the probability depends on $\hat{T}_\ell$ and $\delta$ only through $\hat{T}_\ell/\delta$. In the models without the reference of Eq. (7) (gate only, ATT, PET-M, PET-A, iT+gate), scaling the last linear-layer weights and bias of the leaf-temperature head by $\zeta > 0$ gives $\hat{T}_\ell \mapsto \zeta\hat{T}_\ell$, and setting jointly $\delta \mapsto \zeta\delta$ ($\zeta\delta < 10^3$) in the multiplicative form and $w \mapsto w/\zeta$ in the additive form leaves all $P_\ell$ and the $\mathcal{L}_{\mathrm{cls}}$ of Eq. (11) unchanged. The regression term of Eq. (12), in contrast, generally changes if $\zeta \neq 1$. Thus, in these models, $\delta$ has °C units because the regression term with $\eta > 0$ ties $\hat{T}$ to the observed leaf temperature. In the models that use Eq. (7) (PET-E, FULL, iT-EB), scaling the head by $\zeta$ changes only the residual $r_\ell$ while the reference $T_{a,24} - D_{24}$ in °C stays unchanged, so this symmetry does not hold, and the scale of $\delta$ is fixed jointly by the reference and the regression term. (iii) Even with the scale fixed by the regression term, $\delta$ is not identified separately from the data path. Because the data path $u_\ell$ also comes from the same representation $\boldsymbol{h}$, it can absorb part of the change in $\delta$. $\delta$ is therefore interpreted only as an effective scale conditioned on the data path, and its learned value is reported together with the initial value (§3.5) in §5.9.

**Remark A.2 (Attention bias as a multiplicative prior).** From Eq. (10) and the definition of softmax,

$$a_{ij} = \frac{\exp(q_i k_j^{\top}/\sqrt{d_k})\, e^{R_j}}{\sum_{j'} \exp(q_i k_{j'}^{\top}/\sqrt{d_k})\, e^{R_{j'}}} = \frac{a_{ij}^0\, e^{R_j}}{\sum_{j'} a_{ij'}^0\, e^{R_{j'}}} \tag{A.7}$$

where $a^0$ is the bias-free attention weight and the second equality is obtained by dividing numerator and denominator by $\sum_{j'} \exp\left(q_i k_{j'}^{\top}/\sqrt{d_k}\right)$. Adding the same constant $\nu$ to all $j$ multiplies numerator and denominator by the same $e^{\nu}$, so $R_j \mapsto R_j + \nu$ does not change $a$. The implementation's subtraction of the input-window mean from $R_j$ is such a shift. A global scale factor $\lambda$ on $R_j$, unlike a shift, changes $a_{ij}$, but $\lambda$ is absorbed into $\kappa$ in ATT and FULL and into the output scale of the learned $g_\varphi$ in PET-M.

**Remark A.3 (Cap and alarms).** The cap of Proposition 4 acts only to lower the probability. For samples with $\hat{T} > 0$, $P \leq \sigma(-\hat{T}/\delta) < 1/2$ is enforced, so the multiplicative form structurally compresses the probability in the range where the predicted surface is warm. At the decision threshold $\theta_\ell^*$, alarms are excluded for samples with $\hat{T}_\ell \geq \delta \ln\{(1-\theta_\ell^*)/\theta_\ell^*\}$ (§3.6). In the experiments as well, of the three physics variants of the 2×2 ablation evaluated in the sensor-free condition of the 2025–2026 season (PET-M, gate only, and attention bias only), the two with the multiplicative gate significantly lowered, relative to BASE, the fraction of all alarms issued at times when the observed leaf temperature was 2 °C or higher (§5.3, §5.8). This bound provides a structural interpretation of the lower fraction of alarms at observed leaf temperatures ≥ 2 °C; it does not by itself imply improved detection.

## References

Beucler, T., Pritchard, M., Rasp, S., Ott, J., Baldi, P., Gentine, P., 2021. Enforcing analytic constraints in neural networks emulating physical systems. Phys. Rev. Lett. 126, 098302. https://doi.org/10.1103/PhysRevLett.126.098302

Crawford, T.M., Duchon, C.E., 1999. An improved parameterization for estimating effective atmospheric emissivity for use in calculating daytime downwelling longwave radiation. J. Appl. Meteorol. 38, 474–480. https://doi.org/10.1175/1520-0450(1999)038<0474:AIPFEE>2.0.CO;2

Loshchilov, I., Hutter, F., 2019. Decoupled weight decay regularization. In: International Conference on Learning Representations (ICLR). https://openreview.net/forum?id=Bkg6RiCqY7 (accessed 10 November 2026)

Murphy, A.H., 1973. A new vector partition of the probability score. J. Appl. Meteorol. 12, 595–600. https://doi.org/10.1175/1520-0450(1973)012<0595:ANVPOT>2.0.CO;2